\documentclass[a4paper,fleqn,usenatbib]{mnras}

\usepackage{txfonts}

\usepackage[T1]{fontenc}
\usepackage{ae,aecompl}

\usepackage{graphicx}	
\usepackage{amstext}	
\usepackage{amssymb}	
 
\title[Probing IGM properties using blazars]{Probing the physical properties of the intergalactic medium using blazars}

\author[T. Dalton et al]{
Tony Dalton,$^{1}$\thanks{E-mail:tonydalton@live.ie}
Simon L. Morris,$^{1}$
Michele Fumagalli,$^{2,3}$
and Efrain Gatuzz$^{4}$\\
$^{1}$Centre for Extragalactic Astronomy, Durham University, South Road, Durham DH1 3LE, UK\\
$^{2}$Dipartimento di Fisica `G. Occhialini', Universit\`a degli Studi di Milano-Bicocca, Piazza della Scienza 3, 20126 Milano, Italy\\
$^{3}$INAF - Osservatorio Astronomico di Trieste, via G. B. Tiepolo 11, 34143 Trieste, Italy \\
$^{4}$Max-Planck-Institut f\"ur extraterrestrische Physik, Gie{\ss}enbachstra{\ss}e 1, 85748 Garching, Germany\\
}

\date{Accepted 2021 September 8. Received 2021 September 8; in original form 2021 July 9 }

\pubyear{2015}

\begin{document}
\label{firstpage}
\pagerange{\pageref{firstpage}--\pageref{lastpage}}
\maketitle

\begin{abstract}

We use \textit{Swift} blazar spectra to estimate  the  key intergalactic medium (IGM) properties of hydrogen column density ($\mathit{N}\textsc{hxigm}$), metallicity and temperature over a redshift range of $0.03 \leq z \leq 4.7$,  using a collisional ionisation equilibrium (CIE)  model for the ionised plasma.  We adopted a conservative approach to the blazar continuum model given its intrinsic variability and use a range of power law models. We subjected our results to a number of tests  and found that the $\mathit{N}\textsc{hxigm}$ parameter was robust with respect to individual exposure data and co-added spectra for each source, and between \textit{Swift} and \textit{XMM-Newton} source data. We also found no relation between $\mathit{N}\textsc{hxigm}$ and variations in source flux or intrinsic power laws. Though some objects may have a bulk Comptonisation component which could mimic absorption, it did not alter our overall results. The $\mathit{N}\textsc{hxigm}$ from the combined blazar sample scales as 
$(1 + z)^{1.8\pm0.2}$.  
The mean hydrogen density at $z = 0$  is $n_0 = (3.2\pm{0.5}) \times 10^{-7}$ cm$^{-3}$. The mean IGM temperature over the full redshift range is log($T$/K) $= 6.1\pm{0.1}$, and the mean metallicity is $[X/$H$] = -1.62\pm0.04 (Z \sim 0.02)$. When combining with the results with a gamma-ray burst (GRB) sample, we find the results are consistent over an extended redshift range of $0.03 \leq z \leq 6.3$. Using our  model for blazars and GRBs, we conclude that the IGM contributes substantially to the total absorption seen in both blazar and GRB spectra.

\end{abstract}




\section{Introduction}\label{sec:1}
The main objective of this paper is to estimate key IGM parameters of column density, metallicity, and temperature, using a model for ionised absorption on the line of sight (LOS) to blazars. Our hypothesis is that there is significant absorption in the diffuse IGM, and that this IGM column density increases with redshift. Our approach is different to most other blazar studies which focus primarily on the intrinsic curvature of the X-ray spectral flux, or where works attribute to the host only, any spectral hardening due to excess absorption over our Galaxy \citep[e.g][]{Bottacini2010,Paliya2016,Ricci2017}. Instead, we focus on the possible absorption due to the IGM using a sophisticated highly ionised absorption component in addition to best fit intrinsic curvature models. We test the robustness of our result from a number of perspectives. Finally, we combine our blazar sample with an extended redshift GRB sample to enable cross tracer comparison.

Most baryonic matter resides in the IGM  \citep{McQuinn2016a}. Simulations predict that up to 50\% of the baryons by mass have been shock-heated into a warm-hot phase (WHIM) at $z< 2$, with $T = 10^5 - 10^7$ K and $n_b = 10^{-6} - 10^{-4}$ cm$^{-3}$ where $n_b$ is the baryon density \citep[e.g][]{Cen1999,Cen2006,Dave2007,   Schaye2015}.  \citet{Martizzi2019}, using the IllustrisTNG simulations \footnote{http://www.tng-pro ject.org/}\citep{Piattella2018}, estimated that the cool diffuse IGM constitutes $\sim 39\%$ and the WHIM $\sim 46\%$ of the baryons at redshift $z = 0$.  Observations of the cool diffuse IGM and WHIM are required to trace matter across time and to validate the simulations \citep{Danforth2016}.  

 A significant fraction of the cool gas probed by strong Ly$\alpha$ forest systems (SLFSs): 15 < log$\mathit{N}_{\mathrm{H}\/\ \textsc{i}}$ < 16.2\footnote{Throughout this paper, logarithmic column densities are expressed in units of cm$^{-2}$}; partial Lyman Limit Systems (pLLSs): 16.2 < log$\mathit{N}_{\mathrm{H}\/\ \textsc{i}}$< 17.2; Lyman Limit Systems (LLSs):  17.2 < log$\mathit{N}_{\mathrm{H}\/\ \textsc{i}}$< 19 ; super-LLSs (sLLSs): 19.0 <log$\mathit{N}_{\mathrm{H}\/\ \textsc{i}}$< 20.3; and Damped Ly$\alpha$ Systems (DLAs) log$\mathit{N}_{\mathrm{H}\/\ \textsc{i}}$ >20.3 \citep{Fumagalli2014} has been associated with galaxy haloes and the circum-galactic medium (CGM) \citep{Pieri2014,Fumagalli2013,Fumagalli2016}. Over the last several decades, observations of redshifted Ly$\alpha$ absorption in the spectra of quasars has provided a highly sensitive probe of the cool IGM \citep[e.g.][]{Morris1991,York2000,Harris2016,Fumagalli2020a}.  At higher temperatures, for some time, the expected baryons were not detected in the WHIM, giving rise to the ``missing'' baryon problem \citep{Danforth2005,Danforth2008,Shull2012,Shull2014}. Recent literature points to the CGM as the reservoir for at least a fraction of this missing matter \citep{Tumlinson2011,Tumlinson2013,Werk2013,Lehner2016}. Other claims to have detected the WHIM include possible detection of $\mathrm{O\/\ \textsc{vii}}$ lines, excess dispersion measure over our Galaxy and the host galaxy in Fast Radio Bursts (FRB), using the thermal Sunyaev Zelodovich effect, and X-ray emission from cosmic web filaments  \citep[e.g.][]{Nicastro2018, Macquart2020,Tanimura2020b,Tanimura2020}.

Detection of the WHIM is extremely challenging, as its emission is very faint, it lacks sufficient neutral hydrogen to be seen via Ly$\alpha$ absorption in spectra of distant quasars, and the X-ray absorption signal expected from the WHIM is extremely weak \citep{Nicastro2018,Khabibullin2019}. The vast majority of hydrogen and helium is ionized in the IGM  post reionisation. Therefore, the observation of metals is essential for exploring the IGM properties including density, temperature and metallicity. Absorption-line studies in optical to UV, of individual systems that use the ionization states of abundant heavy elements, have been very successful  \citep[e.g.][]{Shull2014,Raghunathan2016,Selsing2016,Lusso2015}. However, most very highly ionised metals are not observed in optical to UV. Tracing individual features of the IGM metals in X-ray with current instruments is very limited.  Although the X-ray absorption cross-section is mostly dominated by metals, it is typically reported as an equivalent hydrogen column density (hereafter $\textit{N}\textsc{hx}$).

 Extremely energetic objects such as active galactic nuclei (AGN) and GRBs are currently some of the most effective tracers to study the IGM as their X-ray absorption provides information on the total absorbing column density of matter between the observer and the source \citep[e.g.][]{Galama2001a,Watson2007,Watson2011,Wang2013,Schady2017,Nicastro2017,Nicastro2018}. $\textit{N}\textsc{hx}$ consists of contributions from the host local environment, the IGM, and our own Galactic medium. The Galactic component is usually known from studies such as \citet[][hereafter W13]{Kalberla2005,Willingale2013}. 

 One of the main results of earlier studies of the IGM using high redshift tracers is the apparent increase in excess  of $\textit{N}\textsc{hx}$ with redshift \citep[e.g.][]{Behar2011,Watson2011,Campana2012}.  The cause of the $\mathit{N}\textsc{hx}$ rise with redshift seen in high redshift tracers has been the source of much debate over the last two decades. One school of thought argues that the object host accounts for all the excess and evolution \citep[e.g][]{Schady2011,Watson2013,Buchner2017} . The other school of thought argues that while the host can have an absorption contribution, the IGM contributes substantially to the excess absorption and is redshift related \citep[e.g][hereafter D20, D21]{Starling2013a,Arcodia2016,Rahin2019a,Dalton2020,Dalton2021}. The convention in earlier studies using AGN and GRBs was to use solar metallicity for a neutral absorber, as a device used to place all of the absorbing column density measurements on a comparable scale. These studies all noted that the resulting column densities were, therefore, lower limits as GRBs typically have much lower metallicities. D20 used realistic GRB host metallicities to generate improved estimates of $\textit{N}\textsc{hx}$. They confirmed the $\textit{N}\textsc{hx}$ redshift relation.
 
 While GRBs can have significant host absorption, blazars are thought to have negligible X-ray absorption on the LOS within the host galaxy, swept by the kpc-scale relativistic jet \citep[hereafter A18]{Arcodia2018}. This makes blazars ideal candidates for testing the absorption component of the IGM. Despite the suitability of blazars as IGM tracers, A18 is the only previous study to use them to explore IGM absorption as the cause of spectral curvature. Blazars are a special class of radio-loud AGN in which the relativistic plasma emerges from the galaxy core as a jet towards the observer. The broad-band spectra of blazars are characterized by two humps. The first hump produces a peak located between infrared to X-ray frequencies and is attributed to synchrotron processes.The second hump is typically found in X-ray to $\gamma$-ray frequencies and relates to inverse Compton (IC) processes. The seed photons for the IC process can be intrinsic to the jet, emitted through synchrotron processes at low frequencies called Synchrotron Self-Compton (SSC) \citep[e.g][]{Ghisellini1989}. Alternatively, if the seed photons originated from the accretion disc and are reprocessed by the broad-line region and/or the molecular torus, it is referred to as External Compton (EC) \citep[e.g.][]{Sikora1994}. Blazars are conventionally classified as either flat spectrum radio quasars (FSRQs) characterized by strong quasar emission lines and higher radio polarization, or BL Lac objects exhibiting featureless optical spectra \citep{Urry1995}. The distribution of the synchrotron peak frequency is significantly different for the two blazar classes. While the rest-frame energy distribution of FSRQs is strongly peaked at low frequencies ($\leq 10^{14.5}$ Hz), the energy distribution of BL Lacs is shifted to higher values by at least one order of magnitude \citep{Padovani2012}. FSRQs have a much higher median redshift than BL Lacs, and can be found out to high redshift \citep{Sahakyan2020}. Therefore, we focus primarily on FSRQ blazars in this study. 

\begin{table*}
    \renewcommand{\arraystretch}{1.3}
	\centering
	\caption{$\textit{Swift}$ blazar sample. For each blazar, the columns give the name, type, redshift, number of counts in 0.3-10 keV range and count rate ($s^{-1}$). Co-added spectra for each blazar are used which often are observed over a number of years, so we do not provide individual observation information.}
	\label{tab:Table_fullsample}
	\begin{tabular}{cc@{\hspace*{0.7cm}} c@{\hspace*{0.5
	cm}}c@{\hspace*{0.5
	cm}}c@{\hspace*{0.5cm}}c@{\hspace*{0.5cm}}c@{\hspace*{0.5
	cm}}}
		\hline

		Blazar & Type & $z$ & Total counts & Mean count rate ($s^{-1}$) \\
		
		\hline
		Mrk 501 &BL Lac & 0.03 & 4398 &$4.471\pm0.009$  
		\\
		PKS 0521-365 &BL Lac*  & 0.06 & 17076 & $0.673\pm0.004$
		\\
		BL Lac &BL Lac & 0.07 & 78117 & $0.360\pm0.001$
		\\
		1ES 0347-121 &BL Lac & 0.18 & 2318 &$1.158\pm0.018$
		
		\\
		1ES 1216+304 &BL Lac & 0.18 & 26601 &$ 1.404\pm0.007$
		
		\\
		4C +34.47 &FSRQ & 0.21 & 9631 &$0.345\pm0.006$
		
		\\
		1ES 0120+340 &BL Lac & 0.27 & 13358 &$0.987\pm0.007$ 
		\\
		S50716+714 &FSRQ & 0.31 & 69783 &$0.596\pm0.002$
		\\
		PKS 1510-089 &FSRQ & 0.36 & 60630 &$0.264\pm0.001$ 
		\\
		J1031+5053 &BL Lac & 0.36 & 5653 &$1.022\pm0.009$
		\\
		3C 279&FSRQ & 0.54 & 139130 &$0.462\pm0.001$ 
		\\
		1ES 1641+399&FSRQ & 0.59 & 13752 &$0.168\pm0.002$ 
		\\
		PKS 0637-752&FSRQ & 0.64 & 4767 &$0.192\pm0.003$ 
		\\
		PKS 0903-57&FSRQ & 0.70 & 361 &$0.116\pm0.004$ 
		\\
		3C 454.3&FSRQ & 0.86 & 83571 &$1.259\pm0.002$ 
		\\
		PKS 1441+25&FSRQ & 0.94 & 2282 &$0.072\pm0.002$ 
		\\
		4C +04.42&FSRQ & 0.97 & 1831 &$0.108\pm0.003$ 
		\\
		PKS 0208-512&FSRQ & 1.00 & 8218 &$0.085\pm0.008$ 
		\\
		PKS 1240-294&FSRQ & 1.13 & 944 &$0.126\pm0.005$ 
		\\
		PKS 1127-14&FSRQ & 1.18 & 5950 &$0.167\pm0.002$ 
		\\
		NRAO 140&FSRQ & 1.26 & 5673 &$0.305\pm0.004$ 
		\\
		OS 319&FSRQ & 1.40 & 1180 &$0.004\pm0.001$ 
		\\
		PKS 2223-05&FSRQ & 1.40 & 1765 &$0.078\pm0.002$ 
		\\
		PKS 2052-47&FSRQ & 1.49 & 1360 &$0.094\pm0.003$
		\\
		4C 38.41&FSRQ & 1.81 & 20962 &$0.145\pm0.001$ 
		\\
		PKS 2134+004&FSRQ & 1.93 & 1394 &$0.090\pm0.003$ 
		\\
		PKS 0528+134&FSRQ & 2.06 & 8779 &$0.060\pm0.001$ 
		\\
		1ES 0836+710&FSRQ & 2.17 & 
	54675&$0.664\pm0.002$ 
		\\
		PKS 2149-306&FSRQ & 2.35 & 16986 &$0.415\pm0.003$ 
		\\
		J1656-3302 & FSRQ & 2.40 & 1365 &$0.113\pm0.003$ 
		\\
		PKS 1830-211 & FSRQ & 2.50 & 13367 &$0.208\pm0.002$ 
		\\
		TXS0222+185 & FSRQ & 2.69 & 4461&$0.207\pm0.003$ 
		\\
		PKS 0834-20 & FSRQ & 2.75 & 826 &$0.038\pm0.001$ 
		\\
		TXS0800+618 & FSRQ & 3.03 & 647 &$0.057\pm0.002$ 
		\\
		PKS 0537-286 & FSRQ & 3.10 & 4205 &$0.079\pm0.001$ 
		\\
		PKS 2126-158 & FSRQ & 3.27 & 7280 &$0.225\pm0.003$ 
		\\
		S50014+81 & FSRQ & 3.37 & 2051 &$0.112\pm0.002$ 
		\\
		J064632+445116 & FSRQ & 3.39 & 569 &$0.029\pm0.001$ 
		\\
		J013126-100931 & FSRQ & 3.51 & 592 &$0.055\pm0.002$
		\\
		B3 1428+422 & FSRQ & 4.70 & 488 &$0.049\pm0.002$
		\\
	\hline
	\end{tabular}
    
    {*The classification of PKS0521-365 is disputed as being either a FSRQ, BL Lac or even a non-blazar \citep[e.g.][A18]{Urry1995,Zhang2021}}
\end{table*}

The sections that follow are: Section \ref{sec:2} describes the data selection and methodology; Section \ref{sec:3} covers the models for the IGM LOS including key assumptions and parameters, and blazar continuum models; Section \ref{sec:Blazar results} gives the results of blazar spectra fits using collisional IGM models with free IGM key parameters; in Section \ref{sec:Robust} we investigate and test the robustness of the IGM model fits; in Section \ref{sec:combinedGRBblazar} we combine a GRB sample with our blazar sample for cross-tracer analysis. We discuss the results and compare with other studies in Section \ref{sec:Discuss}, and Section \ref{sec:Conclusion} gives our conclusions. We suggest for readers interested in the key findings on IGM parameters from fits only, read Sections \ref{sec:Blazar results}, \ref{sec:combinedGRBblazar} and \ref{sec:Conclusion}. Readers interested in detailed spectra fitting methodology and model assumptions should also read Sections \ref{sec:2} and \ref{sec:3}. Finally, for readers interested in more detailed examination of robustness of the blazar spectra fitting and comparison with other studies, read Sections \ref{sec:Robust} and \ref{sec:Discuss}.

\section{Data selection and methodology}\label{sec:2}
 
 \begin{table}
    
	\centering
	\caption{\textit{XMM-Newton}  sub-sample for individual observation comparison with \textit{Swift} co-added spectrum results. For each blazar, the columns give Observation ID, redshift, total counts and count rate for 0.3-10 keV.}
	\label{tab:XMM_blazar_subsample   }
	\begin{tabular}{lcccr} 
		\hline

		Blazar & Observation & $z$ & Total & Mean count     \\
		& ID & &counts & rate ($s^{-1}$)\\
		
		\hline
		3C 454.3 &mean  & 0.86 & 261165 & 24.86
		\\
		&0401700201  &  &  &  
		\\
		&0401700401  &  &  &  
		\\
		&0401700501  &  &  &  
		\\
		&0103060601  &  &  & 
		\\
		
		PKS 2149-306&0103060401  &2.35 & 39152& 1.91 
		\\
		PKS 2126-158&0103060101  &3.26 & 39504& 2.62 
		\\
		PKS 0537-286& mean &3.10  & 44938 & 1.40 
		\\
		& 0114090101 &  &  &  
		\\
		& 0206350101 &  &  &  
		\\
		1ES 0836+710& 0112620101 &2.17  & 236705 &  10.08
		\\
		TXS 0222+185& mean &2.69 & 528277 &  6.80
		\\
		&0150180101  &  &  &  
		\\
		&0690900101  &  &  &  
		\\
		&0690900201  & &  & 
		\\
		PKS 0528+134&mean  &2.06 &18051  & 0.88
		\\
		&0600121401  &  &  &  
		\\
		& 0600121501 &  &  &  
		\\
		&0401700601  &  &  &  
		\\
		&0103060701  &  &  & 
		\\
        \hline
	\end{tabular}
\end{table}

The total number of confirmed blazars by means of published spectra as of 2020 was 2,968 (1,909 FSRQ and 1,059 BL Lac) per the Roma-BZCAT Multifrequency Catalogue (Roma), which is regarded as the most comprehensive list of blazars \citep{Paggi2020a,massaro2015}. The vast majority of blazars are at $z < 2$, with less than $4\%$ at $z > 2$ and $1.2\%$ at $z > 2.5$ \citep{Sahakyan2020}. As our objective is to examine possible absorption by the IGM in blazar spectra, our percentage coverage is greater at higher redshift than lower redshift i.e. $\sim13\%$ of blazars with $z > 2$ and $\sim25\%$ of blazars at $z > 2.5$ based on the Roma Catalogue numbers. Our sample criteria requires blazars with confirmed redshift whose spectra have high counts for spectral analysis with a spread across redshift up to $z = 4.7$. Table 1 gives the counts for each blazar in the sample. The counts range from 361 to 139,139, with the highest redshift blazars ($z > 3$) accounting for most of the counts below 1000. We reviewed literature which studied large numbers of blazars \citep[e.g.][]{Perlman1998,Donato2005,Eitan2013,Ighina2019,Marcotulli2020} and, within our criteria, selected objects randomly for our sample. Our sample has 14 with $z > 2$, a high fraction of the total available as noted above. To keep a reasonable spread across the redshift range, we randomly selected 9 from $1 < z < 2$, with 17 with $z < 1$.

A key part of our research is the comparative analysis with the GRB sample from D20 and D21 which was taken primarily from the UK \textit{Swift} Science Data Centre\footnote {http://www.swift.ac.uk/xrtspectra}
repository \citep[hereafter \textit{Swift};][]{Burrows2005}. To ensure a homogeneous dataset across Blazars and GRBs, our main sample of 40 blazars is drawn from the $\mathit{Swift}$ 2SXPS Catalogue \citep{Evans2020}, using their XRT data products generator. $\mathit{Swift}$ was designed for GRBs which are thought to explode randomly across the sky and blazars are totally unrelated to these sources. Therefore, $\mathit{Swift}$ provides a  highly unbiased, all sky, serendipitous database of blazars. $\mathit{Swift}$ recovers much of the comparative sensitivity with $\mathit{XMM-Newton}$ even though it has a lower effective area and smaller field of view \citep{Evans2020}.

$\textit{Swift}$ has proven to be an excellent multi-frequency observatory for blazar research. Our sample spectra from the \textit{Swift} repository were taken from the Photon Counting mode with high photon count. High signal-to-noise X-ray spectra are necessary to properly assess the presence of a curved spectrum in distant extra-galactic sources and its components. Therefore, where multiple observations of the same object were available, we use co-added spectra. Our sample is representative of the range from $0.03 < z < 4.7$ and details are available in Table \ref{tab:Table_fullsample}.

In section \ref{sec:Robust}, as part of robustness testing, we use a sub-sample of 7 $\textit{XMM-Newton}$ spectra (Table \ref{tab:XMM_blazar_subsample   }). The $\textit{XMM-Newton}$ European Photon Imaging Camera-PN \citep{Struder2001} spectra were obtained in timing mode, using the thin filter. Data reduction, including background subtraction, was done with the Science Analysis System (SAS2, version 19.1.0) following the standard procedure to obtain the spectra.

While we fit 7 BL Lac, in our analysis of IGM parameters such as the $\textit{N}\textsc{hxigm}$ redshift relation, we use only FSRQ and omit BL Lac. FSRQ are more powerful and therefore more likely to sweep out any host absorbers. Further, their spectra have less intrinsic features which may be degenerate with potential IGM absorption curvature.

We use \textsc{xspec} version 12.11.1 for all our fitting \citep{Arnaud1996}. We use the C-statistic \citep{Cash1979} (Cstat in \textsc{xspec}) with no rebinning. The common practice of rebinning data to use a $\chi^2$ statistic results in loss of energy resolution.  The maximum likelihood C-statistic, based on the Poisson likelihood, does not suffer from these issues. For a spectrum with many counts per bin  C-statistic $\rightarrow \chi^2$, but where the number of counts per bin is small, the value for C-statistic can be substantially smaller than the $\chi^2$ value \citep{Kaastra2017}.  

Given we are studying the IGM with X-ray spectra, we can expect some degeneracies between the parameters. Therefore,  there may be several local probability maxima with multiple, separate, adequate solutions. In these circumstances, the local optimisation algorithms like the Levenberg- Marquardt cannot identify them or jump from one local maximum to the other. Given the issues of goodness of fit and getting out of local probability maxima, we  follow the same method as D21, and use a combination of the \textsc{xspec} \textsc{steppar} function, and confirmation with Markov chain Monte Carlo (MCMC) to validate our fitting and to provide confidence intervals on Cstat.

Approximating a $\chi^2$ criterion, some authors consider fits to be significantly improved by the addition of a component if the reduction in Cstat$^2> 2.71$ for each extra free parameter \citep[e.g. ][]{Ricci2017}.  We follow this method in Section \ref{sec:Blazar results} when fitting continuum only models and full models with an IGM absorption component.

As the bulk of matter in the IGM is ionised and exists outside of gravitationally bound structures (apart from the CGM), in this paper we use a homogeneity assumption.  We use blazars that have LOS orders of magnitude greater than the large scale structure. 

\section{Models for the Blazar LOS}\label{sec:3}
In this section we describe the motivation and expected physical conditions in the IGM that lead to our choice of models, the priors and parameter ranges. We also describe the models and reasoning used for fitting the intrinsic spectra of the blazars. 

\subsection{Galactic absorption}
For Galactic absorption ($\textit{N}\textsc{hxGal})$, we use \textsc{tbabs} \citep[hereafter W00]{Wilms2000} fixed to the values based on  \citet{Willingale2013} i.e. estimated using 21 cm radio emission maps from \citep{Kalberla2005} , including a molecular hydrogen column density component. \textsc{tbabs} calculates the cross-section for X-ray absorption by the ISM as the sum of the cross sections for the gas, grain and molecules in the ISM.  \citet{Asplund2009} is generally regarded as providing the most accurate solar abundances, though this has been updated by \citet{Asplund2021}. However, we used the solar abundances from W00 which take into account dust and H$_2$ in the interstellar medium in galaxies. Below 1 keV, N, O, and Ne are the dominant absorption features.

\subsection{Continuum model}

In the energy range we are studying for the main sample (0.3-10 keV), blazar spectra typically show curvature in soft X-ray. A log-parabolic spectrum (\textsc{logpar} in \textsc{xspec}) can be produced by a log-parabolic distribution of relativistic particles \citep{Paggi2009}. This curved continuum shape could also arise from a power-law particle distribution with a cooled high energy tail \citep{Furniss2013}. 

Similarly, a broken power law intrinsic curvature can be interpreted as relativistic electrons in the jet following a broken power law energy distribution with a low energy cut-off, or an inefficient radiative cooling of lower energy electrons producing few synchrotron photons \citep{Gianni2011}.

\begin{figure}

	\includegraphics[width=\columnwidth]{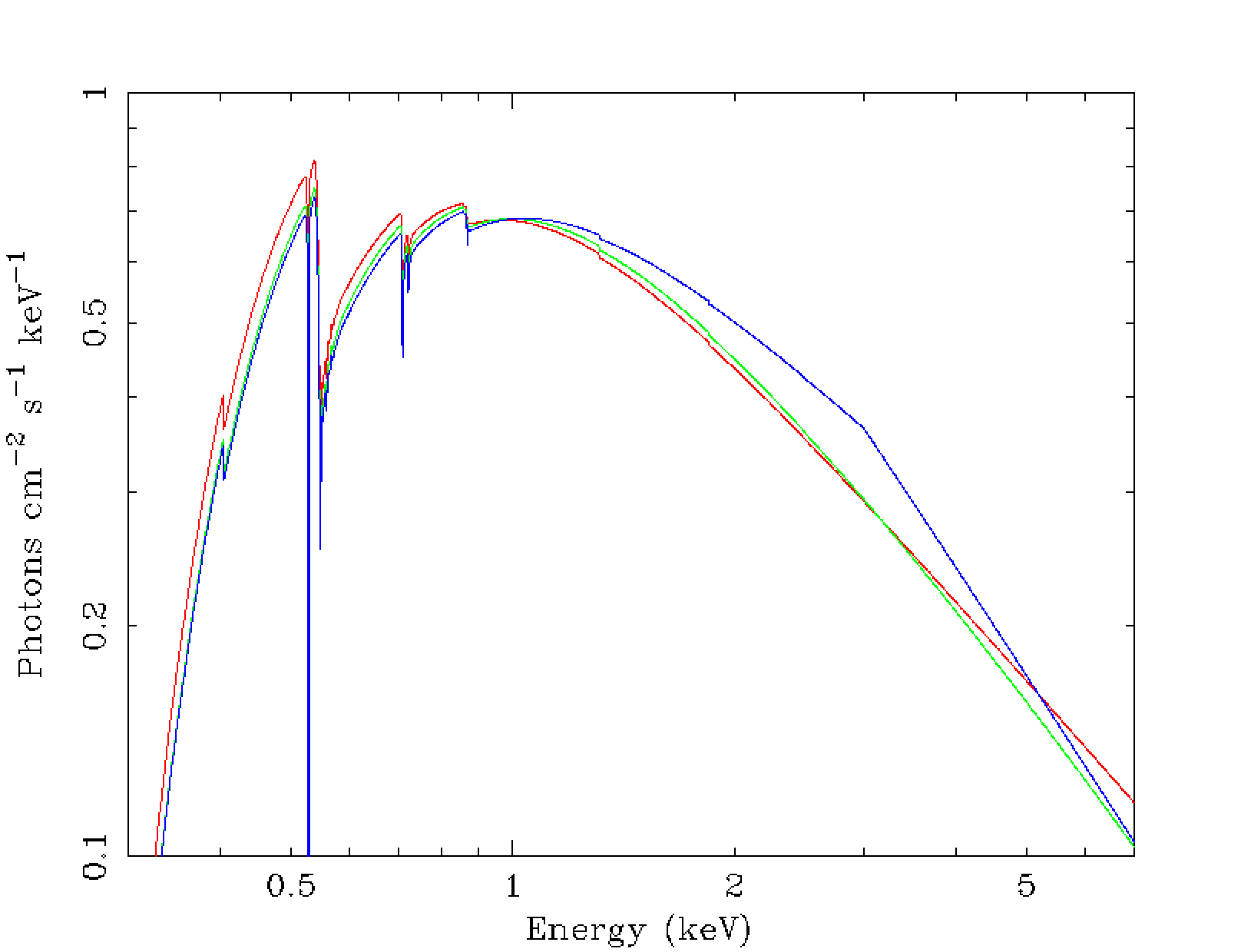}
    \caption{Intrinsic models with Galactic absorption only, in the energy range 0.3 - 2.0keV ($\textit{Swift}$ spectra extend to 10keV), simple power law (red), log-parabolic (green) and broken power law (blue). Below 1 keV, N, O, and Ne are the dominant absorption features.}
    \label{fig:Figure po_models}
\end{figure}

Given that it is very difficult to determine whether spectral softening is caused by absorption or is intrinsic to the blazar emission, we adopted a conservative approach. We first fitted our full sample spectra with three different power laws: simple power law, log-parabolic and broken power law. For all our sample, both log-parabolic and broken power laws provided better fits than the simple power law. We noted the best fit model and then proceeded to add an ionised absorption component to represent potential IGM absorption where we  compare the Cstat results of the full models with the continuum only results. Fig \ref{fig:Figure po_models} shows the impact of using the three different continuum power law models on intrinsic curvature with absorption assumed only from our Galaxy.

\graphicspath{ {./figurespaper3/}  }

\subsection{Ionised IGM}
In earlier studies which examined the hypothesis of absorption causing the observed soft X-ray spectral hardening, only an intrinsic host absorption or some discrete intervening neutral absorbers (DLAs, LLS, etc) were considered. However, in blazars the host contribution is probably negligible, consistent with low levels of optical-UV extinction observed \citep{Paliya2016}.  Further, strong intervening absorption by neutral absorbers ($\textit{N}\textsc{hi}$) is too rare in blazar LOS to account for the observed curvature \citep[e.g][]{Elvis1994,Cappi1997,Fabian2001,Page2005}. Therefore, we omit any absorption contribution from the blazar host or low-column density intervening neutral Lyman absorbers (log($\textit{N}\textsc{hi}) < 21$) in our models. If there is a large known intervening absorber such as a galaxy (e.g. PKS 1830-211) or substantial DLA with log($\textit{N}\textsc{hi}) > 21$ then it was included in the model using \textsc{xspec} \textsc{ztbabs} placed at the redshift of the intervening object. We note that it is possible that further unidentified individual strong absorbers may exist on the LOS to our sample  which would then be included in the integrated $\mathit{N}\textsc{hxigm}$ derived from the fits.

D21 examined different ionisation models to represent diffuse IGM absorption including collional ionised equilibrium  models (CIE): \textsc{hotabs} \citep{Kallman2009}, \textsc{ioneq} \citep{Gatuzz2015} and \textsc{absori} \citep{Done1992}, and photionisation equilibrim (PIE) model \textsc{warmabs} \citep{Kallman2009}. \textsc{Absori} allows one to have both ionisation parameter and temperature as free parameters which would not occur in either pure PIE or CIE \citep{Done2010}.  In order to compare with CIE models, D21 froze the ionisation parameter leaving temperature as a free parameter required for CIE scenarios.
In earlier works on using tracers such as GRBs and blazars for IGM absorption, \textsc{absori} was generally used \citep[e.g.][A18]{Bottacini2010,Behar2011,Starling2013a}. While \textsc{absori} was the best model available when it was developed in 1992, it is not self-consistent, and is limited to 10 metals, all of which are fixed at solar metallicity except Fe \citep{Done1992}. D21 concluded that \textsc{warmabs} and \textsc{hotabs} are the most sophisticated of these models currently available, and the MCMC integrated probability plots were the most consistent with the \textsc{steppar} results. We followed their methodology for the modelling the IGM absorption. Initially, we fitted a sub-sample of 20 blazars with both PIE and CIE absorption separately as extreme scenarios where all the IGM absorption is either in the CIE or PIE phase. Consistently with D21, we found that similar results were obtained for both models and that it is not possible to conclude whether PIE or CIE is the better single model for the IGM at all redshift, though a combination is the most physically plausible scenario. Given the fact that similar results are obtained for both PIE and CIE models, we proceeded only with fitting CIE model \textsc{hotabs}. Therefore, the full models for each blazar in \textsc{xspec} language are (with the addition of \textsc{ztbabs} for known interving objects):\\

\textsc{tbabs*hotabs*logpar}

or

\textsc{tbabs*hotabs*bknpo}\\

We model the IGM assuming a thin uniform plane parallel slab geometry in thermal and ionization equilibrium. This simple approximation is generally used for a homogeneous medium \citep[e.g.][]{Savage2014,Nicastro2017,Khabibullin2019,Lehner2019}. This slab is placed at half the blazar redshift as an approximation of the full LOS medium. The parameters ranges that were applied to the CIE models are taken from D21 and summarised in Table \ref{tab:Table_paramaters}. We follow the same methodology as D21 in using hotabs and the IGM parameter range for IGM density, temperature and metallicity. D21 provides detail on hotabs which  calculates the absorption due to neutral and all ionized species of elements with atomic number $Z \leq 30$. Further, D21 clarifies that the fitting method uses the continuum total absorption to model plasma as opposed to fitting individual line absorption as the required resolution is not available currently in X-ray. As we are modelling and fitting the continuum curvature and not specific lines or edges specifically, scope for degeneracy occurs. In the D21 Supplementary material, transmission figures are gives to show the impact of changes in temperature, metallicity and redshift. In the cool IGM phases, typical metallicity is observed to be $-4 < [X/$H$] < -2$ \citep[e.g.][]{Schaye2003,Simcoe2004,Aguirre2008}. In the warmer phases including the WHIM, the metallicity has been observed to be higher $[X/$H$] \sim -1$ \citep[e.g.][]{Danforth2016,Pratt2018}. As we are modelling the LOS through the cool, warm and hot diffuse IGM, we will set the \textsc{xspec} metallicity parameter range following D21 as $-4 < [X/$H$] < -0.7$.

\begin{table}
	\centering
	\caption{Upper and lower limits for the free parameters in the IGM models. Continuum parameters were also free parameters. The fixed parameters are Galactic  log($\textit{N}\textsc{hx}$), the IGM slab at half the GRB redshift, and any known intervening object log($\textit{N}\textsc{hx}$). }
	\label{tab:Table_paramaters}
	\begin{tabular}{lccr} 
		\hline
		IGM parameter &  range in \textsc{xspec} models\\
		\hline
		column density & $19 \leq$ log($\textit{N}\textsc{hx}$) $\leq 23$ \\
		temperature\ & $4 \leq$ log($T$/K) $\leq 8$  \\
		metallicity & $-4 \leq [X/$H$] \leq -0.7$ \\
		\hline
	\end{tabular}
\end{table}

\graphicspath{ {./figurespaper3/}  }
\begin{figure}

	\includegraphics[width=\columnwidth]{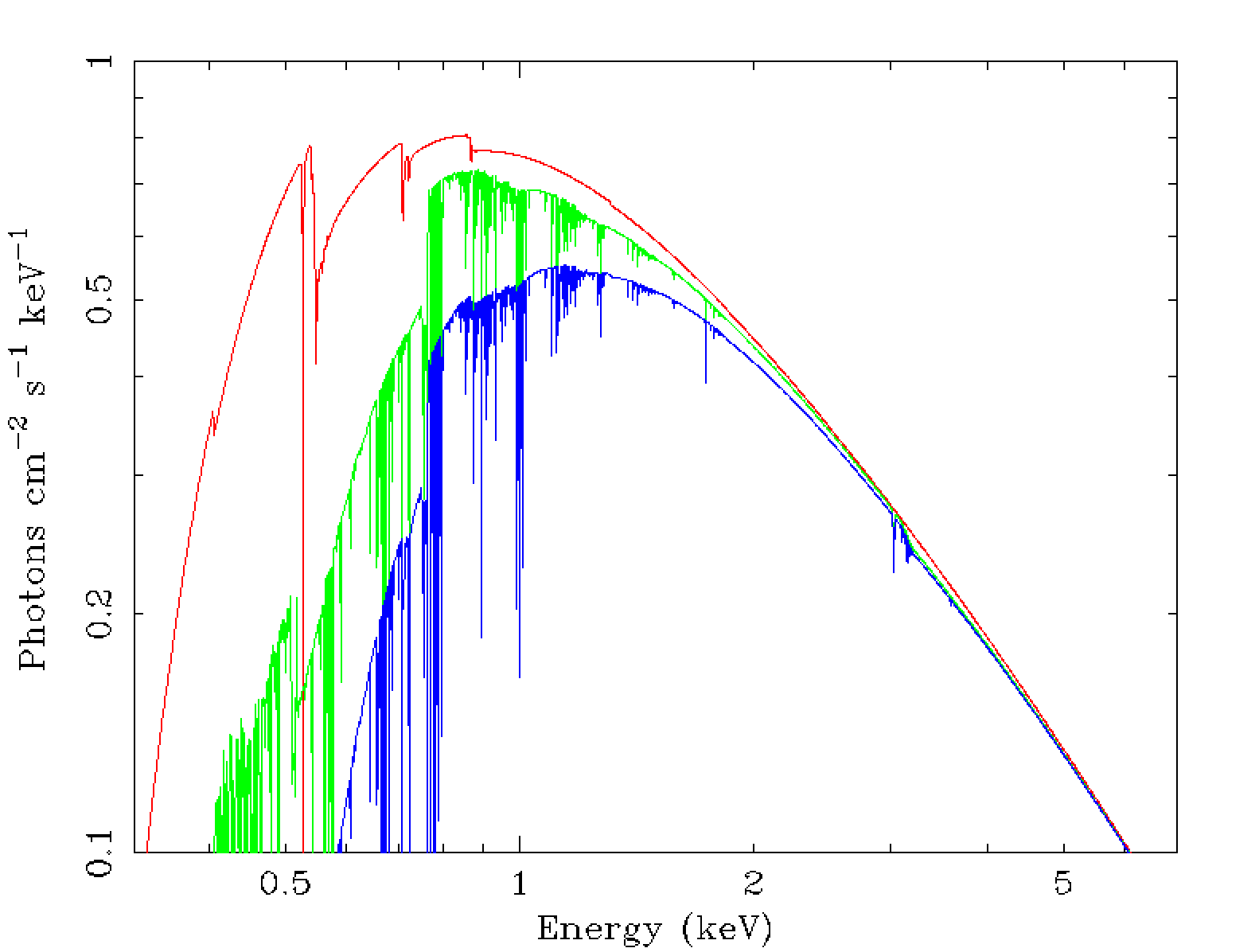}
    \caption{Model components for the LOS absorption to a blazar using a log-parabolic power law, \textsc{hotabs} for IGM CIE absorption in the energy range 0.3 - 2.0keV ($\textit{Swift}$ spectra extend to 10keV). The  model example is for a blazar at redshift $z = 2.69$, with log($\textit{N}\textsc{hxigm}) = 22.28$, $[X/$H$] = -1.59$, and log($T/$K) = 6.2 for the IGM, log($\textit{N}\textsc{hx}) = 21.2$ for our Galaxy. The IGM CIE absorption curve is green, our Galaxy red and the total absorption from both components is the blue curve. }
    \label{fig:Full model}
\end{figure}

\graphicspath{ {./figurespaper3/}  }
 \begin{figure*}
     \centering
     \begin{tabular}{c|c|c}
    \includegraphics[scale=0.2]{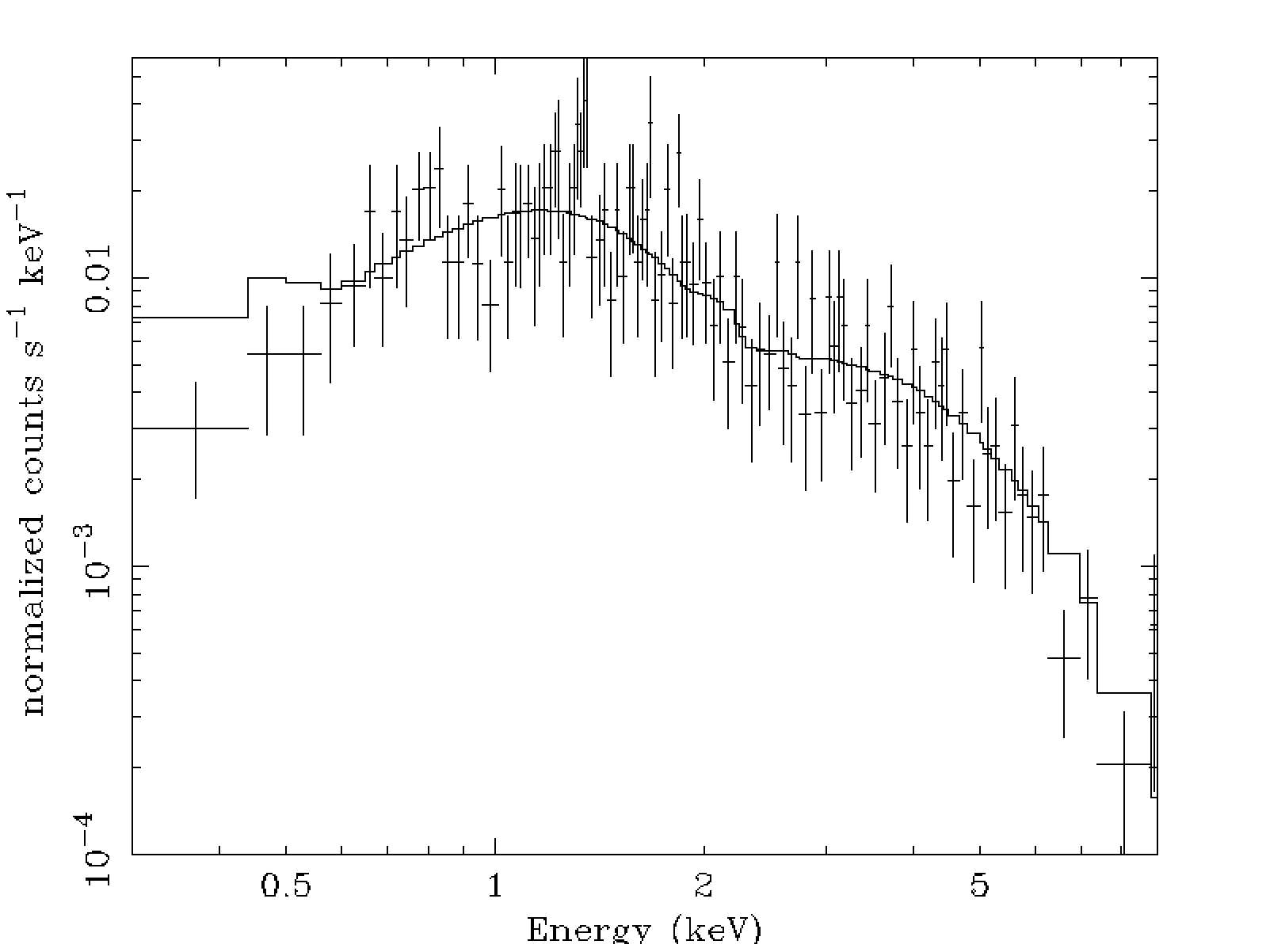} &
    \includegraphics[scale=0.2]{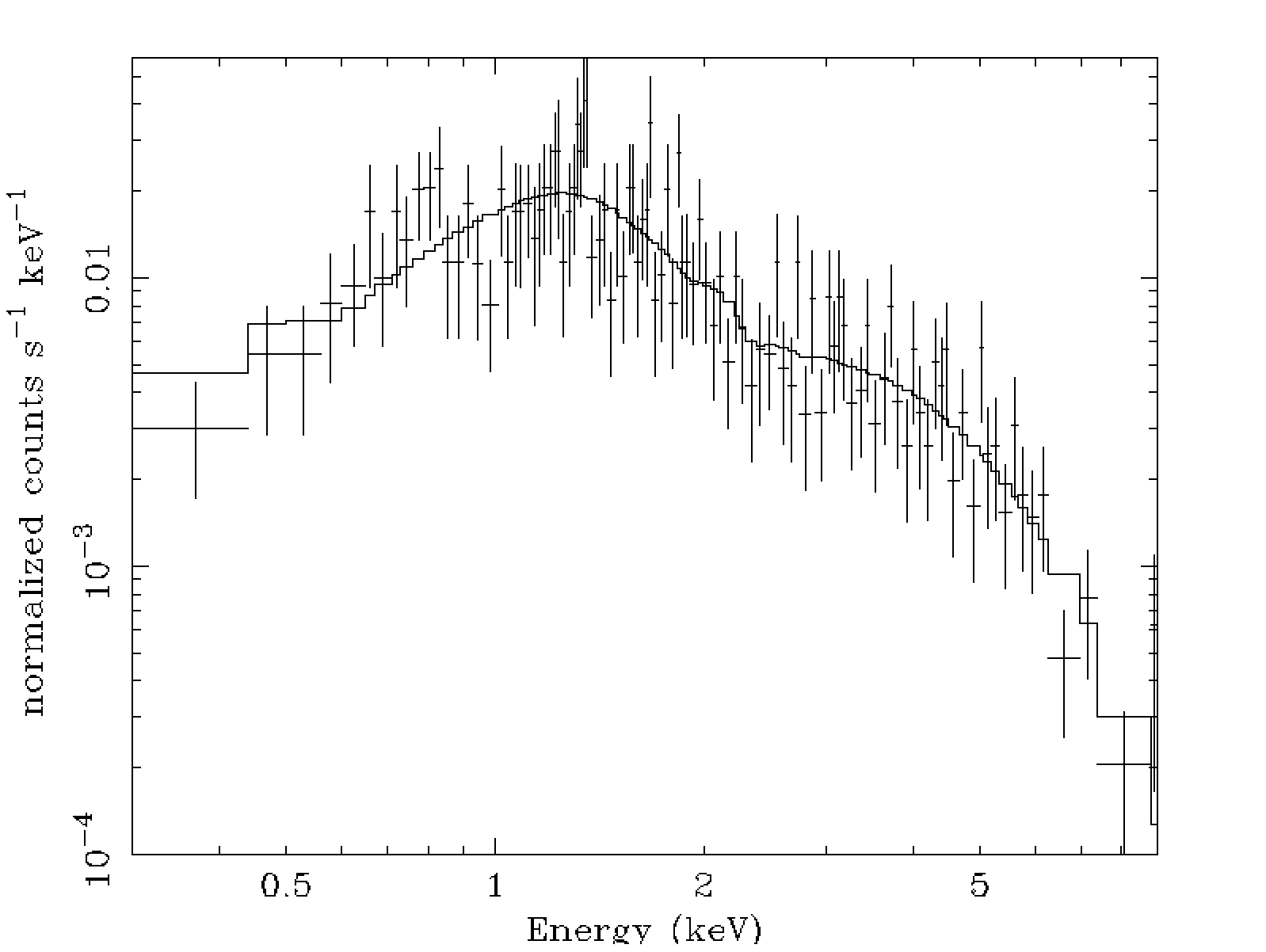} &
    \includegraphics[scale=0.2]{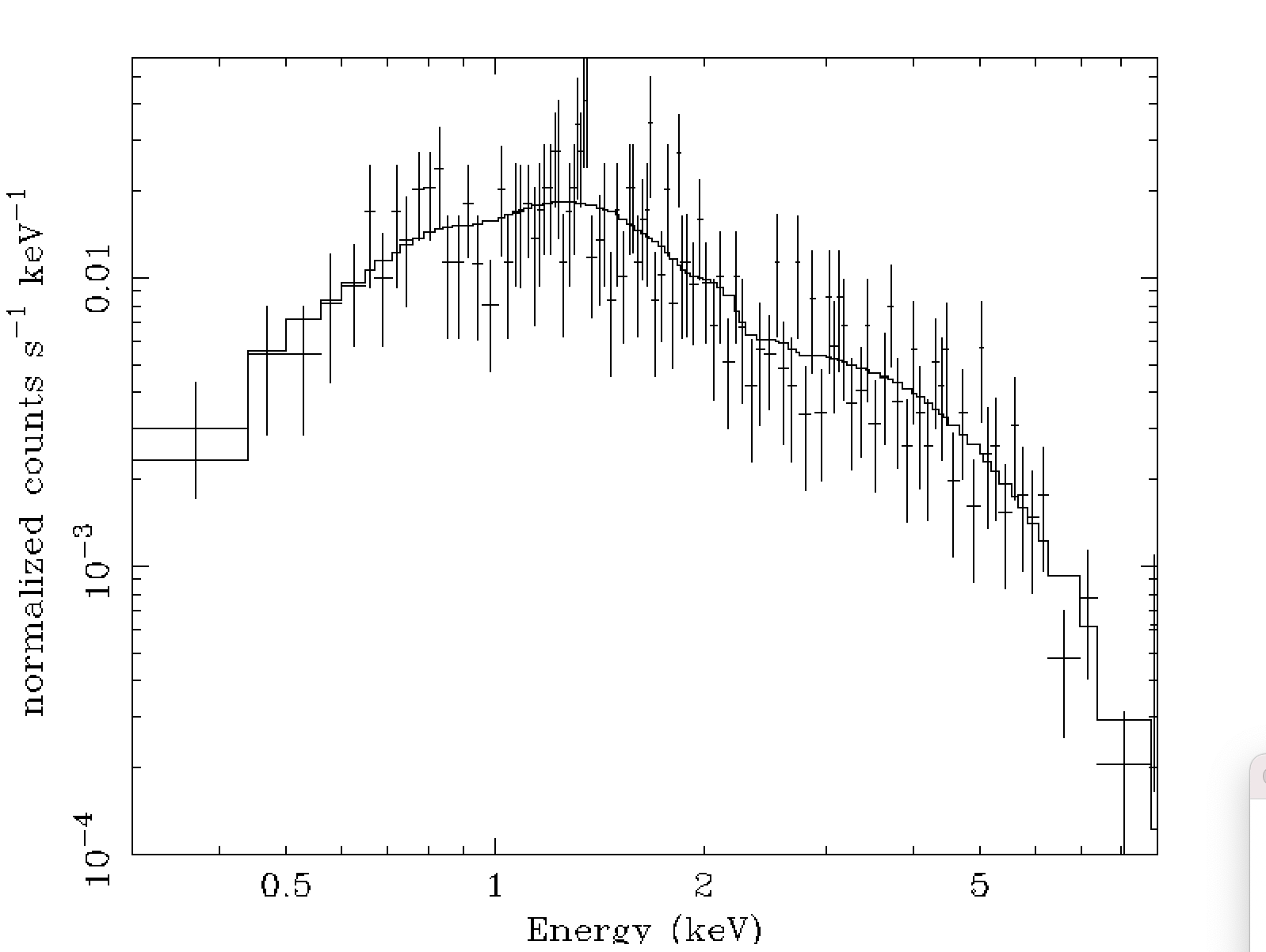}
     \\
    \end{tabular}
    \caption{Impact of using different intrinsic curvature models and additional IGM absorption components for blazar J013126-100931. All fits include Galactic absorption. The left panel is a simple power law. The middle panel is with a broken power law. The right panel is a broken power law with a CIE IGM absorption component.}
        \label{fig:J013126}
\end{figure*}

 Fig.~\ref{fig:Full model} shows an example of the model components for the full LOS absorption using \textsc{hotabs} for IGM CIE absorption. The  model example assumes a blazar at redshift $z = 2.69$, using a log-parabolic power law, \textsc{hotabs} for IGM CIE absorption, log($\textit{N}\textsc{hxigm}) = 22.28$, $[X/$H$] = -1.59$, and log($T/$K) = 6.2 for the IGM, and log($\textit{N}\textsc{hx}) = 21.2$ for our Galaxy. The IGM CIE absorption curve is green, our Galaxy red and the total absorption from both components is the blue curve. In the model, the absorption lines are clearly visible. However, we would expect that these lines would not be detected due to instrument limitations and being smeared out over a large redshift range.

 \graphicspath{ {./figurespaper3/}  }
 \begin{figure*} 
    \centering
    \begin{tabular}{c|c}
    \includegraphics[scale=0.31]{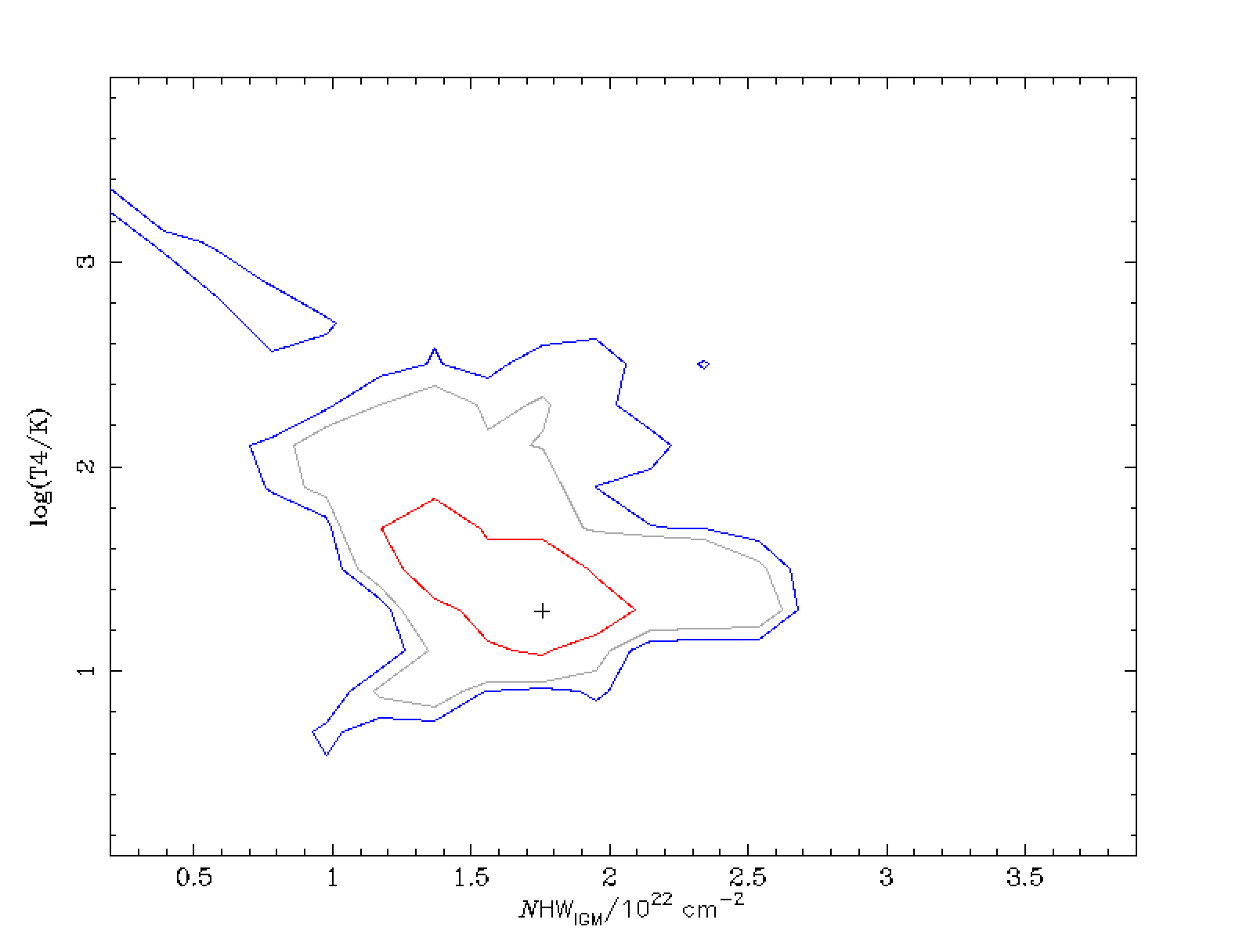} &
    \includegraphics[scale=0.29]{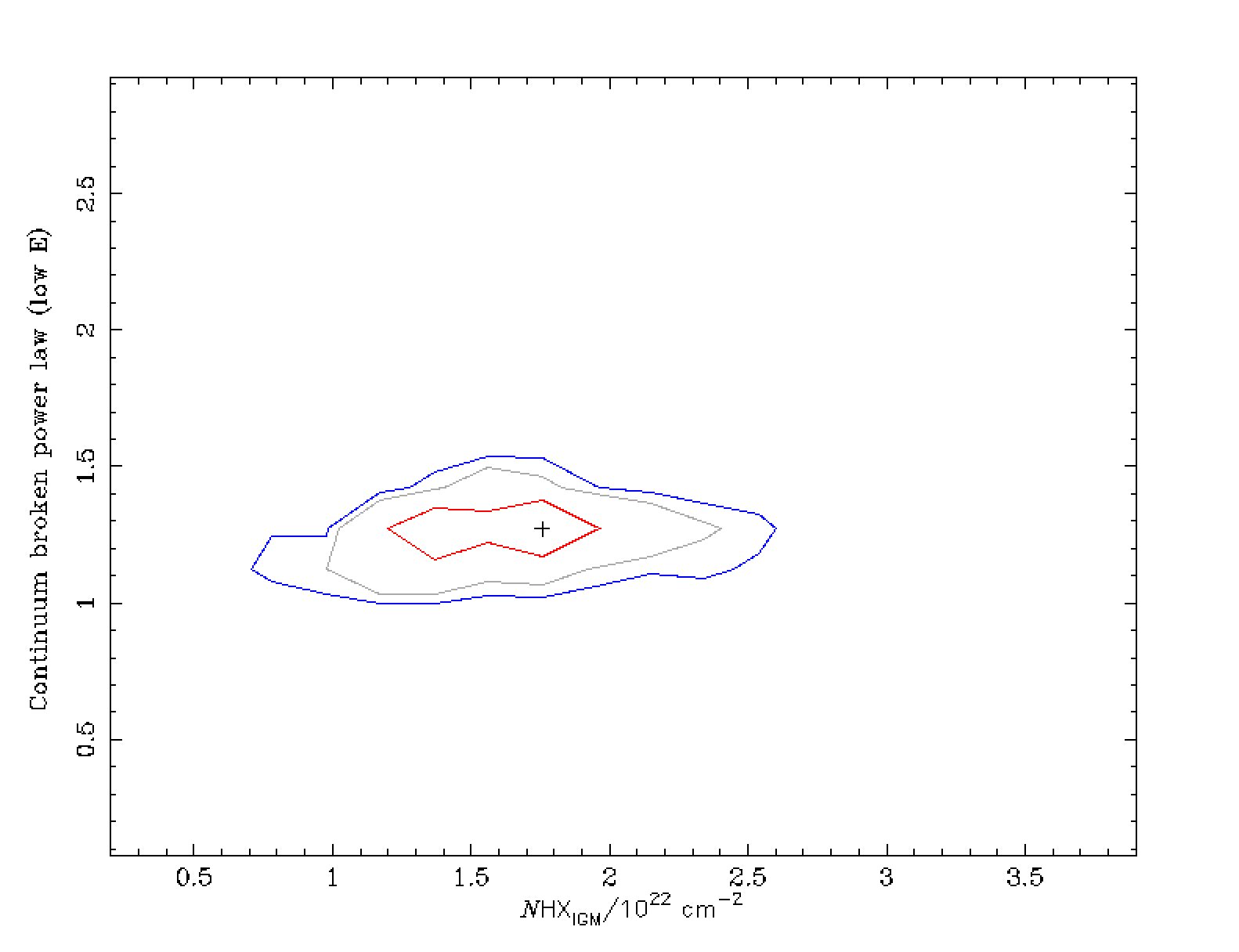}
    \\
    \end{tabular}
    \caption{The left and right panels show the MCMC integrated probability results for J013126-100931 for $\mathit{N}_{\textsc{hxigm}}$ with temperature and broken power law (low energy) respectively. The red, grey and blue contours represent $68\%, 90\%$ and $95\%$ confidence ranges for the two parameters respectively. On the y-axis in the left panel T4 means the log of the temperature is in units of 10$^4$ K.}
        \label{fig:J013126_MCMC}
\end{figure*}

\section{Spectral analysis results}\label{sec:Blazar results}

In Section \ref{sec:Blazar results}, we firstly discuss the impact of using different intrinsic blazar continuum models in Section \ref{subsec:4.1}, then give the results for IGM parameters for the full sample using the CIE IGM absorption model in Section \ref{subsec:4.2}. All spectral fits incude \textsc{tbabs} for Galactic absorption.

\subsection{Spectra fit improvements from alternative continuum models and IGM component }\label{subsec:4.1}

We show the fit results in Fig. \ref{fig:J013126} for J013126-100931 at redshift $z = 3.51$ as an example of typical results.  We initially fitted a simple power law. The left panel in Fig. \ref{fig:J013126} shows residuals at low energy with a Cstat of 273.24 for 330 degrees of freedom (dof). We then tried both a log-parabolic and broken power law. The middle panel shows the fit with a broken power law which had a better result than both log-parabolic or simple power law, Cstat/dof = 262.88/328. An improved fit at soft energy can be seen. We then added a variable CIE IGM component to the broken power law while allowing the intrinsic parameters to also vary shown in the right panel of Fig. \ref{fig:J013126}. The spectral fit is improved compared with the Galactic absorbed broken law only model, with less low energy residual, Cstat/dof = 256.98/326. Fig. \ref{fig:J013126_MCMC} left and right panels show the MCMC integrated probability results for $\mathit{N}\textsc{hxigm}$ with temperature and broken power law index (low energy) respectively. The red, grey and blue contours represent $68\%, 90\%$ and $95\%$ ranges for the two parameters respectively. On the y-axis left panel, log(T4/K) means that 0 is log$(T/$K$) \/\ = \/\ 4$. The contours  in both plots and particularly $\mathit{N}\textsc{hxigm}$ and the low energy power law provide reasonably tight ranges of parameter results at 2$\sigma$ ($95\%$ confidence).

\subsection{IGM parameter results using a CIE IGM model}\label{subsec:4.2}

We now give the results for IGM parameters based on fitting the full sample of 40 blazars using \textsc{hotabs} for CIE IGM. The IGM $\textit{N}\textsc{hx}$, metallicity and temperature parameters are all free, as are the power law parameters. The error bars for all fits are reported with a $90\%$ confidence interval. In the plots of $\textit{N}\textsc{hx}$ and redshift, the green line is the mean hydrogen density of the IGM based on the simple model used in D20 and references therein \citep[e.g.][]{Starling2013,Shull2018}.
\begin{equation} 
\label{eq:simpleIGM}
\mathit{N}_{\textsc{hxigm}} = \frac{n_0 c}{H_0} \int_0^z \frac{(1 + z)^2 dz}{[\Omega _M(1 + z)^3 + \Omega _\Lambda ]^\frac{1}{2}}
\end{equation}
where n$_{0}$ is the hydrogen density at redshift zero, taken as $1.7 \times 10^{-7}$ cm$^{-3}$ \citep{Behar2011}.
This value is based on $90\%$ of the baryons being in the IGM. When giving results for the mean hydrogen density at $z = 0$ ($n_0$), they are derived by rearranging equation \ref{eq:simpleIGM} to give $n_0$. We then used our results for $\mathit{N}\textsc{hxigm}$ and actual redshift for each blazar to get their equivalent $n_0$. Finally, we took the mean of our full sample and the standard error.

\graphicspath{ {./figurespaper3/}  }
 \begin{figure*} 
    \centering
    \begin{tabular}{c|c}
    \includegraphics[scale=0.55]{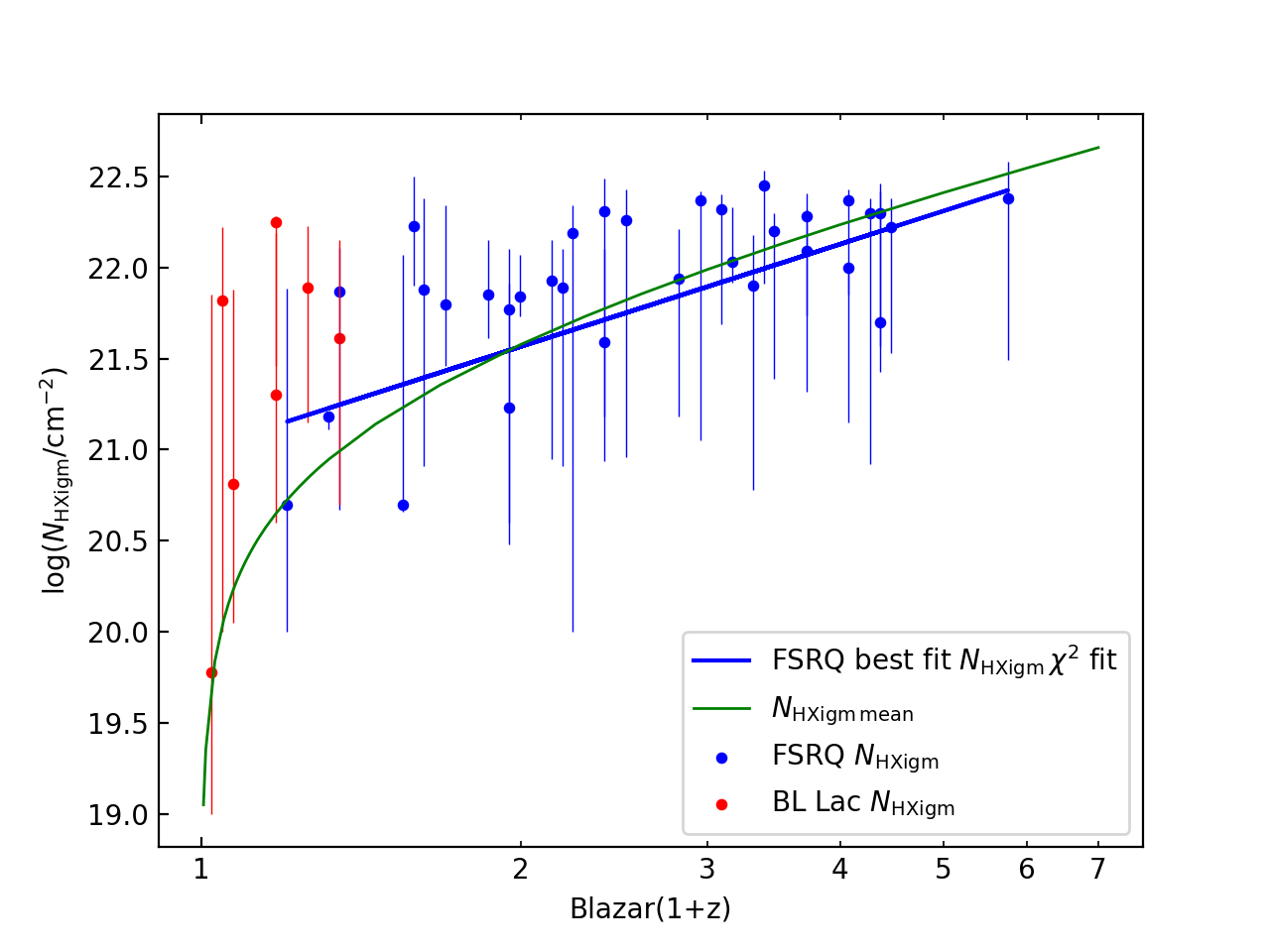} &
    \includegraphics[scale=0.55]{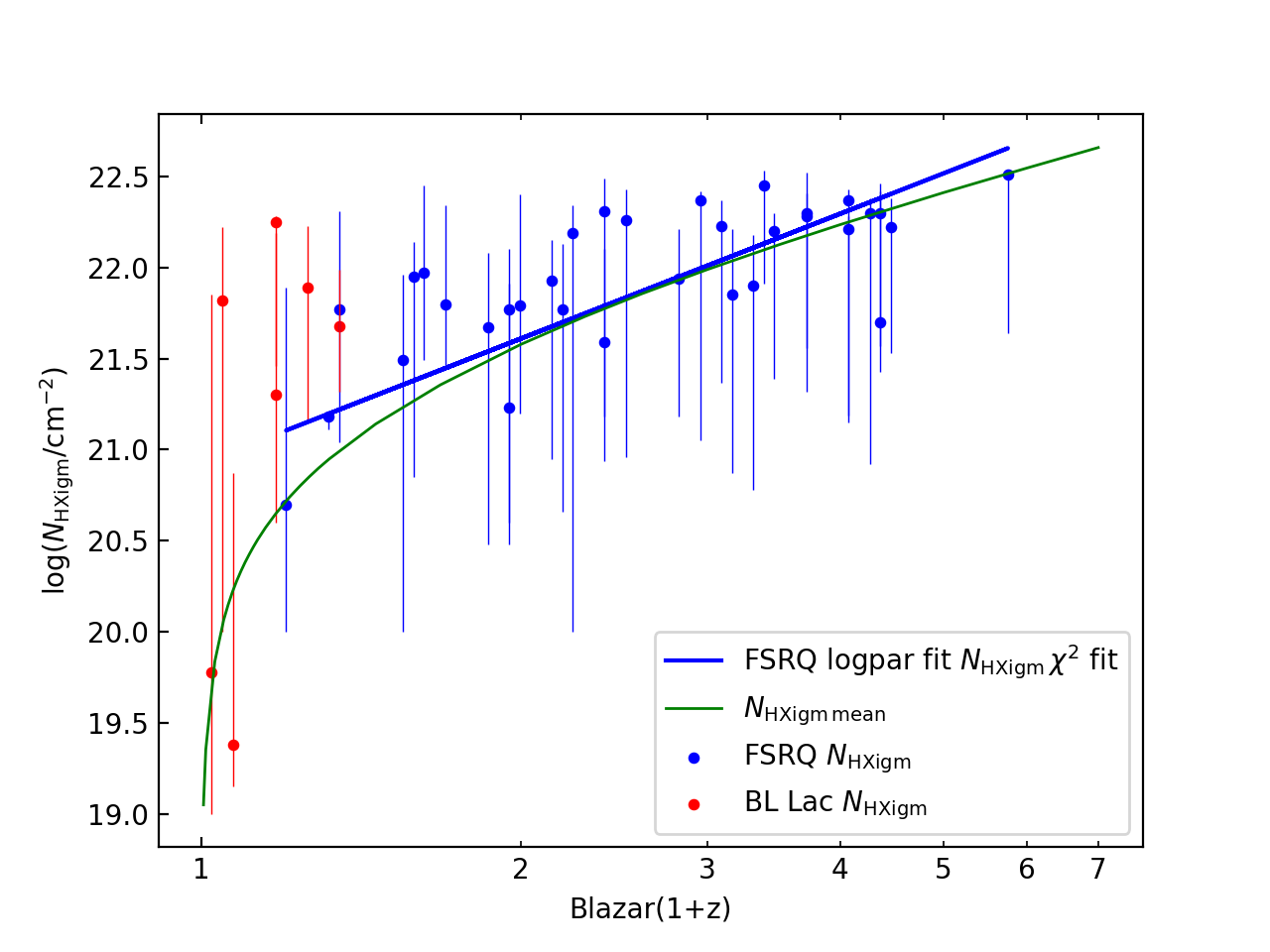}
    \\
    \end{tabular}
    \caption{Results for the IGM $\textit{N}\textsc{hx}$ parameter and redshift using the CIE (\textsc{hotabs}) model. The error bars are reported with a $90\%$ confidence interval. The green line is the simple IGM model using a mean IGM density. Left panel is $\textit{N}\textsc{hx}$ and redshift selecting best Cstat results from the different power law intrinsic models. Right panel is the full sample with the IGM component and a log-parabolic power law only (best fit for 26/40).}
        \label{fig:NHX_z}
\end{figure*}

\graphicspath{ {./figurespaper3/}  }
 \begin{figure*} 
    \centering
    \begin{tabular}{c|c}
    \includegraphics[scale=0.55]{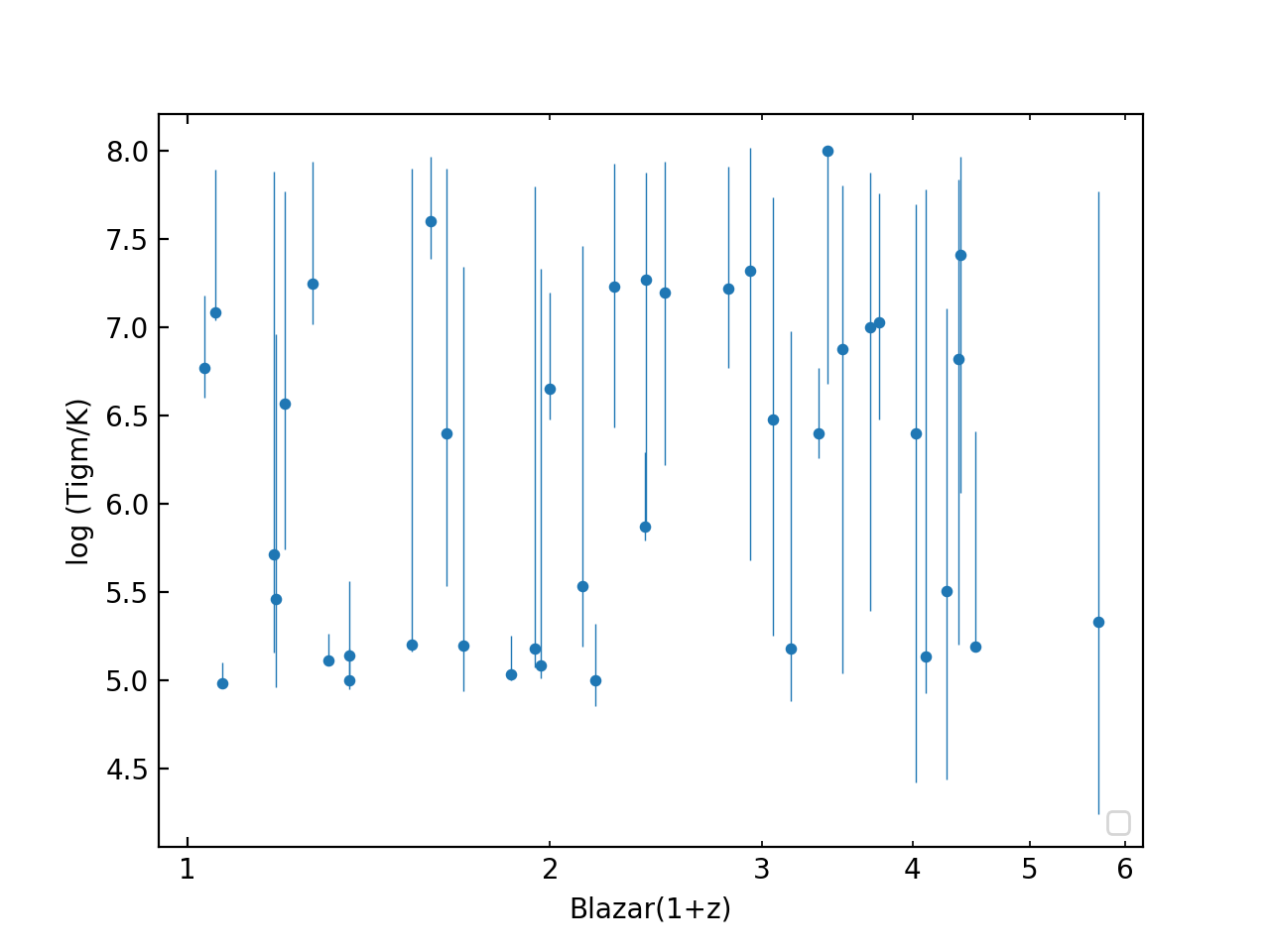} &
    \includegraphics[scale=0.55]{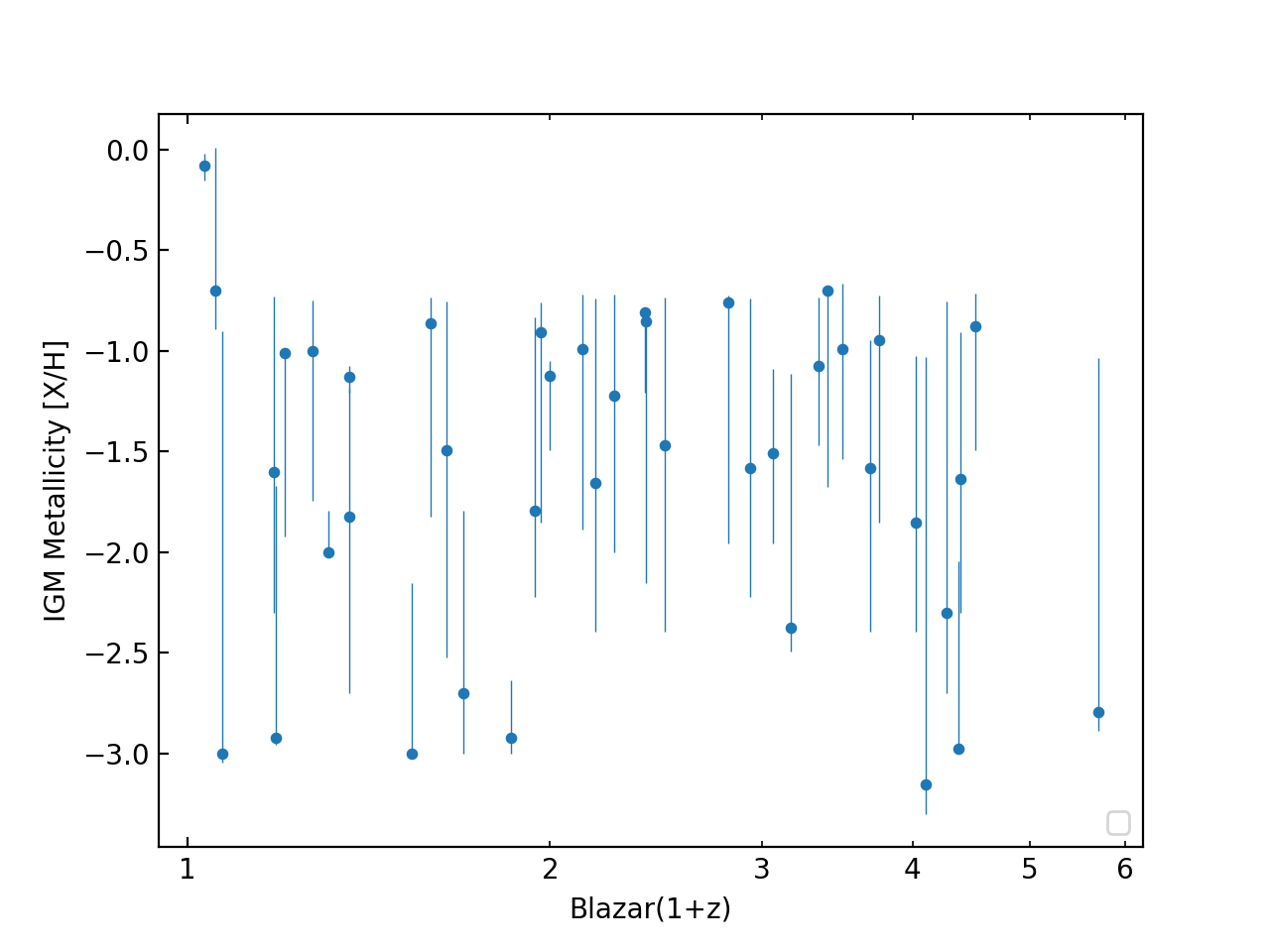}
    \\
    \end{tabular}
    \caption{Results for the IGM  parameters and redshift using the CIE (\textsc{hotabs}) model best fit results from the various models. The error bars are reported with a $90\%$ confidence interval. Left panel is temperature and redshift and right panel is the $[X/\mathrm{H}]$ and redshift. We do not include a $\chi^2$ curve in the plots as the fit was poor due to a large scatter.}
        \label{fig:BestfitIGM_TandZ}
\end{figure*}

We note that in all cases, the addition of the absorption component improved the Cstat fit for all blazars in our sample indicating that the addition of the absorption component is required in the model. Overall, the best fit Cstat results were achieved using the IGM component with a log-parabolic power law (26/40 spectra). Only 6 blazars out of the sample with $z > 1$ (23), had a better fit with a broken power law. The continuum model fit favouring a log-parabolic or broken power law over a simple power law is consistent with prior studies \citep[e.g.][A18]{Bhatta2018,Sahakyan2020,Gaur2020}. Modelling the IGM using \textsc{hotabs} for CIE with parameters $\mathit{N}\textsc{hxigm}$, Z and T free, results in  $\mathit{N}\textsc{hxigm}$ showing similar values and trend with redshift as the mean IGM density model. In Fig. \ref{fig:NHX_z} left panel, we show the results for $\mathit{N}\textsc{hxigm}$ and redshift selecting the best fits from both log-parabolic and broken law (none were better with a simple power law). The right panel of Fig. \ref{fig:NHX_z} are the results using the log-parabolic power law with the IGM component for comparison. Due to the differences between FSRQ (33/40 in our sample, blue) and BL Lac (red), we have plotted both categories coloured separately.

Based on best fit results, a power law fit to the $\mathit{N}\textsc{hxigm}$ versus redshift trend for the FSRQ objects scales as $(1 + z)^{1.8\pm0.2}$, reduced $\chi^2 = 1.69$, (p-value = 0.0011, root mean square (rms) = 0.39). However, given that the FSRQ sample redshift includes blazars as low as $z = 0.31$, a linear $\chi^2$ fit is not appropriate as can be seen from the simple IGM curve. The mean hydrogen density using equation \ref{eq:simpleIGM} at $z = 0$ from the FSRQ sample is $n_0 = (3.2\pm{0.5}) \times 10^{-7}$ cm$^{-3}$. This is higher than the value of $1.7 \times 10^{-7}$ cm$^{-3}$ for the simple  IGM model (green line in Fig \ref{fig:NHX_z}). Taking a sub-sample of FSRQ with $z >1.6$, similar to the GRB sample in D21, gives $n_0 = (2.1\pm0.2) \times 10^{-7}$ cm$^{-3}$. While noting that many of the error bars are large, jointly the blazars would support any model that is proximate to the $\chi^2$ fit which includes the mean IGM density curve. At low redshift, several blazars have higher $\mathit{N}\textsc{hxigm}$ than the simple IGM model. This may be evidence of the CGM in both our Galaxy and the host galaxy providing a minimum column density. While there is no observed significant evolution in neutral hydrogen column density in the CGM, there is evidence of evolution in total hydrogen column density including the partially ionised hydrogen column \citep{Fumagalli2016,Lehner2016}.  Models incorporating more advanced modelling for a warm/hot CGM component  are needed to explore the relative contribution of the IGM and the host CGM to the observed absorption in blazars and GRBs (D21). \citet{Das2021} claim to have detected three distinct phases in our Galaxy CGM, with a hot phase having log($T$/K) $\sim 7.5$ and log($\mathit{N}\textsc{hx}) \sim21$. The BL Lacs dominate the sample at very low redshift and the majority appear  to have high $\mathit{N}\textsc{hxigm}$ although with large error bars. This could be due to the overall model incorrectly describing the BL Lac spectra. The  4 BL Lac objects that most exceed the simple IGM curve are all high energy peaked blazars, known as HBLs. These objects have their peak synchrotron humps at energies that can appear in soft X-ray. Therefore, we have excluded the BL Lac in our derivation of the mean hydrogen density at $z = 0$ and the $\chi^2$ fit for  $\mathit{N}\textsc{hxigm}$ versus redshift trend.

 There is a large range in the fitted temperature $5.0 <$ log($T$/K) $< 8.0$, many with substantial error bars in Fig. \ref{fig:BestfitIGM_TandZ} left panel. The mean temperature over the full redshift range is log($T$/K) $= 6.1\pm{0.1}$. These values are consistent with the generally accepted WHIM.  There is no apparent relation of temperature with redshift. It should be noted, however, that the fits are for the integrated LOS and not representative of any individual absorber temperature. Further, at high redshift, it is possible that the IGM comprises a cooler diffuse gas which is contributing to the absorption but not captured in this CIE model.
 
 The right panel of Fig. \ref{fig:BestfitIGM_TandZ} shows no apparent relation of $[X/$H$]$ with redshift. The mean metallicity over the full redshift range is $[X/$H$] = -1.62\pm0.04$. Metallicity ranges from approximately $[X/$H$] -0.7\/\ (0.2Z\sun)$ to $[X/$H$] -3\/\ (0.001Z\sun)$ with one outlier. This is the BL Lac Mrk 501 at $z = 0.03$. The initial fitting went to the upper metallicity limit of $[X/$H$] < -0.7$. We increased the upper limit to solar and the best fit was with $[X/$H$] = -0.08 (0.8Z\sun)$. Our model may not be appropriate for this object.

\graphicspath{ {./figurespaper3/}  }
 \begin{figure*} 
    \centering
    \begin{tabular}{c|c}
    
    \includegraphics[scale=0.55]{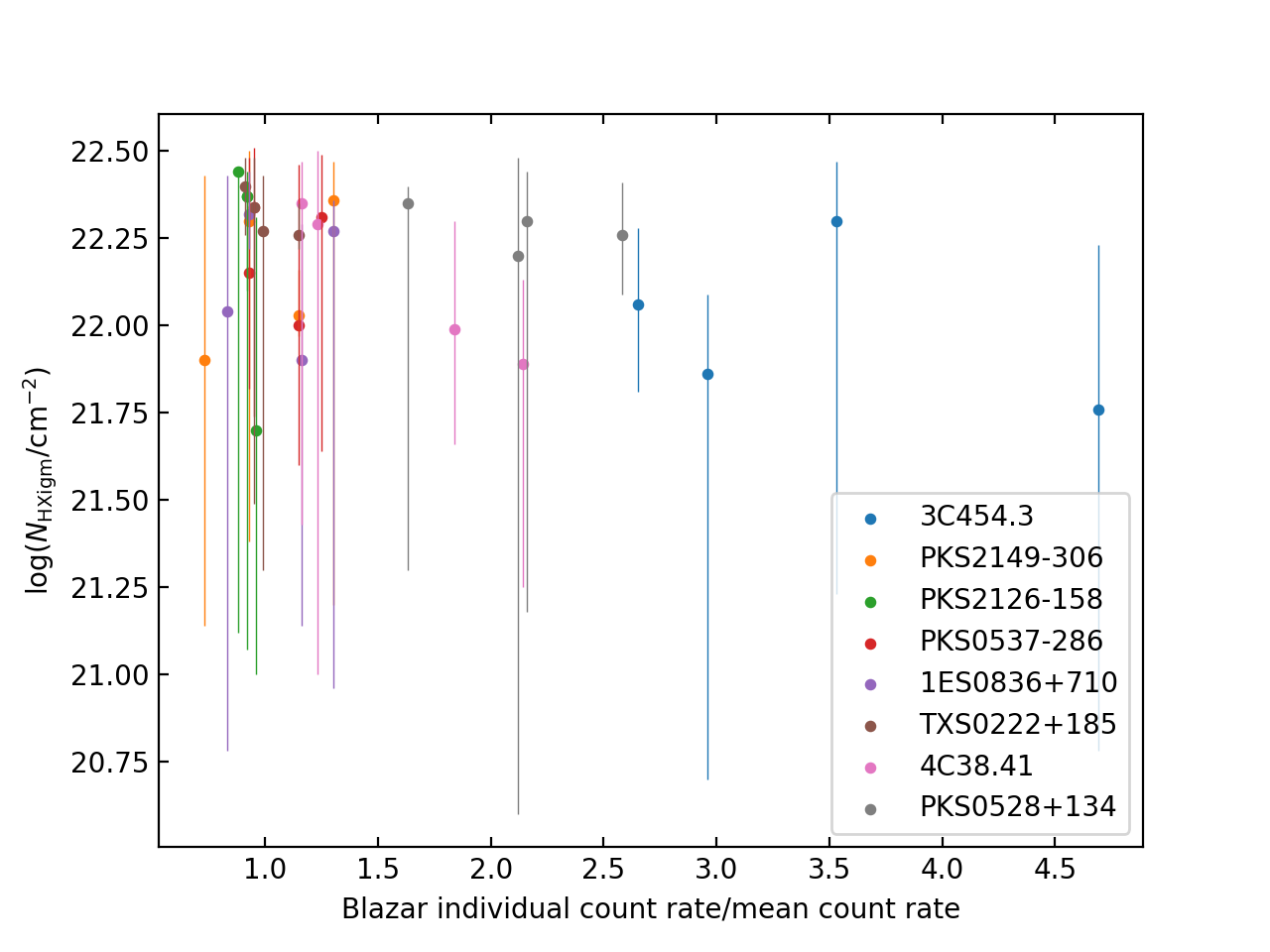} &
    \includegraphics[scale=0.55]{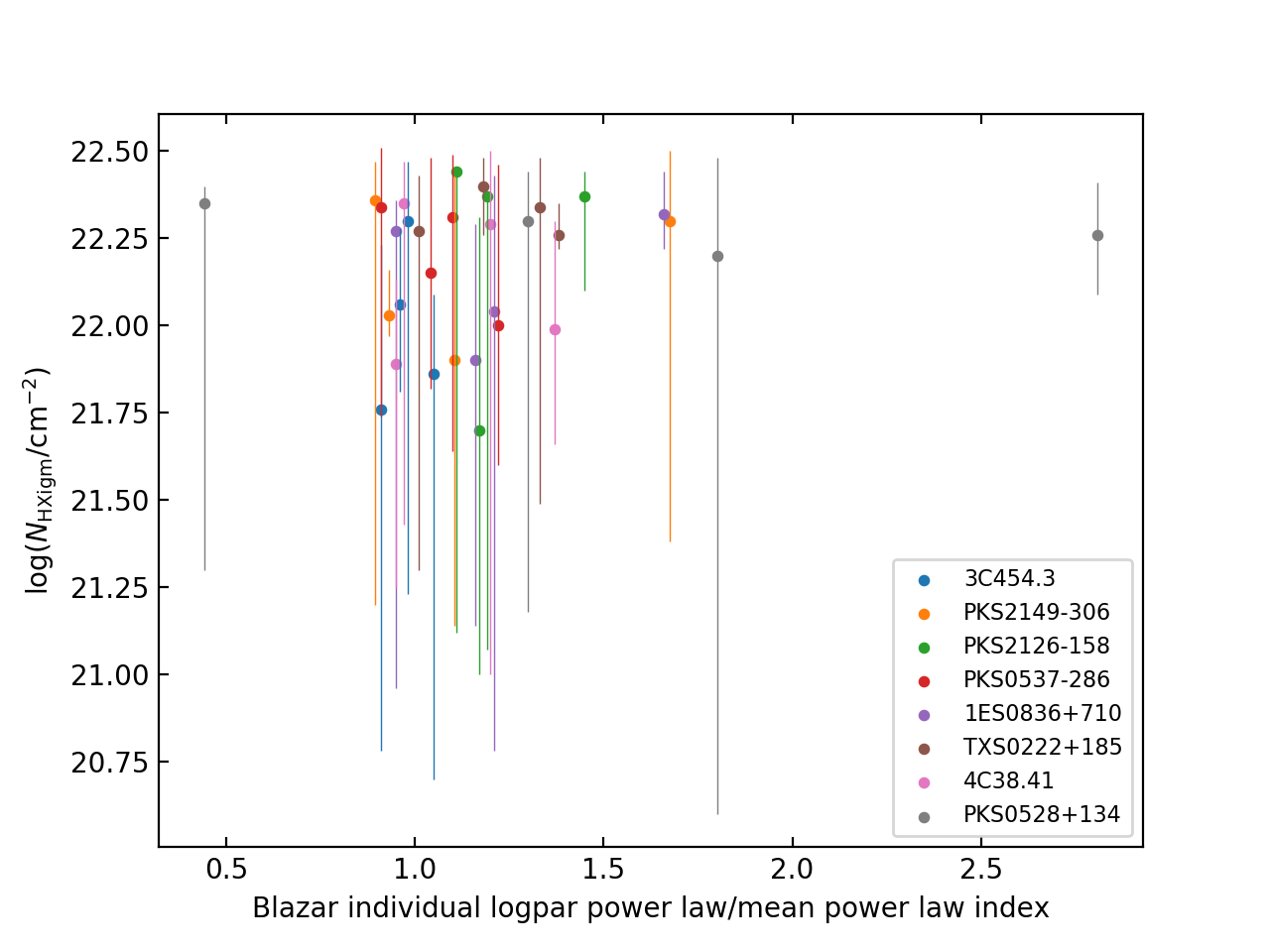} 
    \\
    \end{tabular}
    \caption{Testing for possible relation between $\mathit{N}\textsc{hxigm}$ and flux variability  or spectral slope for four individual observations from the sub-sample of eight blazars. Objects are colour-coded to enable comparison of individual blazar results as well as overall possible relations. Left panel is $\mathit{N}\textsc{hxigm}$ and flux variability given as count rate over mean count rate . Right panel is $\mathit{N}\textsc{hxigm}$ and individual log-parabolic power law/ mean power law index for each object. The error bars are reported with a $90\%$ confidence interval.}
        \label{fig:NHX and flux}
\end{figure*}

In conclusion, with the caveats of low X-ray resolution, a CIE IGM component only and the slab model to represent to full LOS, there are reasonable grounds for arguing that the CIE model using \textsc{hotabs} is plausible for modelling the warm/hot component  of the IGM at all redshifts. In all fits, the Cstat was better than best fits for models with only Galactic absorption. Our CIE IGM component had three free parameters, $\mathit{N}\textsc{hxigm}$, temperature and $[X/$H$]$. There is scope for degeneracy as we model the continuum curvature and not specific absorption features as set out in Section \ref{sec:3} and D21. While there was a large range in Cstat improvements across the sample, the average Cstat improvement for the full sample per free IGM parameter was 3.9, with 20 out of 40 blazars exceeding $\Delta$Cstat$^2> 2.71$.  The model using a log-parabolic continuum model with a CIE IGM absorption appears to be more consistent with the simple IGM curve than the selected best fits from both log-parabolic and broken power laws. Overall, the results for $\mathit{N}\textsc{hxigm}$ using either a log-parabolic or broken power law are statistically indistinguishable indicating that IGM component is independent.  Our temperature and metallicity results are consistent with the expected values from simulations for a warm/hot phase. However, as noted in Section \ref{sec:3}, initial trials with a warm photionised IGM component gave similar results to a collisional ionised model. It is most likely that a combined model would be more physical, but testing this requires better data. We now test the robustness of the results in Section \ref{sec:Robust} and discuss the results further and compare with other studies in Section \ref{sec:Discuss}.

\section{Tests for robustness of IGM parameter results}\label{sec:Robust}

There are several alternative potential explanations for the curvature seen in blazar spectra which may not be partly or wholly attributable to IGM absorption. $\mathit{N}\textsc{hxigm}$ can be degenerate to some degree with spectral slope i.e. a harder spectrum slope can mimic higher $\mathit{N}\textsc{hxigm}$ and vice versa. Further, blazars can be highly variable in flux and spectral slope. Finally, some fits were nearly indistinguishable in terms of visual spectra  ratio and/or Cstat i.e. there may be a concern about an a priori assumption of IGM absorption. Accordingly, we examine our results from a number of perspectives to test their robustness.

\graphicspath{ {./figurespaper3/}  }
 \begin{figure*} 
    \centering
    \begin{tabular}{c|c}
    
    \includegraphics[scale=0.55]{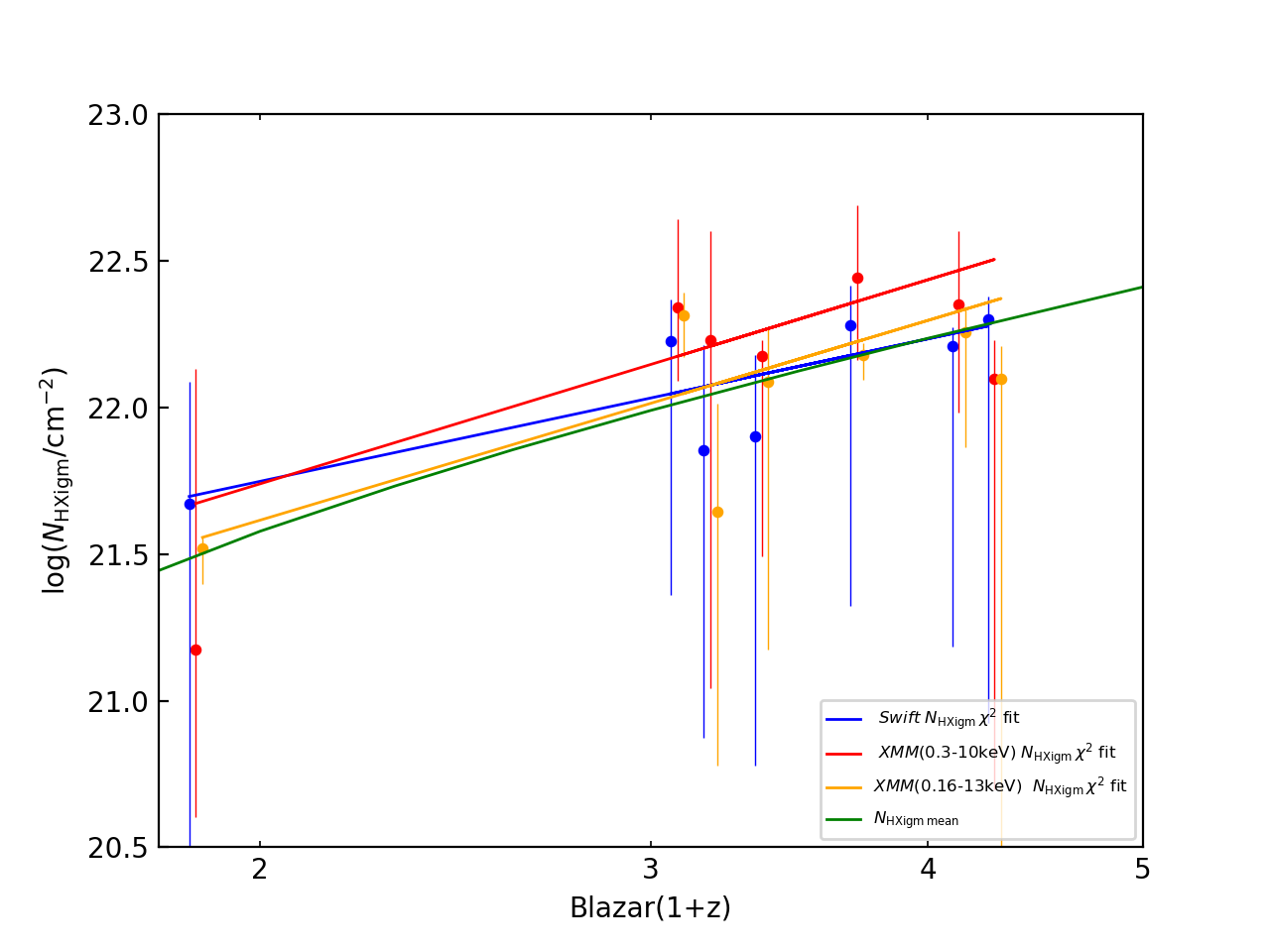} &
    \includegraphics[scale=0.55]{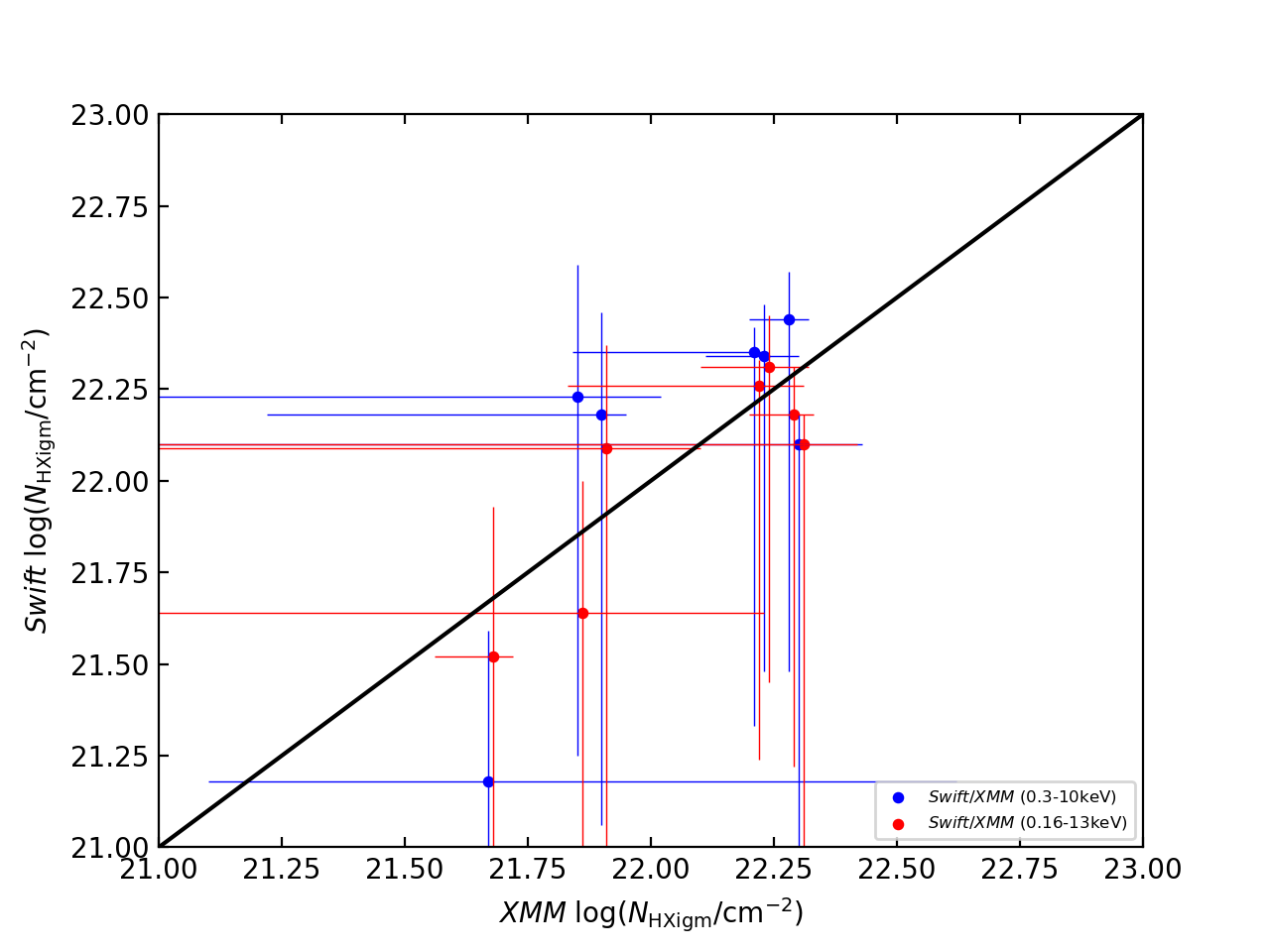} 
    \\
    \end{tabular}
    \caption{Comparing $\mathit{N}\textsc{hxigm}$ and redshift results for $\textit{Swift}$ (blue), $\textit{XMM-Newton}$ (0.3-10keV)(red) and (0.16-13keV)(orange) from a sub-sample of seven blazars (varied slightly on the left panel x-axis to enable error bar visibility). Left panel is $\mathit{N}\textsc{hxigm}$ and redshift. The green line is the simple IGM model using a mean IGM density. Right panel is $\mathit{N}\textsc{hxigm}$ for $\textit{XMM-Newton}$ on the x-axis (varied marginally for visibility) and $\textit{Swift}$ 0.3-10keV (blue) and 0.16-13keV (red) on the y-axis. The error bars are reported with a $90\%$ confidence interval. The black line in the right panel is parity.}
        \label{fig:XMM comparison}
\end{figure*}

\graphicspath{ {./figurespaper3/}  }
 \begin{figure*} 
    \centering
    \begin{tabular}{c|c}
    \includegraphics[scale=0.3]{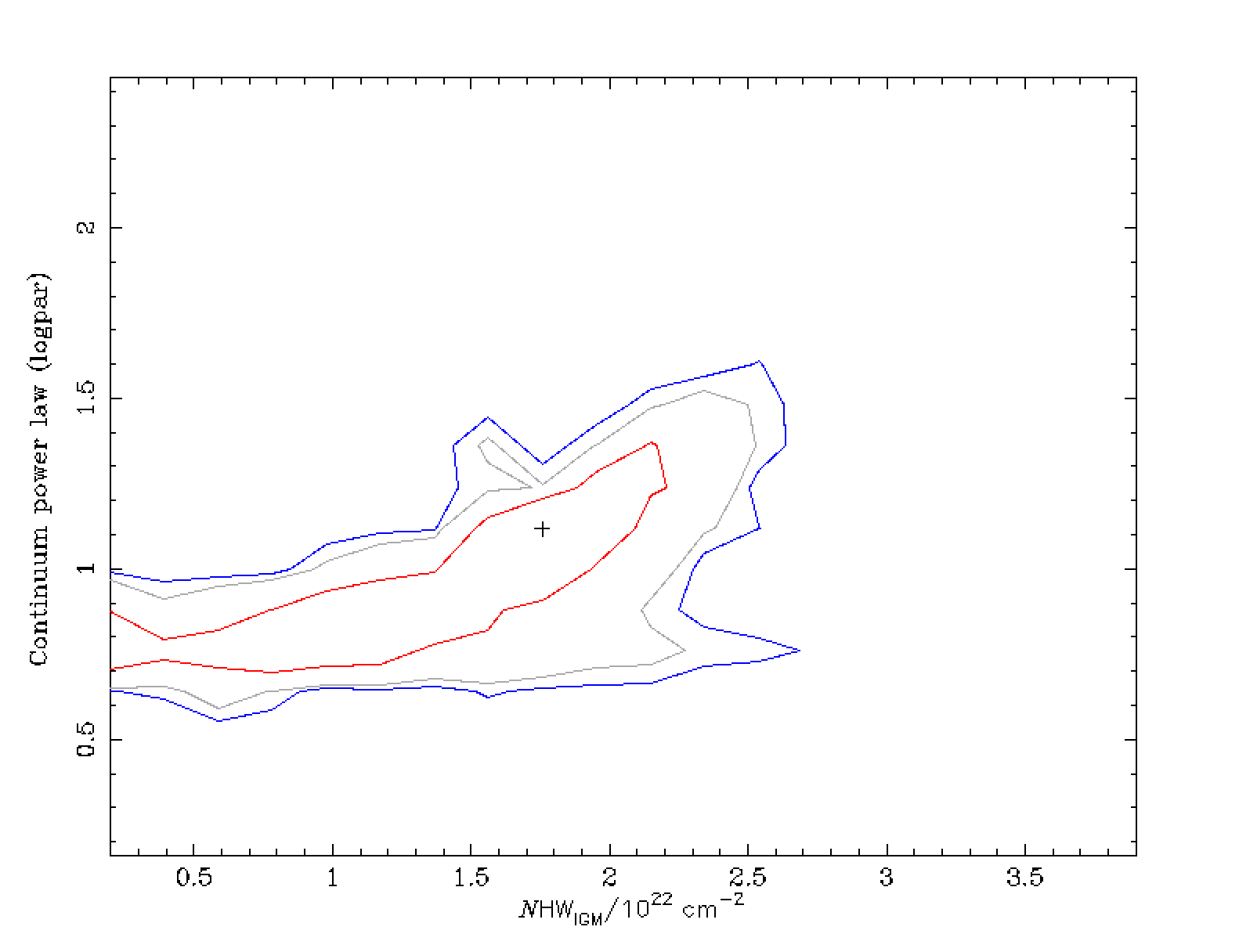} &
    \includegraphics[scale=0.3]{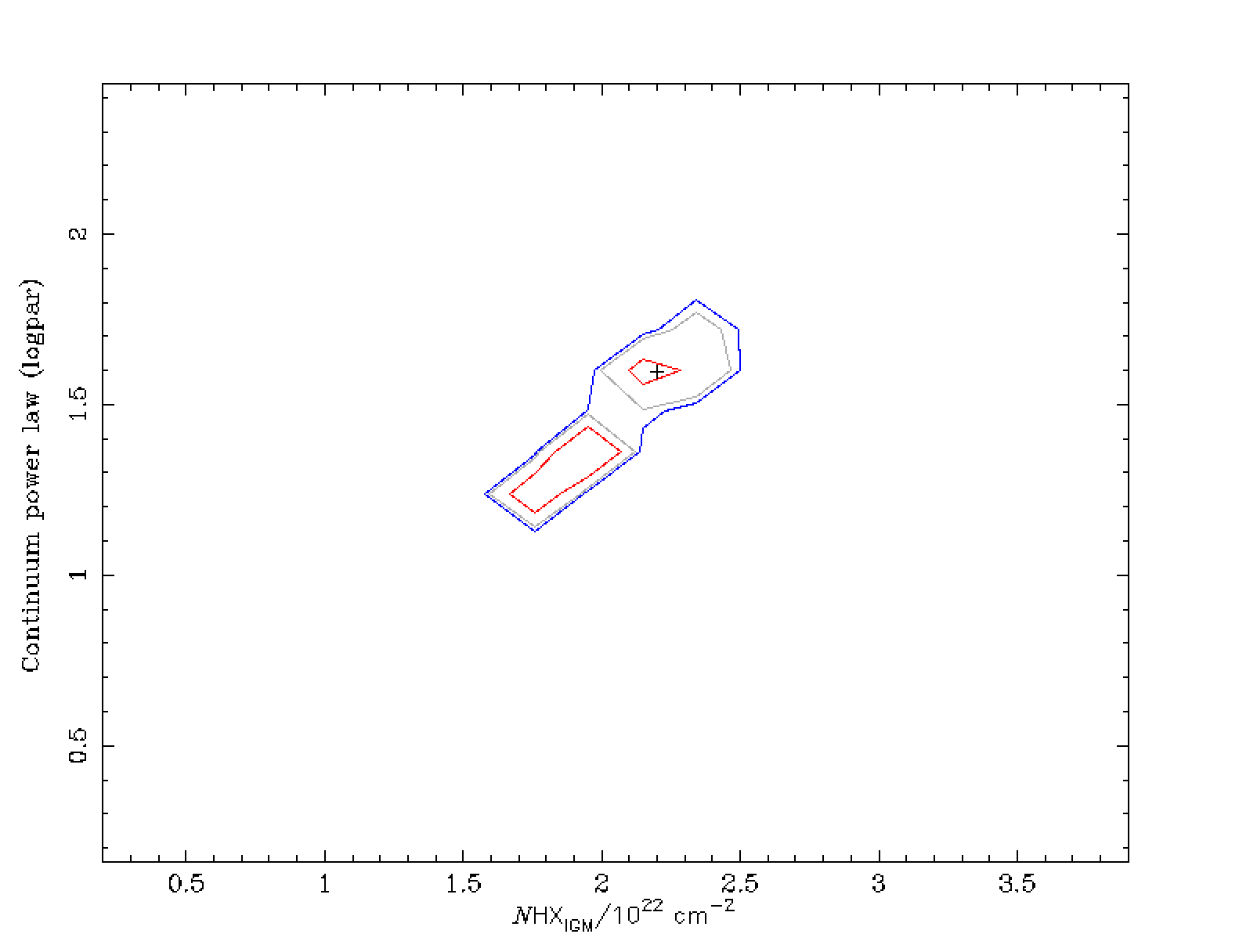}
    \\
    \end{tabular}
    \caption{The left and right panels show the MCMC integrated probability results for PKS 0528+134 $\mathit{N}_{\textsc{hxigm}}$ and log-parabolic power law indices for $\textit{Swift}$ and $\textit{XMM-Newton}$ 0.3-10keV respectively. The red, grey and blue contours represent $68\%, 90\%$ and $95\%$ ranges.}
        \label{fig:PKS0528+134_MCMC}
\end{figure*}

 \subsection{Flux variability}\label{subsec:5.1}
Blazars are known to have spectral variability (associated with flux variability). The physical causes of such variability can involve many processes, such as particle acceleration, injection, cooling, and escape, which contain a number of known and unknown physical parameters \citep{Gaur2020}. Also, a local absorber at the host can show variability, while an IGM absorber should not. An absorber that shows variability within a reasonable short time-frame cannot be on intergalactic length scales, and must therefore be attributed to the host, or to intrinsic variability of the source \citep{Haim2019}.

As noted in Section \ref{sec:2}, we use co-added blazar spectra. To test for possible absorption variability and/or a relation between flux and spectral hardening, we selected a sub-sample of 8 blazars with a redshift range $0.86 \leq z \leq 3.26$ as representative of the full sample. For each, we selected four different individual observations taken at different dates, with high counts but which showed different flux rates to the mean co-added rate. We used the log-parabolic model for all 8 blazars. Table \ref{tab:individual_blazar sub-sample_results} in the Appendix reports the Observation ID, count rate over the mean count rate, $\mathit{N}\textsc{hxigm}$ and log-parabolic power law over the mean power law for each blazar observation.

In Fig. \ref{fig:NHX and flux} left panel, we can see that there is no apparent relation between $\mathit{N}\textsc{hxigm}$ and flux across all the observations, with a $\chi^2$ fit slope approximating zero ($-0.09\pm0.05$). The blazars are individually colour-coded and there is no obvious relation between $\mathit{N}\textsc{hxigm}$ and flux for any individual blazar apart from possibly 4C 38.41. From Table \ref{tab:individual_blazar sub-sample_results}, we can see that all the individual results for $\mathit{N}\textsc{hxigm}$ are consistent with the mean result within the errors for each blazar.

\subsection{Column density and spectral slope degeneracy}\label{subsec:5.2}

Given the scope for degeneracy between $\mathit{N}\textsc{hxigm}$ and spectral hardening, for the sub-sample of 8 blazars we checked for any relation between these two parameters using the same four individual observations for each. As previously noted, due to the large variability frequently observed in blazars, non-simultaneous observations are expected to show different states of the object. Therefore, the log-parabolic parameters and normalizations were allowed free to vary. In Fig. \ref{fig:NHX and  flux} right panel, we show the  $\mathit{N}\textsc{hxigm}$ using log-parabolic power law for all fits for the individual observations. There is no apparent relation between column density and power law indices with a $\chi^2$ slope fit of $0.10\pm0.06$. We again colour code each blazar and there is no apparent relation between $\mathit{N}\textsc{hxigm}$ and power law index variability for any individual object.

Overall, our results are consistent with other studies \citep[e.g.][A18]{Haim2019} which tested for a fixed absorber and noted that it did not change significantly across different observational times.

\graphicspath{ {./figurespaper3/}  }
 \begin{figure*}
     \centering
     \begin{tabular}{c|c|c}
    \includegraphics[scale=0.2]{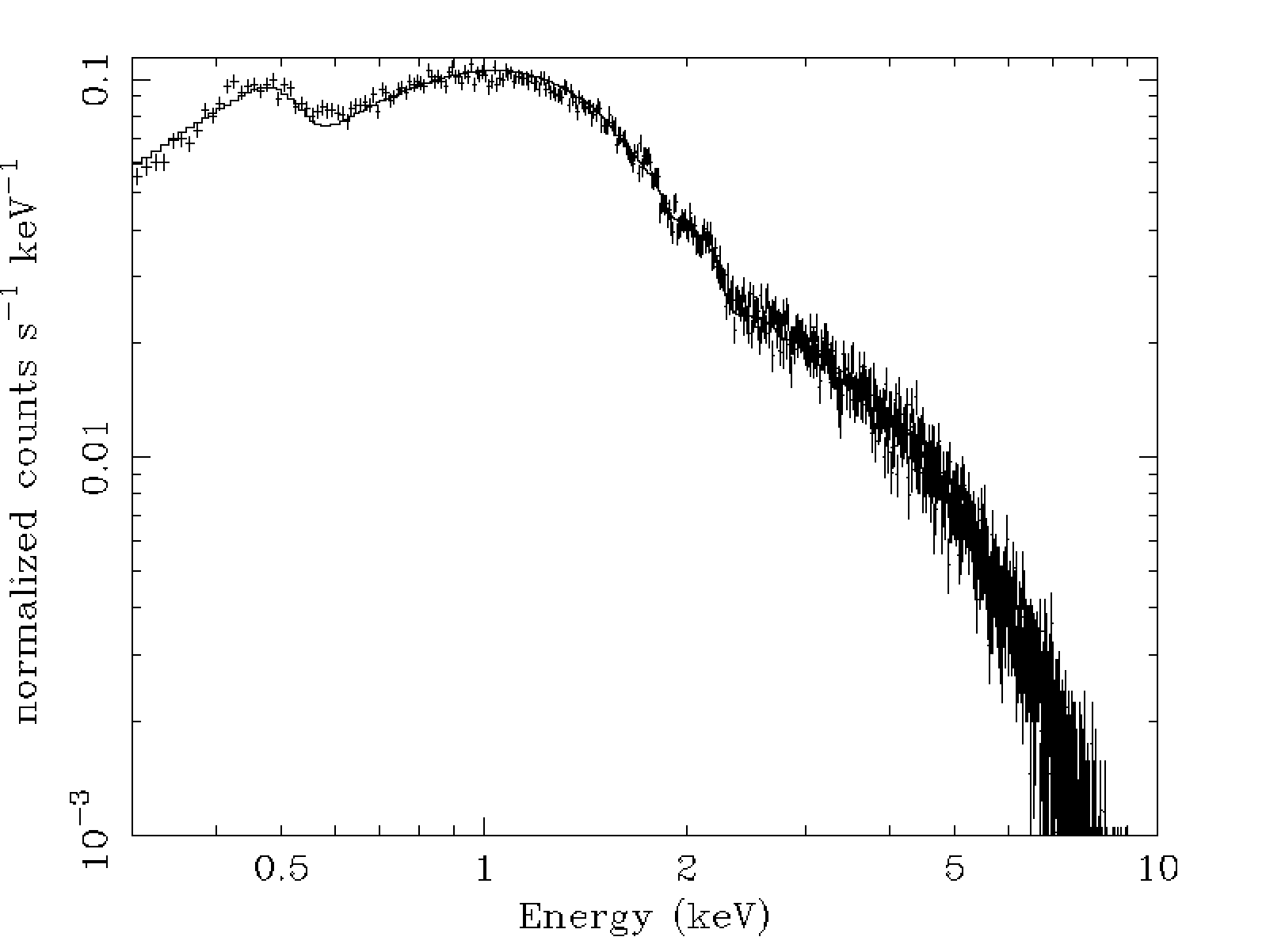} &
    \includegraphics[scale=0.2]{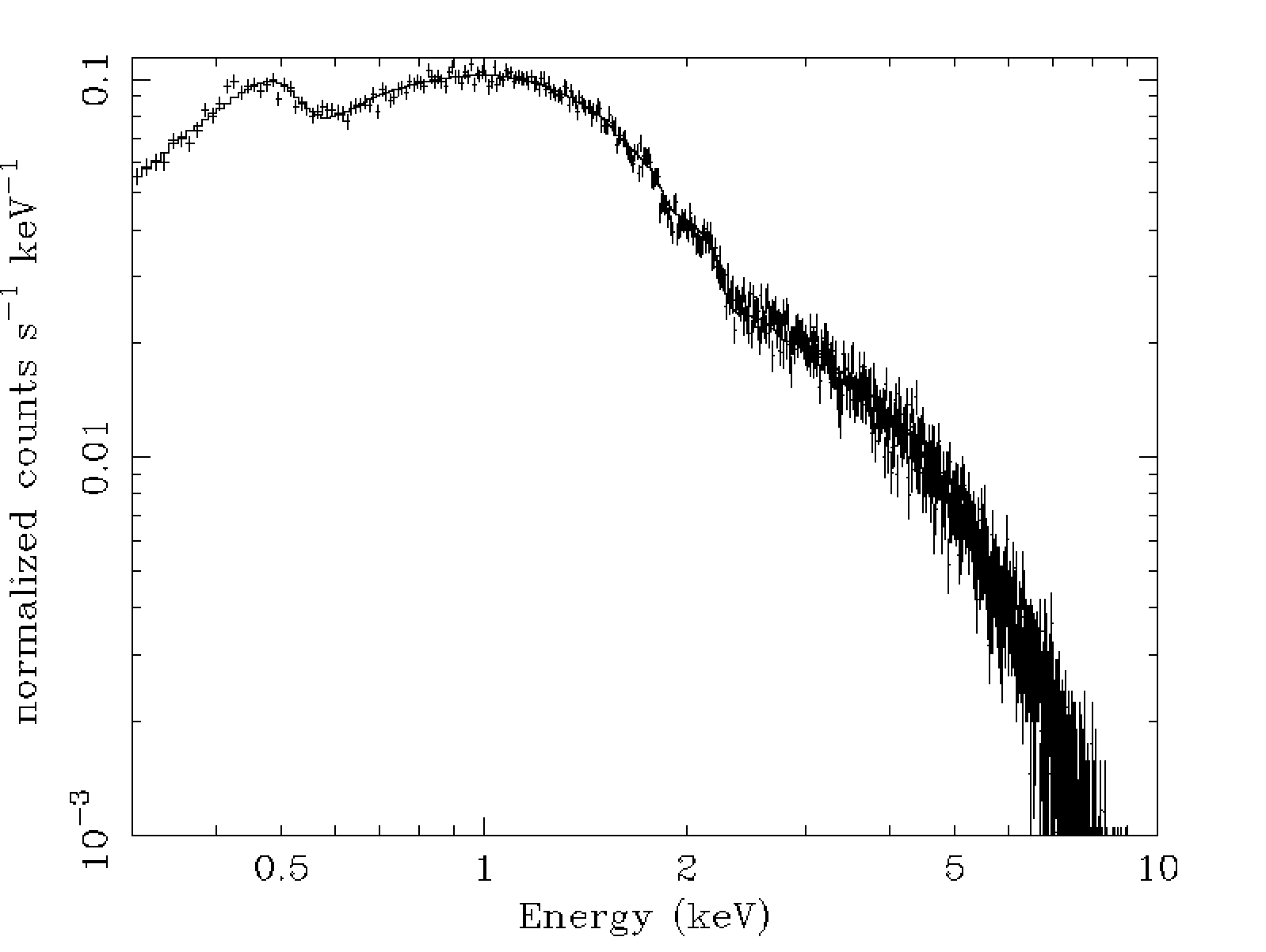} &
    \includegraphics[scale=0.2]{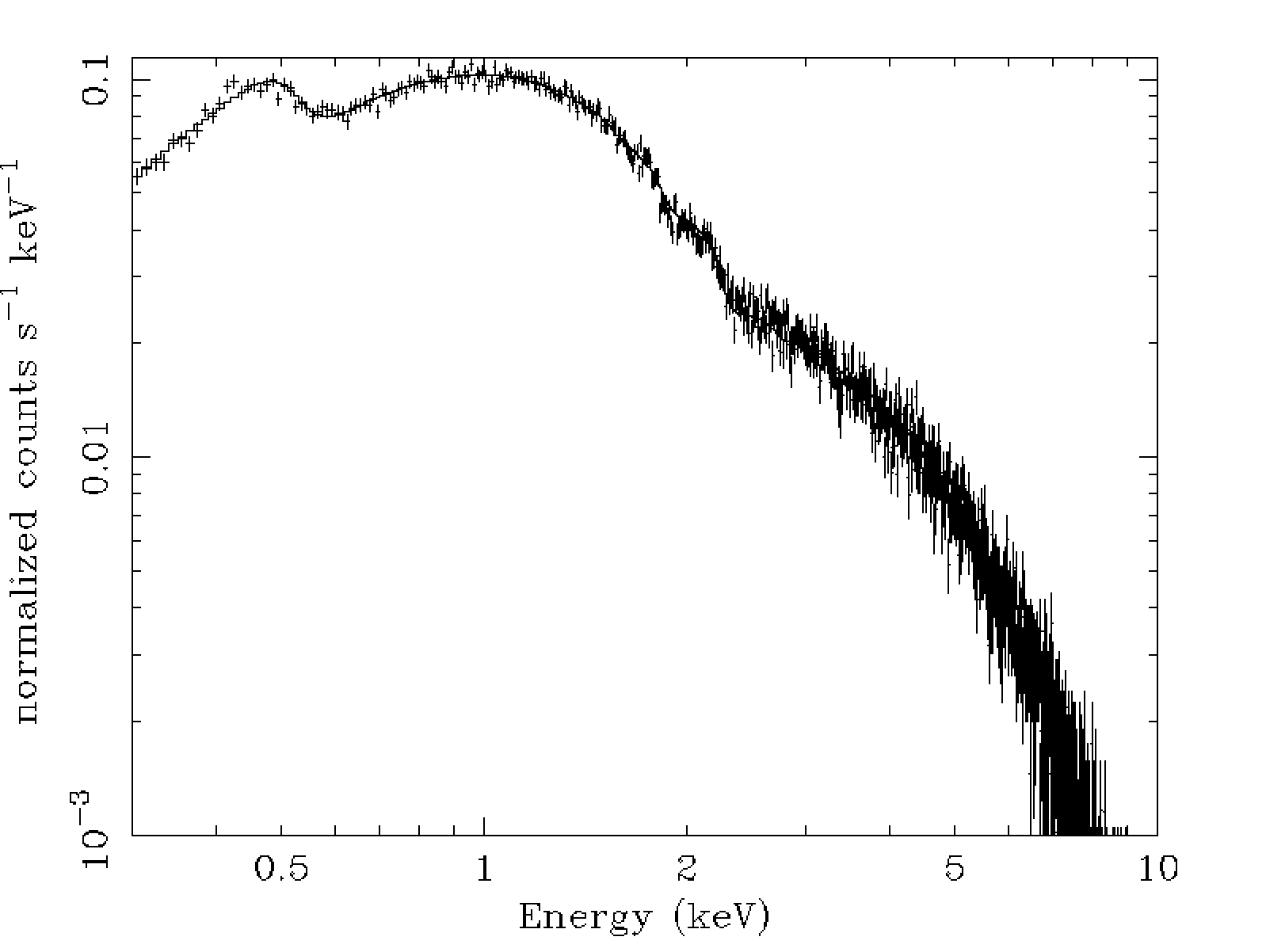}
     \\
    \end{tabular}
    \caption{3C 279 as an example of FSRQ spectrum showing  second hump in soft X-ray. All fits are with a broken power law, the best intrinsic curvature. The left panel is with an IGM component. The middle panel is with a blackbody component. The right panel is with both an IGM absorption component and a blackbody.}
        \label{fig:3C279}
\end{figure*}

\graphicspath{ {./figurespaper3/}  }
\begin{figure}

	\includegraphics[width=\columnwidth]{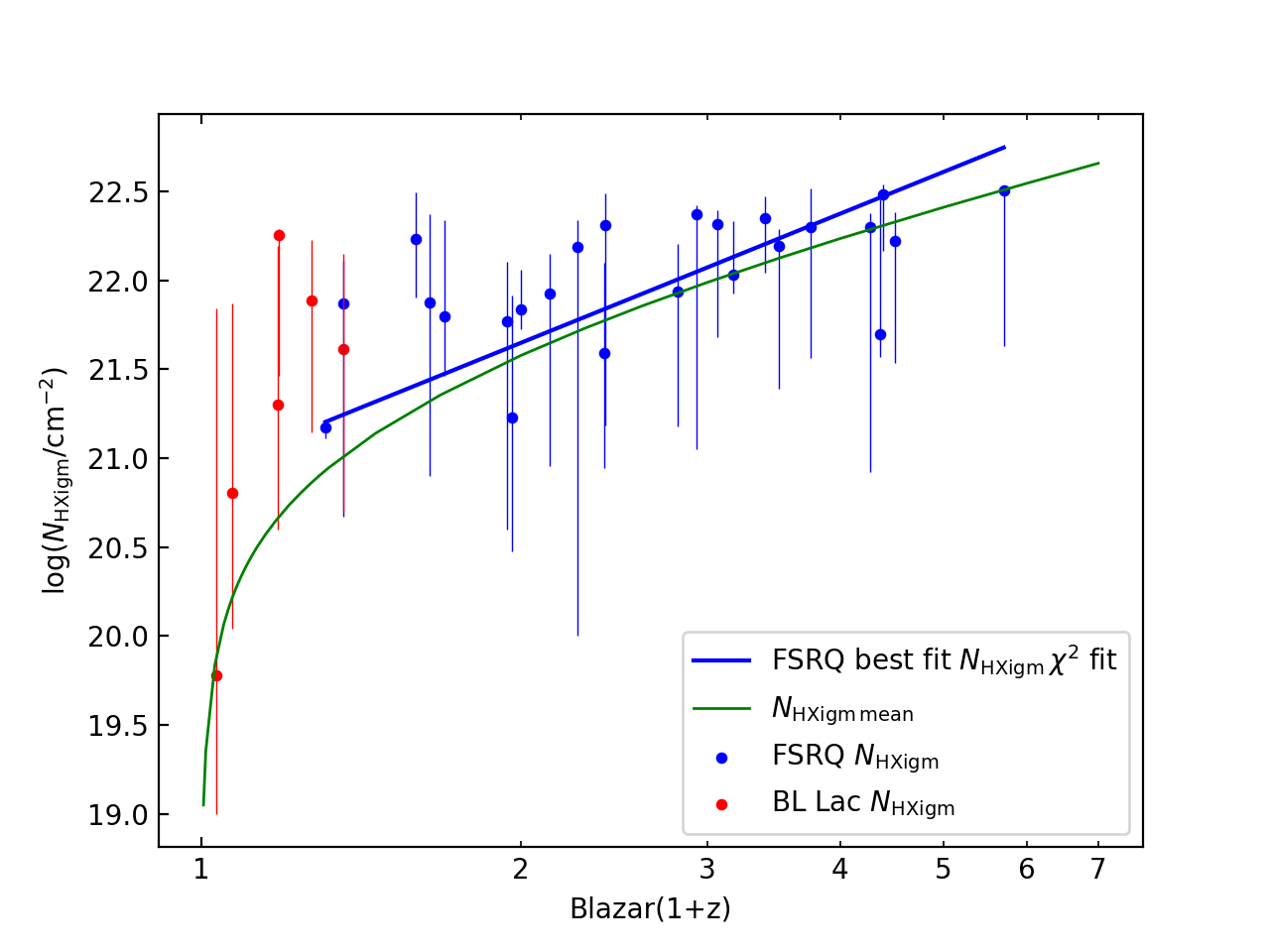}
    \caption{Results for the IGM $\textit{N}\textsc{hx}$ parameter and redshift omitting the 10 blazars where adding a bulk Comptonisation component improved the fit. The error bars are reported with a $90\%$ confidence interval. The green line is the simple IGM model using a mean IGM density.}
    \label{fig:NHX_z_excludeBB}
\end{figure}

\subsection{XMM-Newton spectra comparison}\label{subsec:5.3}

$\mathit{XMM-Newton}$ has excellent low energy response down to 0.15 keV, extreme sensitivity to extended emission, and large effective area facilitating analysis of the soft X-ray properties. We continued our robustness tests using $\textit{XMM-Newton}$ PN spectra for the same sub-sample of blazars with the exception of 4C 38.41 which was not available.  We used the log-parabolic power law again with the CIE IGM component. We chose a selection of both individual spectra and co-added spectra as given in Table \ref{tab:XMM_blazar_subsample   }.  We used the same energy range as $\textit{Swift}$ for consistent comparison (0.3 - 10 keV). We also separately fitted our model to an extended energy range of 0.16 - 13 keV given the excellent sensitivity of $\textit{XMM-Newton}$ over this range.

Table \ref{tab:XMM_blazar sub-sample_results} in the Appendix reports the redshift, and  $\mathit{N}\textsc{hxigm}$ for $\textit{Swift}$ 0.3-10keV, $\textit{XMM-Newton}$ 0.3-10keV and 0.16-13keV respectively for each blazar. The values for $\mathit{N}\textsc{hxigm}$ are consistent for each blazar within the errors. Fig. \ref{fig:XMM comparison} left panel, shows $\mathit{N}\textsc{hxigm}$ for $\textit{Swift}$ and both $\textit{XMM-Newton}$ fits for each blazar. To enable the error bars to be separately visible, we have very slightly changed the redshift of the three spectra for each object. All $\mathit{N}\textsc{hxigm}$ are proximate to the simple IGM curve. We also plot the best $\chi^2$ fit for the $\textit{Swift}$ and both $\textit{XMM-Newton}$ energy ranges. The slopes of both the  $\textit{XMM-Newton}$ fits are very similar with 0.3-10 keV being $2.3\pm1.0$ and 0.16-13 keV being $2.3\pm0.4$ . The slope of the $\chi^2$ fit for  $\textit{Swift}$ is slightly less at $1.6\pm0.7$. Overall, within the errors, the $\textit{Swift}$ and both $\textit{XMM-Newton}$ results are consistent. 

In Fig. \ref{fig:XMM comparison} right panel, we plot $\mathit{N}\textsc{hxigm}$ for $\textit{Swift}$ on the x-axis and $\textit{XMM-Newton}$ 0.3-10keV (blue) and 0.16-13keV (red) on the y-axis The black line in the right panel is parity. We have varied the \textit{Swift} $\mathit{N}\textsc{hxigm}$ marginally to enable error bar visibility. All $\textit{Swift}$ and $\textit{XMM-Newton}$ data-points are proximate to the black parity line, with the possible exception of 3C 454.3 which has a reasonably higher $\mathit{N}\textsc{hxigm}$ with $\textit{Swift}$ than $\textit{XMM-Newton}$ 0.3-10keV, but consistent within the errors. 

Fig. \ref{fig:PKS0528+134_MCMC}  show the MCMC integrated probability results for PKS0528+134 as a typical example of results for $\mathit{N}\textsc{hxigm}$ and log-parabolic power law for $\textit{Swift}$ and $\textit{XMM-Newton}$ 0.3-10keV respectively. While the $\mathit{N}\textsc{hxigm}$ best fit result is similar for both, due to the high resolution of $\textit{XMM-Newton}$, the contours are much tighter. 

Overall, our investigations demonstrate that our results are consistent for observations by both $\textit{Swift}$ and $\textit{XMM-Newton}$. Further, they reinforce the findings from \ref{subsec:5.2} that the IGM absorption results do not vary on a temporal basis.

\subsection{Bulk Comptonisation}
Some blazar spectra have a hump feature at soft X-rays, whose origin is still debated. Bulk Comptonisation (BC) has been suggested as an explanation where cold electrons could up-scatter cooler extreme ultraviolet photons from the disk and/or BLR to soft X-ray energies \citep[e.g][]{Sikora1994,Celotti2007}. This BC related excess emission over the blazar continuum would appear as a hump in soft X-ray. Depending on the energy peak of this hump, it could mimic or mask absorption. If the the hump were to be in the region of $\sim3$ keV, the apparent deficit at softer energies can mimic absorption \citep[and references therein]{Kammoun2018}.  This possible BC related feature has been modelled using a blackbody as a phenomenological representation \citep[e.g][]{Ricci2017}.

Fig. \ref{fig:3C279} shows the spectrum of 3C 279 as an example of a FSRQ showing a second hump in soft X-ray where adding a blackbody component significantly improved the Cstat fit. All fits are with a broken power law which provided the best intrinsic curvature fit. The left panel is with an IGM component (Cstat 992.1/933). The middle panel is with no IGM but an additional blackbody emission component (Cstat 917.06/934), where a slight visual improvement in fit can be seen at $\sim 0.6$ keV. The right panel is with both an IGM and blackbody component (Cstat 916.03/931). 

10 out of 40 in our blazar sample had a second soft X-ray hump where the Cstat was similar for both an IGM or blackbody component, and showed an improvement in Cstat with both components included. There was a large range in Cstat improvement for the model with both the IGM and blackbody components for the 10 blazars, with the average Cstat improvement per additional free blackbody parameter being 8.3, and  8 out of 10 blazars exceeding $\Delta$Cstat$^2 > 2.71$ .

In 8/10, there was a reduction in $\mathit{N}\textsc{hxigm}$ in the combined IGM and BC model ranging up to one dex. 2/10 objects showed increased $\mathit{N}\textsc{hxigm}$ from $100\% - 180\%$. Given this large impact on $\mathit{N}\textsc{hxigm}$, we replotted the $\mathit{N}\textsc{hxigm}$ redshift relation omitting the 10 blazars possibly impacted by a BC component. In Fig. \ref{fig:NHX_z_excludeBB}, we can see the clear $\mathit{N}\textsc{hxigm}$ redshift relation remains. In fact the power law fit to the $\mathit{N}\textsc{hxigm}$ versus redshift trend for the FSRQ objects scales as $(1 + z)^{2.4\pm0.2}$ (p-value = 0.01, rms = 0.36) compared to $(1 + z)^{1.8\pm0.2}$ from the full sample (Section \ref{sec:Blazar results}). The hydrogen equivalent density at $z = 0$ is similar at $n_0 = (3.5\pm{0.7}) \times 10^{-7}$ cm$^{-3}$. Without the BC component, 5/10 favoured a log-parabloic power law over a broken power law. When the blackbody component was added, this changed to 9/10 favouring a broken power law.

\graphicspath{ {./figurespaper3/}  }
 \begin{figure*} 
    \centering
    \begin{tabular}{c|c}
    \includegraphics[scale=0.55]{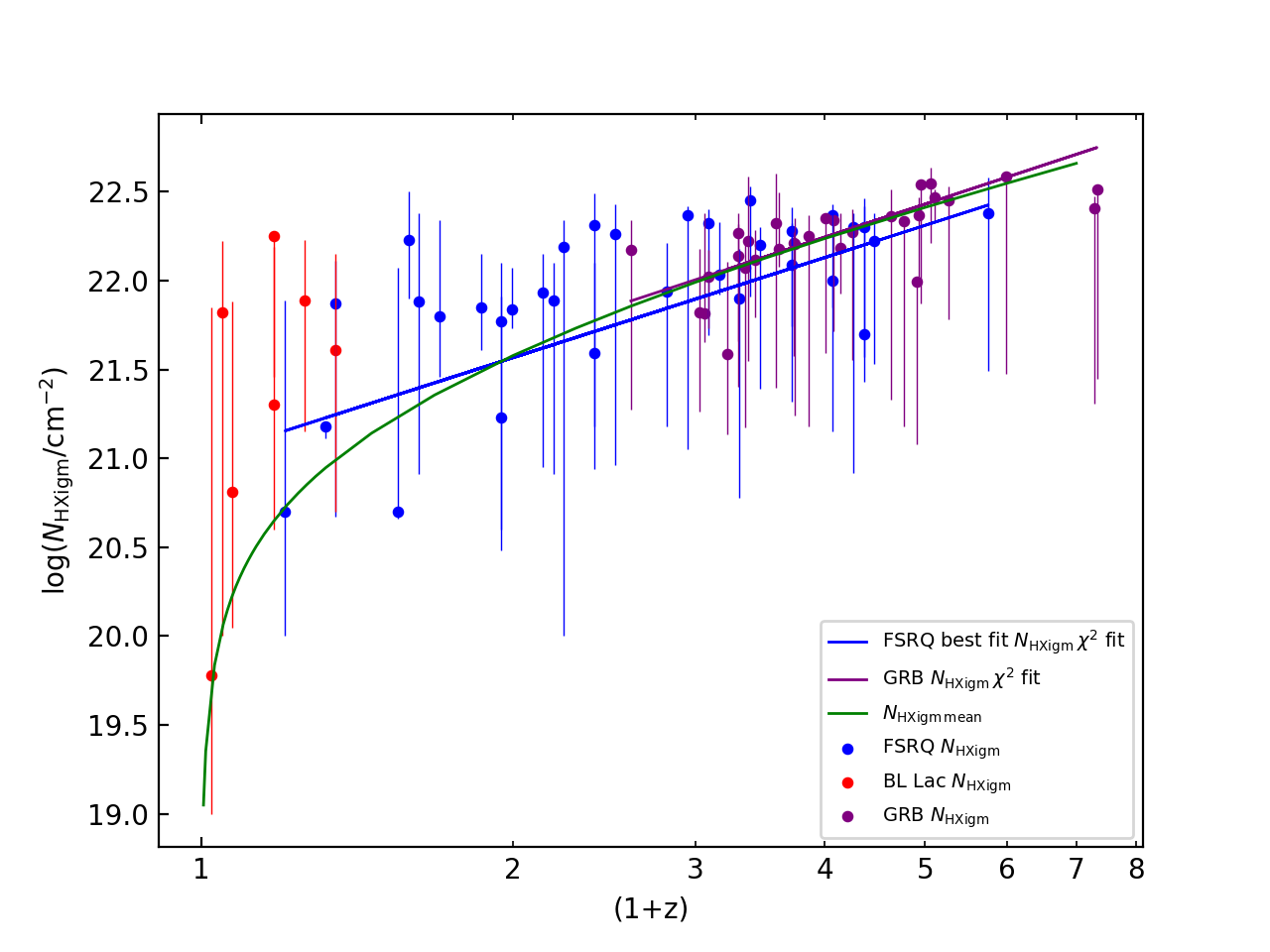} &
    \includegraphics[scale=0.55]{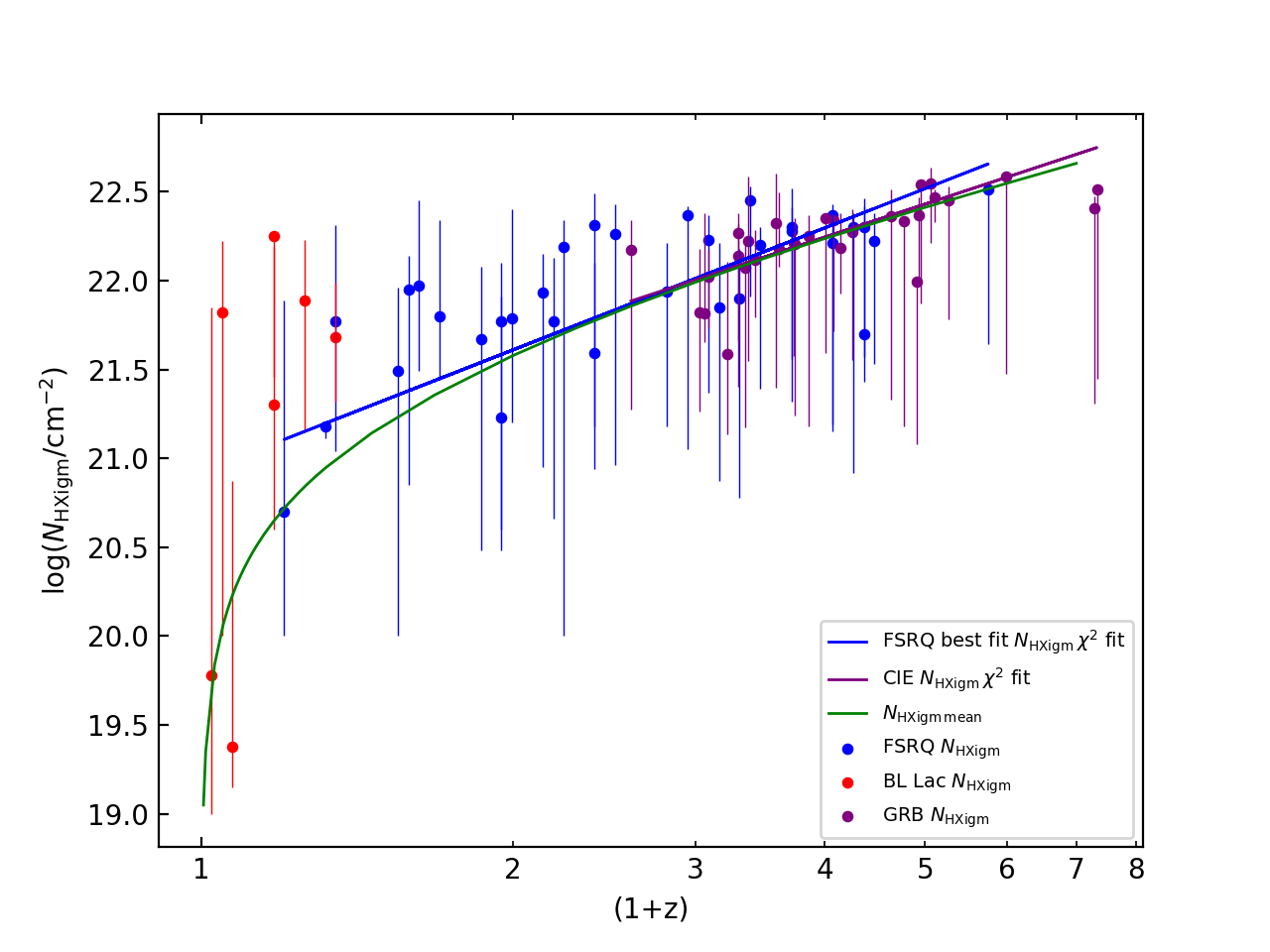}
    \\
    \end{tabular}
    \caption{Results for the IGM $\textit{N}\textsc{hx}$ parameter and redshift using the combined GRB sample from D21 (purple) and our Blazar sample (FSRQ - blue and BL Lac - red) using the CIE (\textsc{hotabs}) model. The blue and purple lines are $\chi^{2}$ fits to the respective FSRQ and GRB samples. The error bars are reported with a $90\%$ confidence interval. The green line is the simple IGM model using a mean IGM density. Left panel is $\textit{N}\textsc{hx}$ and redshift selecting best Cstat results for blazars from all three power law intrinsic models. Right panel is the full sample with the IGM component and a log-parabolic power law only (best fit for 26/40) for blazars. }
        \label{fig:GRB_blazar}
\end{figure*}

\graphicspath{ {./figurespaper3/}  }
 \begin{figure*} 
    \centering
    \begin{tabular}{c|c}
    \includegraphics[scale=0.55]{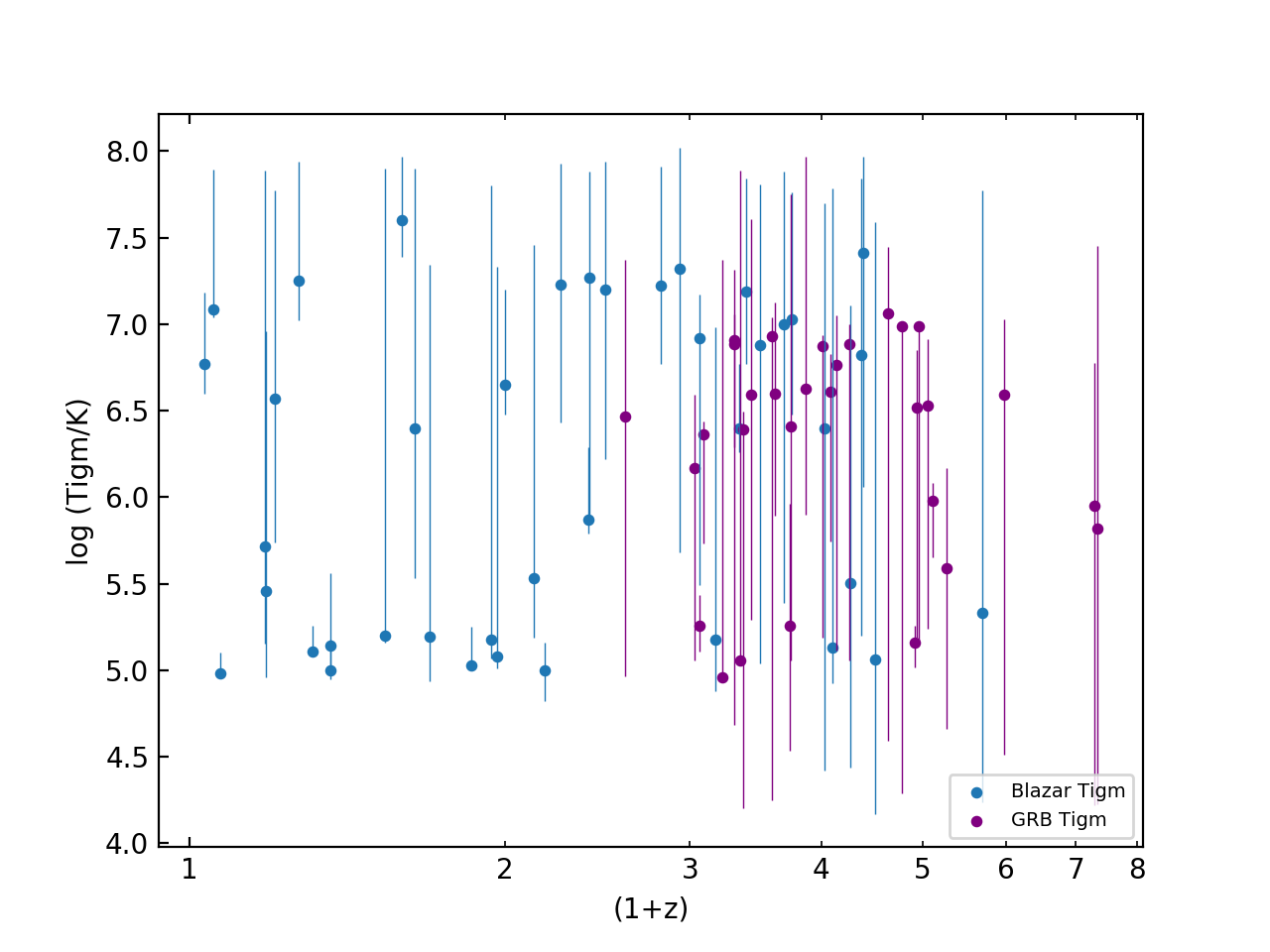} &
    \includegraphics[scale=0.55]{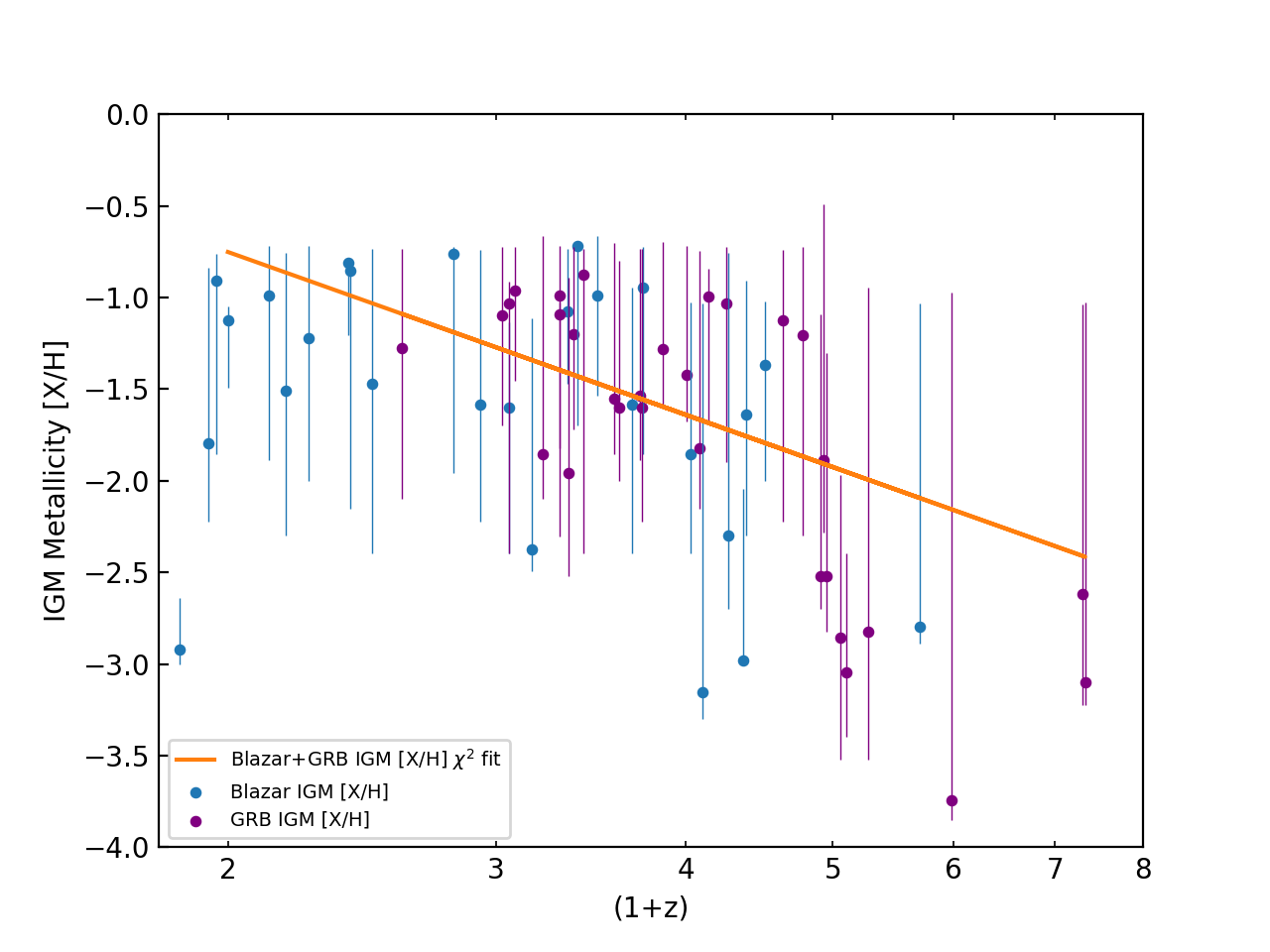}
    \\
    \end{tabular}
    \caption{IGM parameter results using the combined GRB sample from D21 (purple) and our Blazar sample (blue). The error bars are reported with a $90\%$ confidence interval. Left panel is temperature and redshift and right panel is the $[X/\mathrm{H}]$ and redshift. We do not include a $\chi^2$ curve in the temperature plot as the fit was poor due to a large scatter. In the right panel, we show GRB and blazar combined  $\chi^2$ curve for $z >0.9$ showing a possible redshift relation above this redshift.}
        \label{fig:GRB_blazarIGM_TandZ}
\end{figure*}

Based on our investigations, it appears possible that in some cases BC could mimic absorption. On the other hand, depending on where the energy peak of the blackbody-like feature occurs, it could also mask actual absorption, appearing as an excess at soft X-ray. BC itself is still not generally accepted as the cause of this feature. The majority of our sample show improved fit statistics for an IGM component. Further, the clear $\mathit{N}\textsc{hxigm}$ redshift relation remains with the BC impacted blazars removed from the sample.

\section{Combined Blazar and GRB sample analysis}\label{sec:combinedGRBblazar}

In this section, we combine the GRB sample from D21 with our full \textit{Swift} blazar sample in a multiple tracer analysis across a redshift range $0.03 \leq z \leq 6.3$. In this paper, we used D21 results from their fits using \textsc{hotabs} for CIE IGM for consistency. D21 isolated the IGM LOS contribution to the total absorption for the GRBs by assuming that the GRB host absorption was equal to ionised corrected intrinsic neutral column estimated from the Ly$\alpha$ host absorption. They used a realistic host metallicity, dust corrected where available, in generating the host absorption model.

As can be seen in Fig. \ref{fig:GRB_blazar}, our results for blazars for a power law fit to the $\mathit{N}\textsc{hxigm}$ versus redshift trend scaling as $(1 + z)^{1.8\pm0.2}$ (best cstat fit) is consistent with the GRBs in D21 which scale as $(1 + z)^{1.9\pm0.2}$ over the extended redshift. In Fig. \ref{fig:GRB_blazar} left panel, we use the best Cstat fit results for blazars and in the right panel the log-parabolic only continuum results, in combination with the D21 GRB sample. The slopes of the $\chi^2$ fits for GRB and blazar best cstat $\mathit{N}\textsc{hxigm}$ versus redshift are aligned, but the slope for the blazars using log-parabolic continuum power law in combination with the GRBs slightly better traces the simple IGM density curve over the full redshift. 

The mean hydrogen density at $z = 0$ from the combined GRB and blazar samples is $n_0 = (2.2\pm{0.1}) \times 10^{-7}$ cm$^{-3}$. This is marginally higher than the value of $1.7 \times 10^{-7}$ cm$^{-3}$ for the simple  IGM model (green line in Fig \ref{fig:GRB_blazar}) and for the GRB only sample in D21 $n_0 = (1.8\pm{0.1}) \times 10^{-7}$ cm$^{-3}$. In section \ref{sec:Blazar results}, we reported the mean hydrogen density at $z = 0$ from the blazar FSRQ sample as $n_0 = (3.2\pm{0.5}) \times 10^{-7}$ cm$^{-3}$. A possible explanation may be due to our assumption that there is no absorption in the blazar host due to the relativistic jet. In D21, it was also apparent that the lower redshift GRBs appeared to have higher $\mathit{N}\textsc{hxigm}$ that the simple IGM curve. As speculated in Section \ref{subsec:4.2}, this may be a sign of CGM absorption from a hot phase as proposed by \citep{Das2021}. At higher redshift, the IGM contribution to $\mathit{N}\textsc{hxigm}$ dominates any host contribution.

In Fig. \ref{fig:GRB_blazarIGM_TandZ}, we show the combined GRB and blazar sample results for IGM temperature and metallicity. In the left panel, we can see that there is no detectable overall trend with redshift. Some objects in the blazar sample, appear to have a higher maximum temperature  than the GRB sample, up to log($T/$K) $\sim7.5$. \citet{Das2021} have indicated that the Galaxy hot phase has a temperature of log($T/$K) $\sim7.5$. The mean temperature over the full redshift range for the combined samples is log($T$/K) $= 6.2\pm{0.1}$ under the assumption of a CIE scenario.

In Fig. \ref{fig:GRB_blazarIGM_TandZ} right panel, we can see the IGM metallicity results for the combined GRB and blazar sample for $0.9 \leq z \leq 7$. We omitted blazars with  $z < 0.9$ as they showed very large scatter in best fitted metallicity and with substantial errors. There appears to be a possible relation with redshift scaling as $(1+z)^{-2.9\pm0.5}$  (orange $\chi^2$ line). The reduced $\chi^2$ is 1.9,  with a p-value = 0.00014 and rms = 0.60,  indicating a modestly statistical relation. Visually, the metallicity redshift relation, if any, is not clear, or may indicate that a linear model may not be appropriate.  For the GRB only sample, D21 reported a power law fit to the $[X/$H$]$ and redshift trend as $(1+z)^{-5.2\pm1.0}$, ranging from $[X/$H] $= -1\/\ (Z = 0.1Z\sun)$ at $z \sim 2$ to $[X/$H] $ \sim -3\/\ (Z = 0.001Z\sun)$ at high redshift $z > 4$. They speculated that at low redshift, the higher metallicity warm-hot phase is dominant with $Z \sim0.1~Z\sun$, while at higher redshift the low metallicity IGM away from knots and filaments is dominant. In the combined blazar and GRB sample for $z > 0.9$, the possible redshift metallicity relation is less pronounced.
While there is a large range in the lower error bars, some of the blazars at lower redshift have upper metallicity error bars approaching our upper limit of $[X/$H$] = -0.7$. 

Overall, the IGM parameter results from our blazar sample are consistent with the GRB sample from D21. Therefore, the combined sample gives improved robustness to our reported results for the IGM.

\section{Discussion and comparison with other studies}\label{sec:Discuss}

The cause of spectral flattening seen in blazar spectra has been the subject of study and debate for some time. In early works, a cold local host absorber received favour \citep[e.g.][]{Cappi1997}.  Due to the low levels of optical-UV extinction seen in such blazars \citep[e.g][]{Elvis1994, Sikora1994}, subsequent studies favoured an intrinsic curvature explanation with models including log-parabolic, broken power law and variations on this. We accommodated this element of intrinsic curvature by fitting our sample with best fit from a simple power law, log-parabolic and broken power law. Further, when adding the IGM component, we allowed all parameters to vary. In Section \ref{sec:Robust}, we explored any possible relation between IGM absorption and both spectral flux and power-law hardening. No such relation was apparent. 

Several studies have tentatively explored absorption scenarios, either neutral or a warm absorption component, as a minor part of their work, typically placing the absorber at the blazar redshift  \citep[e.g.][]{Bottacini2010,Paliya2016,Ricci2017,Marcotulli2020,Haim2019}. \citet{Bottacini2010} found they could not constrain their photo-ionised model parameter better than an upper limit. They used \textsc{absori} which is not as sophisticated as \textsc{warmabs}, the photo-ionised equivalent of our CIE model using \textsc{hotabs}. However, they reported that they found no evidence of absorption variability, consistent with our results. \citet{Ricci2017} used \textsc{zxipcf}, a photoionisation model to test a scenario of a warm absorber in the blazar host with metallicity fixed to solar. They found that only a small number of fits improved with the added warm absorber component.  This differs from our model where we allow both metallicity and temperature to vary, as solar metallicity is highly unlikely to occur in the diffuse IGM \citep[e.g.][]{Schaye2003,Aguirre2008,Shull2012,Shull2014}, and we place the absorber at an intermediate redshift.

The relation between spectral flattening and redshift has been reported by several authors \citep[e.g.][]{Yuan2006,Behar2011}. The hypothesis of discrete intervening absorbers (DLA, LLS, sLLS, etc) has been investigated to explain this redshift relation \citep[e.g.][]{Wolf1987}. Several studies concluded that the absorption from such cool neutral intervening systems is rare and insufficient to be the cause of  observed spectral curvature \citep[and references therein]{Gianni2011}, leaving the diffuse IGM as the alterntive for non-localised intervening absorption.

Arguments against intervening IGM absorption are that intrinsic curvature is present in very low redshift blazars, or that absorption edges or lines are not observed in such very low redshift blazars \citep[and references therein]{Watson2012}, leading them to conclude that all the spectral flattening is due to intrinsic curvature at all redshifts. In general, most studies are focused on the blazar engine as the main or only cause of intrinsic curvature and leave out IGM absorption based on lack of significantly improved statistical fits. We would argue that it is highly likely that spectral curvature is due to a combination of both intrinsic factors as well as absorption, particularly at soft X-ray, given our findings for $\mathit{N}{\textsc{hxigm}}$ and the redshift relation. If the apparent absorption was actually intrinsic to the blazars, then there would have to be some explanation of the relation to redshift which is absent.

Detection of the WHIM is proving very challenging due to very weak emission and absorption.  \citet{Nicastro2018} claim to have observed the WHIM in absorption. However, with only 1 to 2 strong $\mathrm{O\/\ \textsc{vii}}$ absorbers predicted to exist per unit redshift, the column densities they report are an order of magnitude lower than the simple IGM model, or the results for $\mathit{N}{\textsc{hxigm}}$ reported by D21. Our results for the FSRQ sample $\mathit{N}{\textsc{hxigm}}$ are consistent with the simple IGM model, though we note that many of the BL Lacs showed high $\mathit{N}{\textsc{hxigm}}$. \citet{Haim2019} searched for absorption lines as signals of localised IGM absorption in RBS 315 at $z = 2.69$, one of the brightest FSRQ known. They could find no such line absorption and concluded that, if blazar curvature is at least partly attributable to the IGM, it is not localised but smeared over redshift, consistent with our hypothesis and findings.

A18 is, to our knowledge, the only previous study that was dedicated to exploring the IGM as part of the cause of spectral flattening in blazar spectra, and to use blazars to investigate IGM properties. They used an \textsc{absori} based \textsc{xpec} model (\textsc{igmabs}) for the IGM absorption. They jointly fitted four blazars with IGM parameters of density, temperature and ionisation tied together. They reported that excess absorption is the preferred explanation over intrinsic curvature and that it is related to redshift. They give an IGM average density of $n_0 = 1.0^{+0.53}_{-0.72} \times 10^{-7}$ cm$^{-3}$ and temperature log$(T/$K) $= 6.45^{+0.51}_{-2.12}$. Some caveats to their results are based on their solar metallicity assumption for the IGM and fixing the intrinsic power law parameters for some of their sample, which they adopted due to computational limits of their model which they noted would probably lead to upper limit measures for the IGM. Taking account of these factors, their results are broadly consistent with our results for IGM for $n_0$ and T, but not for our mean IGM metallicity of $[X/$H$] = -1.62\pm0.04 (\sim 0.02Z\sun$). Their derived metallicity was $Z_\textrm{IGM} = 0.59^{+0.31}_{-0.42}Z\sun$, obtained from the ratio of their $n_0 \sim1$ result (based on solar metallicity) to the simple IGM model taken from \citep{Behar2011} $n_0 = 1.7 \times 10^{-7}$ cm$^{-3}$. \citet{Campana2015} used simulations for intervening IGM absorption to AGN and GRBs. For GRBs, they reported log$(T/$K) $\sim  5 - 7$. To calculate the metal column density of the intervening IGM material, 100 LOS to distant sources were used through a $100 h^{-1}$ comoving Mpc Adaptive Mesh Refinement cosmological simulation \citep{Pallottini2013a}. The contribution by each cell was summed, with an absorbing column density weighted for its effective temperature dependent value. Metallicity was obtained by requiring that only $1\%$ of their GRB and AGN sample fall below the simulated hydrogen column density redshift curve. Their mean metallicity $Z = 0.03Z\sun$ is consistent with our results.

A18 combined their results for blazars with GRBs and AGN from other studies. However, all those studies were based on the assumption that all absorption in excess of our Galaxy was at the host redshift, neutral and at solar metallicity. Our combined tracer results in Section \ref{sec:combinedGRBblazar} are more realistic as we use the GRBs from D21 which more accurately isolate the IGM absorption assuming that the GRB host absorption was equal to ionised corrected intrinsic neutral column estimated from the Ly$\alpha$ host absorption. D21 also used more realistic host metallicity, dust corrected where available in generating the host absorption model as opposed to the conventional solar assumption.

\section{Conclusions}\label{sec:Conclusion}
  We used blazars as tracers of the IGM with the main aim to probe the key parameters of column density, metallicity and temperature using a sophisticated software model for collisionally ionised plasma. We used co-added spectra from $\mathit{Swift}$ for 40 blazars as our tracers with a redshift range of $0.03 \leq z \leq 4.7$. Our focus is on FSRQ blazars as they are available over a broad redshift range, and the rest-frame energy distribution of FSRQs is strongly peaked at low frequencies, below soft X-ray, unlike BL Lacs. We adopted a conservative approach to the blazar continuum model and use three different intrinsic power law models. As blazars are thought to have a kpc-scale relativistic jet on our line of sight, we excluded any host absorption in our models. We fixed the Galactic absorption to known values and attributed the  excess to the IGM. We  model  the  IGM  assuming  a  thin  uniform  plane  parallel  slab  geometry  in collisional ionisation equilibrium to represent a LOS  through  a  homogeneous  isothermal medium. We used \textsc{xspec} fitting with both the CIE IGM component and all power law parameters free to vary, and use \textsc{steppar} and MCMC to generate best fits to the blazar spectra. 
  
  We subjected our results to a number of robustness tests using a sub-sample: comparison of individual observation fit results with our co-added spectra for possible temporal absorption variability; testing for a relation between column density and flux; investigating spectral slope degeneracy with column density; comparing results from using $\textit{XMM-Newton}$ with energy range 0.3-10 keV (as for $\textit{Swift}$) and 0.16-13keV; and exploring the impact of using a blackbody like additional component to represent bulk comptonisation which could mimic absorption.
  
  Finally, we combined our sample with the GRB sample from D21 to report results for an extended redshift range using the two different types of tracers.

Our main findings and conclusions are:
\begin{enumerate}
  \item the best fit Cstat results for our blazar sample were achieved using an IGM component with a log-parabolic power law (26/40 spectra) and appear to be more consistent with the simple model IGM curve than the selected best fits from both log-parabolic and broken power law.
  
  \item Using blazars to model the IGM as being in highly ionised collisional equilibrium  with free parameters for density, temperature and metallicity (as well as continnum parameters) appears to give plausible IGM property results. A power law fit to $\mathit{N}\textsc{hxigm}$ versus redshift trend for the FSRQ objects scales as $(1 + z)^{1.8\pm0.2}$. The mean hydrogen density at $z = 0$ from the FSRQ sample is $n_0 = (3.2\pm{0.5}) \times 10^{-7}$ cm$^{-3}$, higher than the value of $1.7 \times 10^{-7}$ cm$^{-3}$ for the simple  IGM model (Fig \ref{fig:NHX_z}). Nearly all blazar fits are proximate to both the $\chi^2$ fit and mean IGM density curve. 
  
  \item At low redshift, several blazars have higher $\mathit{N}\textsc{hxigm}$ than the simple IGM model. BL Lacs dominate the sample at very low redshift and the majority appear  have high fitted $\mathit{N}\textsc{hxigm}$. This may be due to CGM absorption.
 
 \item The IGM temperature range is $5.0 <$ log($T$/K) $< 8.0$, with no apparent redshift relation in the Fig. \ref{fig:BestfitIGM_TandZ} left panel. The mean temperature over the full redshift range is log($T$/K) $= 6.1\pm{0.1}$. These values are consistent with the generally accepted WHIM range indicating that very highly ionised metals are plausible absorbers over the LOS.

 \item The right panel of Fig. \ref{fig:BestfitIGM_TandZ} shows no apparent relation of $[X/$H$]$ with redshift (however, see Section \ref{sec:combinedGRBblazar} for possible metallicity redshift relation using combined blazar and GRB samples). The mean metallicity over the full redshift range is $[X/$H$] = -1.62\pm0.04 (Z \sim 0.02)$. Metallicity ranges from $[X/$H$] -0.7\/\ (0.2Z\sun)$ to $[X/$H$] -3\/\ (0.001Z\sun)$ with one outlier.

 \item   There was a large range in Cstat improvements across the sample, with the average Cstat improvement per free IGM parameter of 3.9. In our models the IGM contributes substantially to the total absorption seen in blazar spectra, and it rises with redshift. We provide evidence that a complete blazar model should also account for absorption by intervening IGM material.

 \item In Fig. \ref{fig:NHX and flux} left panel, there is no apparent relation between $\mathit{N}\textsc{hxigm}$ and flux across all the observations. All the individual results for $\mathit{N}\textsc{hxigm}$ for each blazar are consistent with the mean result within the errors (Table \ref{tab:individual_blazar sub-sample_results}).
 
\item There is no apparent relation between column density and power law index.   Further, there was no temporal variation in IGM parameter results per blazar using observations over time.

\item For $\textit{Swift}$ 0.3-10 keV, $\textit{XMM-Newton}$ 0.3-10 keV and 0.16-13 keV respectively, the values for $\mathit{N}\textsc{hxigm}$ using log-parabolic power laws are consistent for each blazar within the errors. All $\mathit{N}\textsc{hxigm}$ are proximate to the simple IGM curve, Fig. \ref{fig:XMM comparison} left panel.  The slopes of both the  $\textit{XMM-Newton}$ energy ranges are very similar. The slope of the $\chi^2$ fit for $\textit{Swift}$  is less steep but consistent with $\textit{XMM-Newton}$ within the error. The $\textit{XMM-Newton}$ results reinforce the findings that the IGM absorption results do not vary on a temporal basis.

\item Bulk Comptonisation  has been proposed as a cause of the hump feature at soft X-rays seen is some blazars. 10 out of 40 in our blazar sample had a second soft X-ray hump where the Cstat was similar for both an IGM or blackbody component. Based on our investigations, it appears possible that in some cases, BC could mimic absorption. On the other hand, depending on where the energy peak of the blackbody like feature occurs, it could also mask actual absorption, appearing as an excess at soft X-ray. We found that after omitting from the sample the blazars with possible BC, the $\mathit{N}\textsc{hxigm}$ relation with redshift remains and the results are consistent with those from our full sample.

\item Combining our blazar sample with the GRB sample from D21 gives consistent results for the IGM properties over an extended redshift range from $0.03 \leq z \leq 6.3$. The mean hydrogen density at $z = 0$ from the combined GRB and blazar samples is $n_0 = (2.2\pm{0.1}) \times 10^{-7}$ cm$^{-3}$. This is marginally higher than the value of $1.7 \times 10^{-7}$ cm$^{-3}$ for the simple  IGM model, but lower than the blazar only sample ($n_0 = (3.2\pm{0.1}) \times 10^{-7}$ cm$^{-3}$), perhaps indicating that the blazar relativisitc jet may not fully sweep out absorbing material in the host. Our blazar model assumes there is no host absorption which may be true for most FSRQ which are highly luminous, and hence probably more effective in removing host absorbing gas,  but for the less luminous BL Lacs this may not completely happen. The mean temperature over the full redshift range is log($T$/K) $= 6.1\pm{0.1}$, and the mean metallicity over the full redshift range is $[X/$H$] = -1.62\pm0.04 (Z \sim 0.02)$. These values are consistent with the generally accepted WHIM range indicating that very highly ionised metals are plausible absorbers over the LOS. There was no apparent temperature redshift relation. However, we found a possible relation for metallicity and redshift to be $(1+z)^{-2.9\pm0.5}$.

This study is based on observations of blazar X-ray spectra, and provides results on the IGM parameters. The combination of blazars with the GRB sample gives consistent and more robust results for the IGM properties by using multiple tracer types. The IGM property constraints will only be validated when observations are available from instruments with large effective area, high energy resolution, and a low energy threshold in the soft X-ray energy band (e.g. Athena). We will continue our IGM exploration using other tracers in an upcoming paper and will combine the future results with those from this paper, D21 and D20.

\end{enumerate}

\section*{Acknowledgements}
 We thank the anonymous referee for useful comments and suggestions. This work made use of data supplied by the UK Swift Science Data Centre at the University of Leicester; XMM–Newton, an ESA science mission with instruments and contributions directly funded by ESA member states and NASA. S.L. Morris also acknowledges support from STFC (ST/P000541/1).
This project has received funding from the European Research Council (ERC) under the European Union's Horizon 2020 research and innovation programme (grant agreement No 757535) and by Fondazione Cariplo (grant No 2018-2329).

\section*{Data Availability}
Supplementary data including spectra fit, parameter results for the continuum models and IGM, and MCMC integrated probability plots are available on request. Please contact Tony Dalton. $\textit{Swift}$ spectral data for blazars are available from https://www.swift.ac.uk/2SXPS. 



\bibliographystyle{mnras}
\bibliography{references.bib} 

\begin{thebibliography}{}
\makeatletter
\relax
\def\mn@urlcharsother{\let\do\@makeother \do\$\do\&\do\#\do\^\do\_\do\%\do\~}
\def\mn@doi{\begingroup\mn@urlcharsother \@ifnextchar [ {\mn@doi@}
  {\mn@doi@[]}}
\def\mn@doi@[#1]#2{\def\@tempa{#1}\ifx\@tempa\@empty \href
  {http://dx.doi.org/#2} {doi:#2}\else \href {http://dx.doi.org/#2} {#1}\fi
  \endgroup}
\def\mn@eprint#1#2{\mn@eprint@#1:#2::\@nil}
\def\mn@eprint@arXiv#1{\href {http://arxiv.org/abs/#1} {{\tt arXiv:#1}}}
\def\mn@eprint@dblp#1{\href {http://dblp.uni-trier.de/rec/bibtex/#1.xml}
  {dblp:#1}}
\def\mn@eprint@#1:#2:#3:#4\@nil{\def\@tempa {#1}\def\@tempb {#2}\def\@tempc
  {#3}\ifx \@tempc \@empty \let \@tempc \@tempb \let \@tempb \@tempa \fi \ifx
  \@tempb \@empty \def\@tempb {arXiv}\fi \@ifundefined
  {mn@eprint@\@tempb}{\@tempb:\@tempc}{\expandafter \expandafter \csname
  mn@eprint@\@tempb\endcsname \expandafter{\@tempc}}}

\bibitem[\protect\citeauthoryear{Aguirre, Dow‐Hygelund, Schaye  \&
  Theuns}{Aguirre et~al.}{2008}]{Aguirre2008}
Aguirre A.,  Dow‐Hygelund C.,  Schaye J.,   Theuns T.,  2008, \mn@doi [ApJ]
  {10.1086/592554}, 689, 851

\bibitem[\protect\citeauthoryear{Arcodia, Campana  \& Salvaterra}{Arcodia
  et~al.}{2016}]{Arcodia2016}
Arcodia R.,  Campana S.,   Salvaterra R.,  2016, \mn@doi [A{\&}A]
  {10.1051/0004-6361/201628326}, 590, 1

\bibitem[\protect\citeauthoryear{Arcodia, Campana, Salvaterra  \&
  Ghisellini}{Arcodia et~al.}{2018}]{Arcodia2018}
Arcodia R.,  Campana S.,  Salvaterra R.,   Ghisellini G.,  2018, A{\&}A, 590,
  A82

\bibitem[\protect\citeauthoryear{Arnaud}{Arnaud}{1996}]{Arnaud1996}
Arnaud K.,  1996, Astron. Data Anal. Softw. Syst., 101, 17

\bibitem[\protect\citeauthoryear{Asplund, Grevesse, Sauval  \& Scott}{Asplund
  et~al.}{2009}]{Asplund2009}
Asplund M.,  Grevesse N.,  Sauval A.~J.,   Scott P.,  2009, \mn@doi [Annual
  Review of Astronomy and Astrophysics] {10.1007/s10509-010-0288-z}, 47

\bibitem[\protect\citeauthoryear{Asplund, Amarsi  \& Grevesse}{Asplund
  et~al.}{2021}]{Asplund2021}
Asplund M.,  Amarsi A.~M.,   Grevesse N.,  2021,
  https://arxiv.org/abs/2105.01661

\bibitem[\protect\citeauthoryear{Behar, Dado, Dar  \& Laor}{Behar
  et~al.}{2011}]{Behar2011}
Behar E.,  Dado S.,  Dar A.,   Laor A.,  2011, \mn@doi [ApJ]
  {10.1088/0004-637X/734/1/26}, 734, 26

\bibitem[\protect\citeauthoryear{Bhatta, Mohorian  \& Bilinsky}{Bhatta
  et~al.}{2018}]{Bhatta2018}
Bhatta G.,  Mohorian M.,   Bilinsky I.,  2018, \mn@doi [A{\&}A]
  {10.1051/0004-6361/201833628}, 619, 1

\bibitem[\protect\citeauthoryear{Bottacini et~al.,}{Bottacini
  et~al.}{2010}]{Bottacini2010}
Bottacini E.,  et~al., 2010, \mn@doi [A{\&}A] {10.1051/0004-6361/200913260},
  509, 1

\bibitem[\protect\citeauthoryear{Buchner, Schulze  \& Bauer}{Buchner
  et~al.}{2017}]{Buchner2017}
Buchner J.,  Schulze S.,   Bauer F.~E.,  2017, \mn@doi [MNRAS]
  {10.1093/mnras/stw2423}, 464, 4545

\bibitem[\protect\citeauthoryear{Burrows et~al.,}{Burrows
  et~al.}{2005}]{Burrows2005}
Burrows D.~N.,  et~al., 2005, \mn@doi [Space Science Reviews]
  {10.1007/s11214-005-5097-2}, 120, 165

\bibitem[\protect\citeauthoryear{Campana et~al.,}{Campana
  et~al.}{2012}]{Campana2012}
Campana S.,  et~al., 2012, \mn@doi [MNRAS] {10.1111/j.1365-2966.2012.20428.x},
  421, 1697

\bibitem[\protect\citeauthoryear{Campana, Salvaterra, Ferrara  \&
  Pallottini}{Campana et~al.}{2015}]{Campana2015}
Campana S.,  Salvaterra R.,  Ferrara A.,   Pallottini A.,  2015, \mn@doi
  [A{\&}A] {10.1051/0004-6361/201425083}, 575, A43

\bibitem[\protect\citeauthoryear{Cappi, Matsuoka, Comastri, Brinkmann, Elvis,
  Palumbo  \& Vignali}{Cappi et~al.}{1997}]{Cappi1997}
Cappi M.,  Matsuoka M.,  Comastri A.,  Brinkmann W.,  Elvis M.,  Palumbo G.,
  Vignali C.,  1997, ApJ, 478, 492

\bibitem[\protect\citeauthoryear{Cash}{Cash}{1979}]{Cash1979}
Cash W.,  1979, ApJ, 228, 939

\bibitem[\protect\citeauthoryear{Celotti, Ghisellini  \& Fabian}{Celotti
  et~al.}{2007}]{Celotti2007}
Celotti A.,  Ghisellini G.,   Fabian A.~C.,  2007, \mn@doi [MNRAS]
  {10.1111/j.1365-2966.2006.11289.x}, 375, 417

\bibitem[\protect\citeauthoryear{Cen \& Ostriker}{Cen \&
  Ostriker}{1999}]{Cen1999}
Cen R.,  Ostriker j.~p.,  1999, AJ, 514, 1

\bibitem[\protect\citeauthoryear{Cen \& Ostriker}{Cen \&
  Ostriker}{2006}]{Cen2006}
Cen R.,  Ostriker J.~P.,  2006, \mn@doi [ApJ] {10.1086/506505}, 650, 560

\bibitem[\protect\citeauthoryear{Dalton \& Morris}{Dalton \&
  Morris}{2020}]{Dalton2020}
Dalton T.,  Morris S.~L.,  2020, \mn@doi [MNRAS] {10.1093/mnras/staa1321}, 495,
  2342

\bibitem[\protect\citeauthoryear{Dalton, Morris  \& Fumagalli}{Dalton
  et~al.}{2021}]{Dalton2021}
Dalton T.,  Morris S.~L.,   Fumagalli M.,  2021, \mn@doi [MNRAS]
  {10.1093/mnras/stab335}, 502, 5981

\bibitem[\protect\citeauthoryear{Danforth \& Shull}{Danforth \&
  Shull}{2005}]{Danforth2005}
Danforth C.~W.,  Shull J.~M.,  2005, \mn@doi [ApJ] {10.1086/429285}, 624, 555

\bibitem[\protect\citeauthoryear{Danforth \& Shull}{Danforth \&
  Shull}{2008}]{Danforth2008}
Danforth C.~W.,  Shull J.~M.,  2008, ApJ, 679, 194

\bibitem[\protect\citeauthoryear{Danforth et~al.,}{Danforth
  et~al.}{2016}]{Danforth2016}
Danforth C.~W.,  et~al., 2016, \mn@doi [ApJ] {10.3847/0004-637X/817/2/111},
  817, 111

\bibitem[\protect\citeauthoryear{Das, Mathur, Gupta  \& Krongold}{Das
  et~al.}{2021}]{Das2021}
Das S.,  Mathur S.,  Gupta A.,   Krongold Y.,  2021,
  https://arxiv.org/abs/2106.13243

\bibitem[\protect\citeauthoryear{Dav{\'{e}} \& Oppenheimer}{Dav{\'{e}} \&
  Oppenheimer}{2007}]{Dave2007}
Dav{\'{e}} R.,  Oppenheimer B.~D.,  2007, \mn@doi [MNRAS]
  {10.1111/j.1365-2966.2006.11177.x}, 374, 427

\bibitem[\protect\citeauthoryear{Donato, Sambruna  \& Gliozzi}{Donato
  et~al.}{2005}]{Donato2005}
Donato D.,  Sambruna R.~M.,   Gliozzi M.,  2005, \mn@doi [A{\&}A]
  {10.1051/0004-6361:20034555}, 433, 1163

\bibitem[\protect\citeauthoryear{Done}{Done}{2010}]{Done2010}
Done C.,  2010, \mn@doi [Accretion Processes In Astrophysics: XXI Canary
  Islands Winter School Of Astrophysics] {10.1017/CBO9781139343268.007}, arXiv:
  100, 184

\bibitem[\protect\citeauthoryear{Done, Mulchaey, Mushotzky  \& Arnaud}{Done
  et~al.}{1992}]{Done1992}
Done C.,  Mulchaey J.,  Mushotzky R.,   Arnaud K.,  1992, ApJ, 395, 275

\bibitem[\protect\citeauthoryear{Eitan \& Behar}{Eitan \&
  Behar}{2013}]{Eitan2013}
Eitan A.,  Behar E.,  2013, ApJ, 774, 29

\bibitem[\protect\citeauthoryear{Elvis et~al.,}{Elvis et~al.}{1994}]{Elvis1994}
Elvis M.,  et~al., 1994, \mn@doi [The Astrophysical Journal Supplement Series.]
  {10.1086/312966}

\bibitem[\protect\citeauthoryear{Evans et~al.,}{Evans et~al.}{2020}]{Evans2020}
Evans P.~A.,  et~al., 2020, \mn@doi [ApJS] {10.3847/1538-4365/ab7db9}, 247, 54

\bibitem[\protect\citeauthoryear{Fabian, Celotti, Iwasawa  \&
  Ghisellini}{Fabian et~al.}{2001}]{Fabian2001}
Fabian A.~C.,  Celotti A.,  Iwasawa K.,   Ghisellini G.,  2001, \mn@doi [MNRAS]
  {10.1046/j.1365-8711.2001.04348.x}, 324, 628

\bibitem[\protect\citeauthoryear{Fumagalli}{Fumagalli}{2014}]{Fumagalli2014}
Fumagalli M.,  2014, Memorie della Societa Astronomica Italiana, 85, 355

\bibitem[\protect\citeauthoryear{Fumagalli, O'Meara, Prochaska  \&
  Worseck}{Fumagalli et~al.}{2013}]{Fumagalli2013}
Fumagalli M.,  O'Meara J.~M.,  Prochaska J.~X.,   Worseck G.,  2013, \mn@doi
  [ApJ] {10.1088/0004-637X/775/1/78}, 775

\bibitem[\protect\citeauthoryear{Fumagalli, O'Meara  \& {Xavier
  Prochaska}}{Fumagalli et~al.}{2016}]{Fumagalli2016}
Fumagalli M.,  O'Meara J.~M.,   {Xavier Prochaska} J.,  2016, \mn@doi [MNRAS]
  {10.1093/mnras/stv2616}, 455, 4100

\bibitem[\protect\citeauthoryear{Fumagalli, Fotopoulou, Avenue  \&
  Bs}{Fumagalli et~al.}{2020}]{Fumagalli2020a}
Fumagalli M.,  Fotopoulou S.,  Avenue T.,   Bs B.,  2020, \mn@doi [MNRAS]
  {10.1093/mnras/staa2388}, 498, 1951

\bibitem[\protect\citeauthoryear{Furniss, Fumagalli, Falcone  \&
  Williams}{Furniss et~al.}{2013}]{Furniss2013}
Furniss A.,  Fumagalli M.,  Falcone A.,   Williams D.~A.,  2013, \mn@doi [ApJ]
  {10.1088/0004-637X/770/2/109}, 770

\bibitem[\protect\citeauthoryear{Galama \& Wijers}{Galama \&
  Wijers}{2001}]{Galama2001a}
Galama T.~J.,  Wijers R. A. M.~J.,  2001, \mn@doi [ApJ] {10.1086/319162}, 549,
  L209

\bibitem[\protect\citeauthoryear{Gatuzz, Garc{\'{i}}a, Kallman, Mendoza  \&
  Gorczyca}{Gatuzz et~al.}{2015}]{Gatuzz2015}
Gatuzz E.,  Garc{\'{i}}a J.,  Kallman T.~R.,  Mendoza C.,   Gorczyca T.~W.,
  2015, \mn@doi [ApJ] {10.1088/0004-637X/800/1/29}, 800

\bibitem[\protect\citeauthoryear{Gaur}{Gaur}{2020}]{Gaur2020}
Gaur H.,  2020, \mn@doi [Galaxies] {10.3390/GALAXIES8030062}, 8

\bibitem[\protect\citeauthoryear{Ghisellini \& Maraschi}{Ghisellini \&
  Maraschi}{1989}]{Ghisellini1989}
Ghisellini G.,  Maraschi L.,  1989, ApJ, 340, 181

\bibitem[\protect\citeauthoryear{Giann{\'{i}}, {De Rosa}, Bassani, Bazzano,
  Dean  \& Ubertini}{Giann{\'{i}} et~al.}{2011}]{Gianni2011}
Giann{\'{i}} S.,  {De Rosa} A.,  Bassani L.,  Bazzano A.,  Dean T.,   Ubertini
  P.,  2011, \mn@doi [MNRAS] {10.1111/j.1365-2966.2010.17725.x}, 411, 2137

\bibitem[\protect\citeauthoryear{Haim, Behar  \& Mushotzky}{Haim
  et~al.}{2019}]{Haim2019}
Haim S.~B.,  Behar E.,   Mushotzky R.~F.,  2019, \mn@doi [ApJ]
  {10.3847/1538-4357/ab340f}, 882, 130

\bibitem[\protect\citeauthoryear{Harris et~al.,}{Harris
  et~al.}{2016}]{Harris2016}
Harris D.~W.,  et~al., 2016, \mn@doi [AJ] {10.3847/0004-6256/151/6/155}, 151, 1

\bibitem[\protect\citeauthoryear{Ighina, Caccianiga, Moretti, Belladitta,
  {Della Ceca}, Ballo  \& Dallacasa}{Ighina et~al.}{2019}]{Ighina2019}
Ighina L.,  Caccianiga A.,  Moretti A.,  Belladitta S.,  {Della Ceca} R.,
  Ballo L.,   Dallacasa D.,  2019, \mn@doi [MNRAS] {10.1093/mnras/stz2340},
  489, 2732

\bibitem[\protect\citeauthoryear{Kaastra}{Kaastra}{2017}]{Kaastra2017}
Kaastra J.~S.,  2017, \mn@doi [A{\&}A] {10.1051/0004-6361/201629319}, 605, 2

\bibitem[\protect\citeauthoryear{Kalberla, Burton, Hartmann, Arnal, Bajaja,
  Morras  \& P{\"{o}}ppel}{Kalberla et~al.}{2005}]{Kalberla2005}
Kalberla P.~M.,  Burton W.~B.,  Hartmann D.,  Arnal E.~M.,  Bajaja E.,  Morras
  R.,   P{\"{o}}ppel W.~G.,  2005, {The Leiden/Argentine/Bonn (LAB) survey of
  Galactic HI: Final data release of the combined IDS and IAR surveys with
  improved stray-radiation corrections}, \mn@doi{10.1051/0004-6361:20041864}

\bibitem[\protect\citeauthoryear{Kallman, Bautista, Goriely, Mendoza, Miller,
  Palmeri, Quinet  \& Raymond}{Kallman et~al.}{2009}]{Kallman2009}
Kallman T.~R.,  Bautista M.~A.,  Goriely S.,  Mendoza C.,  Miller J.~M.,
  Palmeri P.,  Quinet P.,   Raymond J.,  2009, \mn@doi [ApJ]
  {10.1088/0004-637X/701/2/865}, 701, 865

\bibitem[\protect\citeauthoryear{Kammoun, Nardini, Risaliti, Ghisellini, Behar
  \& Celotti}{Kammoun et~al.}{2018}]{Kammoun2018}
Kammoun E.~S.,  Nardini E.,  Risaliti G.,  Ghisellini G.,  Behar E.,   Celotti
  A.,  2018, \mn@doi [Monthly Notices of the Royal Astronomical Society:
  Letters] {10.1093/mnrasl/slx164}, 473, L89

\bibitem[\protect\citeauthoryear{Khabibullin \& Churazov}{Khabibullin \&
  Churazov}{2019}]{Khabibullin2019}
Khabibullin I.,  Churazov E.,  2019, \mn@doi [MNRAS] {10.1093/mnras/sty2992},
  482, 4972

\bibitem[\protect\citeauthoryear{Lehner, O'Meara, Howk, Prochaska  \&
  Fumagalli}{Lehner et~al.}{2016}]{Lehner2016}
Lehner N.,  O'Meara J.~M.,  Howk J.~C.,  Prochaska J.~X.,   Fumagalli M.,
  2016, \mn@doi [ApJ] {10.3847/1538-4357/833/2/283}, 833, 283

\bibitem[\protect\citeauthoryear{Lehner, Wotta, Howk, O'Meara, Oppenheimer  \&
  Cooksey}{Lehner et~al.}{2019}]{Lehner2019}
Lehner N.,  Wotta C.~B.,  Howk J.~C.,  O'Meara J.~M.,  Oppenheimer B.~D.,
  Cooksey K.~L.,  2019, \mn@doi [ApJ] {10.3847/1538-4357/ab41fd}, 887, 5

\bibitem[\protect\citeauthoryear{Lusso, Worseck, Hennawi, Prochaska, Vignali,
  Stern  \& O'Meara}{Lusso et~al.}{2015}]{Lusso2015}
Lusso E.,  Worseck G.,  Hennawi J.~F.,  Prochaska J.~X.,  Vignali C.,  Stern
  J.,   O'Meara J.~M.,  2015, \mn@doi [MNRAS] {10.1093/mnras/stv516}, 449, 4204

\bibitem[\protect\citeauthoryear{Macquart et~al.,}{Macquart
  et~al.}{2020}]{Macquart2020}
Macquart J.~P.,  et~al., 2020, \mn@doi [https://arxiv.org/abs/2005.13161]
  {10.1038/s41586-020-2300-2}, pp 1--54

\bibitem[\protect\citeauthoryear{Marcotulli et~al.,}{Marcotulli
  et~al.}{2020}]{Marcotulli2020}
Marcotulli L.,  et~al., 2020, \mn@doi [ApJ] {10.3847/1538-4357/ab65f5}, 889,
  164

\bibitem[\protect\citeauthoryear{Martizzi et~al.,}{Martizzi
  et~al.}{2019}]{Martizzi2019}
Martizzi D.,  et~al., 2019, \mn@doi [MNRAS] {10.1093/mnras/stz1106}, 486, 3766

\bibitem[\protect\citeauthoryear{Massaro, Maselli, Leto, Marchegiani, Perri,
  Giommi  \& Piranomonte}{Massaro et~al.}{2015}]{massaro2015}
Massaro E.,  Maselli A.,  Leto C.,  Marchegiani P.,  Perri M.,  Giommi P.,
  Piranomonte S.,  2015, \mn@doi [Astrophysics and Space Science]
  {10.1007/s10509-015-2254-2}, 357, 1

\bibitem[\protect\citeauthoryear{McQuinn}{McQuinn}{2016}]{McQuinn2016a}
McQuinn M.,  2016, \mn@doi [Annual Review of Astronomy and Astrophysics]
  {10.1146/annurev-astro-082214-122355}, 54, 313

\bibitem[\protect\citeauthoryear{Morris, Weymann, Savage  \& Gilliland}{Morris
  et~al.}{1991}]{Morris1991}
Morris S.,  Weymann R.,  Savage B.,   Gilliland R.,  1991, \mn@doi [ApJ]
  {10.1017/CBO9781107415324.004}, 377, L21

\bibitem[\protect\citeauthoryear{Nicastro, Krongold, Mathur  \& Elvis}{Nicastro
  et~al.}{2017}]{Nicastro2017}
Nicastro F.,  Krongold Y.,  Mathur S.,   Elvis M.,  2017, Astronomische
  Nachrichten, 338, 281

\bibitem[\protect\citeauthoryear{Nicastro et~al.,}{Nicastro
  et~al.}{2018}]{Nicastro2018}
Nicastro F.,  et~al., 2018, Nature, 558, 406

\bibitem[\protect\citeauthoryear{Padovani, Giommi  \& Rau}{Padovani
  et~al.}{2012}]{Padovani2012}
Padovani P.,  Giommi P.,   Rau A.,  2012, \mn@doi [Monthly Notices of the Royal
  Astronomical Society: Letters] {10.1111/j.1745-3933.2012.01234.x}, 422, 48

\bibitem[\protect\citeauthoryear{Page, Reeves, O'Brien  \& Turner}{Page
  et~al.}{2005}]{Page2005}
Page K.~L.,  Reeves J.~N.,  O'Brien P.~T.,   Turner M.~J.,  2005, \mn@doi
  [MNRAS] {10.1111/j.1365-2966.2005.09550.x}, 364, 195

\bibitem[\protect\citeauthoryear{Paggi, Massaro, Vittorini, Cavaliere,
  D'Ammando, Vagnetti  \& Tavani}{Paggi et~al.}{2009}]{Paggi2009}
Paggi A.,  Massaro F.,  Vittorini V.,  Cavaliere A.,  D'Ammando F.,  Vagnetti
  F.,   Tavani M.,  2009, \mn@doi [A{\&}A] {10.1051/0004-6361/200912237}, 504,
  821

\bibitem[\protect\citeauthoryear{Paggi, Bonato, Raiteri, Villata, {De Zotti}
  \& Carnerero}{Paggi et~al.}{2020}]{Paggi2020a}
Paggi A.,  Bonato M.,  Raiteri C.~M.,  Villata M.,  {De Zotti} G.,   Carnerero
  M.~I.,  2020, \mn@doi [A{\&}A] {10.1051/0004-6361/202038430}, 641, A62

\bibitem[\protect\citeauthoryear{Paliya, Parker, Fabian  \& Stalin}{Paliya
  et~al.}{2016}]{Paliya2016}
Paliya V.~S.,  Parker M.~L.,  Fabian A.~C.,   Stalin C.~S.,  2016, \mn@doi
  [ApJ] {10.3847/0004-637x/825/1/74}, 825, 74

\bibitem[\protect\citeauthoryear{Pallottini, Ferrara  \& Evoli}{Pallottini
  et~al.}{2013}]{Pallottini2013a}
Pallottini A.,  Ferrara A.,   Evoli C.,  2013, \mn@doi [MNRAS]
  {10.1093/mnras/stt1249}, 434, 3293

\bibitem[\protect\citeauthoryear{Perlman, Padovani, Giommi, Sambruna, Jones,
  Tzioumis  \& Reynolds}{Perlman et~al.}{1998}]{Perlman1998}
Perlman E.~S.,  Padovani P.,  Giommi P.,  Sambruna R.,  Jones L.~R.,  Tzioumis
  A.,   Reynolds J.,  1998, \mn@doi [AJ] {10.1086/300283}, 115, 1253

\bibitem[\protect\citeauthoryear{Piattella}{Piattella}{2018}]{Piattella2018}
Piattella O.~F.,  2018, \mn@doi [Lecture notes in cosmology (Berlin:
  Springer.)] {10.1007/978-3-319-95570-4}, p.~417

\bibitem[\protect\citeauthoryear{Pieri et~al.,}{Pieri et~al.}{2014}]{Pieri2014}
Pieri M.,  et~al., 2014, \mn@doi [MNRAS] {10.1093/mnras/stu577}, 441, 1718

\bibitem[\protect\citeauthoryear{Pratt, Stocke, Keeney  \& Danforth}{Pratt
  et~al.}{2018}]{Pratt2018}
Pratt C.~T.,  Stocke J.~T.,  Keeney B.~A.,   Danforth C.~W.,  2018, \mn@doi
  [ApJ] {10.3847/1538-4357/aaaaac}, 855, 18

\bibitem[\protect\citeauthoryear{Raghunathan, Clowes, Campusano,
  S{\"{o}}chting, Graham  \& Williger}{Raghunathan
  et~al.}{2016}]{Raghunathan2016}
Raghunathan S.,  Clowes R.~G.,  Campusano L.~E.,  S{\"{o}}chting I.~K.,  Graham
  M.~J.,   Williger G.~M.,  2016, \mn@doi [MNRAS] {10.1093/mnras/stw2095}, 463,
  2640

\bibitem[\protect\citeauthoryear{Rahin \& Behar}{Rahin \&
  Behar}{2019}]{Rahin2019a}
Rahin R.,  Behar E.,  2019, \mn@doi [ApJ] {10.3847/1538-4357/ab3e34}, 885, 47

\bibitem[\protect\citeauthoryear{Ricci et~al.,}{Ricci et~al.}{2017}]{Ricci2017}
Ricci C.,  et~al., 2017, \mn@doi [ApJS] {10.3847/1538-4365/aa96ad}, 233, 17

\bibitem[\protect\citeauthoryear{Sahakyan, Israyelyan, Harutyunyan, Khachatryan
   \& Gasparyan}{Sahakyan et~al.}{2020}]{Sahakyan2020}
Sahakyan N.,  Israyelyan D.,  Harutyunyan G.,  Khachatryan M.,   Gasparyan S.,
  2020, \mn@doi [MNRAS] {10.1093/mnras/staa2477}, 498, 2594

\bibitem[\protect\citeauthoryear{Savage, Kim, Wakker, Keeney, Shull, Stocke  \&
  Green}{Savage et~al.}{2014}]{Savage2014}
Savage B.~D.,  Kim T.~S.,  Wakker B.~P.,  Keeney B.,  Shull J.~M.,  Stocke
  J.~T.,   Green J.~C.,  2014, \mn@doi [ApJ] {10.1088/0067-0049/212/1/8}, 212

\bibitem[\protect\citeauthoryear{Schady}{Schady}{2017}]{Schady2017}
Schady P.,  2017, Royal Society open science, 4, 170304

\bibitem[\protect\citeauthoryear{Schady, Savaglio, Kr{\"{u}}hler, Greiner,   \&
  Rau}{Schady et~al.}{2011}]{Schady2011}
Schady P.,  Savaglio S.,  Kr{\"{u}}hler T.,  Greiner J.,    Rau A.,  2011,
  A{\&}A, 525, A113

\bibitem[\protect\citeauthoryear{Schaye, Aguirre, Kim, Theuns, Rauch  \&
  Sargent}{Schaye et~al.}{2003}]{Schaye2003}
Schaye J.,  Aguirre A.,  Kim T.,  Theuns T.,  Rauch M.,   Sargent W. L.~W.,
  2003, \mn@doi [ApJ] {10.1086/378044}, 596, 768

\bibitem[\protect\citeauthoryear{Schaye et~al.,}{Schaye
  et~al.}{2015}]{Schaye2015}
Schaye J.,  et~al., 2015, \mn@doi [MNRAS] {10.1093/mnras/stu2058}, 446, 521

\bibitem[\protect\citeauthoryear{Selsing, Fynbo, Christensen  \&
  Krogager}{Selsing et~al.}{2016}]{Selsing2016}
Selsing J.,  Fynbo J. P.~U.,  Christensen L.,   Krogager J.-K.,  2016, \mn@doi
  [A{\&}A] {10.1051/0004-6361/201527096}, 585, a87

\bibitem[\protect\citeauthoryear{Shull \& Danforth}{Shull \&
  Danforth}{2018}]{Shull2018}
Shull J.~M.,  Danforth C.~W.,  2018, ApJ, 852, L11

\bibitem[\protect\citeauthoryear{Shull, Smith  \& Danforth}{Shull
  et~al.}{2012}]{Shull2012}
Shull J.~M.,  Smith B.~D.,   Danforth C.~W.,  2012, \mn@doi [ApJ]
  {10.1088/0004-637X/759/1/23}, 759, 23

\bibitem[\protect\citeauthoryear{Shull, Danforth  \& Tilton}{Shull
  et~al.}{2014}]{Shull2014}
Shull J.~M.,  Danforth C.~W.,   Tilton E.~M.,  2014, \mn@doi [ApJ]
  {10.1088/0004-637X/796/1/49}, 796

\bibitem[\protect\citeauthoryear{Sikora, Begelman  \& Rees}{Sikora
  et~al.}{1994}]{Sikora1994}
Sikora M.,  Begelman M.,   Rees M.,  1994, ApJ, 421, 153

\bibitem[\protect\citeauthoryear{Simcoe, Sargent  \& Rauch}{Simcoe
  et~al.}{2004}]{Simcoe2004}
Simcoe R.~A.,  Sargent W. L.~W.,   Rauch M.,  2004, \mn@doi [ApJ]
  {10.1086/382777}, 606, 92

\bibitem[\protect\citeauthoryear{Starling, Willingale, Tanvir, Scott, Wiersema,
  O'Brien, Levan  \& Stewart}{Starling et~al.}{2013a}]{Starling2013a}
Starling R.~L.,  Willingale R.,  Tanvir N.~R.,  Scott A.~E.,  Wiersema K.,
  O'Brien P.~T.,  Levan A.~J.,   Stewart G.~C.,  2013a, \mn@doi [MNRAS]
  {10.1093/mnras/stt400}, 431, 3159

\bibitem[\protect\citeauthoryear{Starling, Willingale, Tanvir, Scott, Wiersema,
  O'Brien, Levan  \& Stewart}{Starling et~al.}{2013b}]{Starling2013}
Starling R.~L.,  Willingale R.,  Tanvir N.~R.,  Scott A.~E.,  Wiersema K.,
  O'Brien P.~T.,  Levan A.~J.,   Stewart G.~C.,  2013b, \mn@doi [MNRAS]
  {10.1093/mnras/stt400}, 431, 3159

\bibitem[\protect\citeauthoryear{Str{\"{u}}der et~al.,}{Str{\"{u}}der
  et~al.}{2001}]{Struder2001}
Str{\"{u}}der L.,  et~al., 2001, A{\&}A, 365, L18

\bibitem[\protect\citeauthoryear{Tanimura, Aghanim, Bonjean, Malavasi  \&
  Douspis}{Tanimura et~al.}{2020a}]{Tanimura2020b}
Tanimura H.,  Aghanim N.,  Bonjean V.,  Malavasi N.,   Douspis M.,  2020a,
  \mn@doi [A{\&}A] {10.1051/0004-6361/201937158}, 637

\bibitem[\protect\citeauthoryear{Tanimura, Aghanim, Kolodzig, Douspis  \&
  Malavasi}{Tanimura et~al.}{2020b}]{Tanimura2020}
Tanimura H.,  Aghanim N.,  Kolodzig A.,  Douspis M.,   Malavasi N.,  2020b,
  \mn@doi [A{\&}A] {10.1051/0004-6361/202038521}, 643, 1

\bibitem[\protect\citeauthoryear{Tumlinson et~al.,}{Tumlinson
  et~al.}{2011}]{Tumlinson2011}
Tumlinson J.,  et~al., 2011, \mn@doi [Science] {10.1126/science.1209840}, 334,
  948

\bibitem[\protect\citeauthoryear{Tumlinson et~al.,}{Tumlinson
  et~al.}{2013}]{Tumlinson2013}
Tumlinson J.,  et~al., 2013, \mn@doi [ApJ] {10.1088/0004-637X/777/1/59}, 777

\bibitem[\protect\citeauthoryear{Urry \& Padovani}{Urry \&
  Padovani}{1995}]{Urry1995}
Urry C.,  Padovani P.,  1995, PASP, 107, 803

\bibitem[\protect\citeauthoryear{Wang}{Wang}{2013}]{Wang2013}
Wang J.,  2013, \mn@doi [ApJ] {10.1088/0004-637X/776/2/96}, 776, 96

\bibitem[\protect\citeauthoryear{Watson}{Watson}{2011}]{Watson2011}
Watson D.,  2011, \mn@doi [A{\&}A] {10.1051/0004-6361/201117120}, 533, 16

\bibitem[\protect\citeauthoryear{Watson \& Jakobsson}{Watson \&
  Jakobsson}{2012}]{Watson2012}
Watson D.,  Jakobsson P.,  2012, \mn@doi [ApJ] {10.1088/0004-637X/754/2/89},
  754

\bibitem[\protect\citeauthoryear{Watson, Hjorth, Fynbo, Jakobsson, Foley,
  Sollerman  \& Wijers}{Watson et~al.}{2007}]{Watson2007}
Watson D.,  Hjorth J.,  Fynbo J. P.~U.,  Jakobsson P.,  Foley S.,  Sollerman
  J.,   Wijers R. A. M.~J.,  2007, \mn@doi [ApJ] {10.1086/518310}, 660, L101

\bibitem[\protect\citeauthoryear{Watson et~al.,}{Watson
  et~al.}{2013}]{Watson2013}
Watson D.,  et~al., 2013, \mn@doi [ApJ] {10.1088/0004-637X/768/1/23}, 768, 23

\bibitem[\protect\citeauthoryear{Werk, Prochaska, Thom, Tumlinson, Tripp,
  O'Meara  \& Peeples}{Werk et~al.}{2013}]{Werk2013}
Werk J.~K.,  Prochaska J.~X.,  Thom C.,  Tumlinson J.,  Tripp T.~M.,  O'Meara
  J.~M.,   Peeples M.~S.,  2013, \mn@doi [ApJ] {10.1088/0067-0049/204/2/17},
  204

\bibitem[\protect\citeauthoryear{Willingale, Starling, Beardmore, Tanvir  \&
  O'brien}{Willingale et~al.}{2013}]{Willingale2013}
Willingale R.,  Starling R.~L.,  Beardmore A.~P.,  Tanvir N.~R.,   O'brien
  P.~T.,  2013, \mn@doi [MNRAS] {10.1093/mnras/stt175}, 431, 394

\bibitem[\protect\citeauthoryear{Wilms, Allen  \& McCray}{Wilms
  et~al.}{2000}]{Wilms2000}
Wilms J.,  Allen A.,   McCray R.,  2000, \mn@doi [ApJ] {10.1086/317016}, 542,
  914

\bibitem[\protect\citeauthoryear{Wolf}{Wolf}{1987}]{Wolf1987}
Wolf E.,  1987, \mn@doi [Nature] {10.1038/326363a0}, 326, 363

\bibitem[\protect\citeauthoryear{York et~al.,}{York et~al.}{2000}]{York2000}
York D.~G.,  et~al., 2000, AJ, 120, 1579

\bibitem[\protect\citeauthoryear{Yuan, Fabian, Worsley  \& McMahon}{Yuan
  et~al.}{2006}]{Yuan2006}
Yuan W.,  Fabian A.~C.,  Worsley M.~A.,   McMahon R.~G.,  2006, \mn@doi [MNRAS]
  {10.1111/j.1365-2966.2006.10175.x}, 368, 985

\bibitem[\protect\citeauthoryear{Zhang, Gupta, Gaur, Wiita, An, Lu, Fan  \&
  Xu}{Zhang et~al.}{2021}]{Zhang2021}
Zhang Z.,  Gupta A.~C.,  Gaur H.,  Wiita P.~J.,  An T.,  Lu Y.,  Fan S.,   Xu
  H.,  2021, https://arxiv.org/abs/2101.05977

\makeatother
\end{thebibliography}

\section*{SUPPORTING INFORMATION}



\appendix
\clearpage
\section{Tables reporting fit results for the main blazar sample and sub-samples}

\begin{table*}
    \renewcommand{\arraystretch}{1.3}
	\centering
	\caption{$\textit{Swift}$ blazar sample. For each blazar, the columns give the name, type, redshift, IGM and continuum  best cstat fitted parameter results: $\mathit{N}\textsc{hxigm}$, $[X/$H$]$, temperature, log parabolic power law and $\beta$, or broken power law low energy pewer law (PO1), Energy break ($E_b$) and high energy PO2, Cstat$/$dof}
	\label{tab:Table_fullsampleresults}
	\begin{tabular}{cc@{\hspace*{0.7cm}} c@{\hspace*{0.5
	cm}}c@{\hspace*{0.5
	cm}}c@{\hspace*{0.5cm}}c@{\hspace*{0.5cm}}c@{\hspace*{0.5
	cm}}c@{\hspace*{0.5cm}}c@{\hspace*{0.5cm}}c}
		\hline

		Blazar & Type & $z$ & log$(\frac{\mathit{N}\textsc{hxigm}}{\textrm{cm}^{-2}})$ & $[X/\textrm{H}]$ & log$(\frac{\textrm{T}}{ K})$ & PO or PO1 & $\beta$ or $E_b$ & PO2 & Cstat$/$dof \\
		
		\hline
		Mrk 501 &BL Lac & 0.03 & $19.78^{+2.07}_{-0.78}$ &$-0.08^{+0.06}_{-0.07}$  & $6.77^{+0.41}_{-0.17}$ & $1.92^{+0.06}_{-0.05}$& $0.23^{+0.14}_{-0.16}$  &  & 474.61/525 
		\\
		PKS 0521-365 &BL Lac & 0.06 & $21.82^{+0.40}_{-1.82}$ &$-0.70^{+0.70}_{-0.19}$ & $7.08^{+0.81}_{-0.05}$ & $1.47^{+0.03}_{-0.03}$ & $0.15^{+0.05}_{-0.08}$  &  & 752.59/759
		\\
		BL Lac &BL Lac & 0.07 & $20.81^{+1.07}_{-0.77}$ &$-2.99^{+2.10}_{-0.05}$ & $4.98^{+0.12}_{-0.01}$ & $1.39^{+0.05}_{-0.06}$ & $2.01^{+0.17}_{-0.23}$  & $1.73^{+0.03}_{-0.04}$ & 874.68/917
		\\
		1ES 0347-121 &BL Lac & 0.18 & $21.30^{+0.90}_{-0.70}$ &$-1.60^{+0.90}_{-0.70}$ & $5.72^{+2.16}_{-0.56}$ & $1.97^{+0.18}_{-0.11}$ & $0.05^{+0.15}_{-0.21}$  &  & 375.95/443
		\\
		1ES 1216+304 &BL Lac & 0.18 & $22.25^{+0.04}_{-0.79}$ &$-2.92^{+1.25}_{-0.04}$ & $5.46^{+1.50}_{-0.05}$ & $1.54^{+0.06}_{-0.02}$ & $0.65^{+0.03}_{-0.03}$  & & 681.87/724
		\\
		4C +34.47 &FSRQ & 0.21 & $20.70^{+1.19}_{-0.70}$ &$-1.01^{+0.01}_{-0.91}$ & $6.57^{+1.20}_{-0.93}$ & $1.60^{+0.10}_{-0.06}$ & $-0.15^{+0.08}_{-0.14}$  &  & 531.50/539
		\\
		1ES 0120+340 &BL Lac & 0.27 & $21.89^{+0.34}_{-0.74}$ &$-1.00^{+0.25}_{-0.75}$ & $7.25^{+0.69}_{-0.23}$ & $1.71^{+0.03}_{-0.03}$ & $0.24^{+0.06}_{-0.06}$  &  & 589.57/698
		\\
		S50716+714 &FSRQ & 0.31 & $21.18^{+0.15}_{-0.01}$ &$-2.00^{+0.20}_{-0.00}$ & $5.11^{+0.15}_{-0.01}$ & $1.87^{+0.03}_{-0.10}$ & $0.13^{+0.11}_{-0.04}$  &  & 824.04/822
		\\
		PKS 1510-089 &FSRQ & 0.36 & $21.87^{+0.24}_{-0.12}$ &$-1.82^{+0.64}_{-0.88}$ & $5.14^{+0.42}_{-0.11}$ & $2.10^{+0.20}_{-0.47}$ & $0.98^{+0.10}_{-0.06}$  & $1.33^{+0.02}_{-0.03}$ & 868.77/922
		\\
		J1031+5053&BL Lac & 0.36 & $21.63^{+0.54}_{-0.91}$ &$-1.13^{+0.06}_{-0.08}$ & $5.00^{+0.18}_{-0.05}$ & $3.95^{+0.01}_{-1.42}$ & $0.81^{+0.11}_{-0.11}$  & $2.23^{+0.04}_{-0.11}$ & 463.45/525
		\\
		3C 279&FSRQ & 0.54 & $20.70^{+1.37}_{-0.04}$ &$-2.99^{+0.85}_{-0.01}$ & $5.20^{+0.27}_{-0.04}$ & $1.49^{+0.01}_{-0.02}$ & $2.64^{+0.19}_{-0.26}$  & $1.66^{+0.03}_{-0.03}$ & 992.21/933
		\\
		1ES 1641+399&FSRQ & 0.59 & $22.23^{+0.27}_{-0.33}$ &$-0.86^{+0.13}_{-0.96}$ & $7.60^{+0.37}_{-0.21}$ & $1.59^{+0.04}_{-0.03}$ & $2.79^{+0.62}_{-0.37}$  & $1.74^{+0.81}_{-0.47}$ & 681.57/739
		\\
		PKS 0637-752&FSRQ & 0.64 & $21.88^{+0.50}_{-0.97}$ &$-1.50^{+0.74}_{-1.03}$ & $6.40^{+1.50}_{-0.87}$ & $2.08^{+0.31}_{-0.24}$ & $1.10^{+0.33}_{-0.29}$  & $1.63^{+0.09}_{-0.08}$ & 535.08/629
		\\
		PKS 0903-57&FSRQ & 0.70 & $21.80^{+0.54}_{-0.34}$ &$-2.70^{+0.90}_{-0.30}$ & $5.20^{+2.15}_{-0.26}$ & $0.54^{+0.20}_{-0.18}$ & $1.09^{+0.28}_{-0.24}$  &  & 371.01/433
		\\
		3C 454.3&FSRQ & 0.86 & $21.85^{+0.30}_{-0.24}$ &$-2.92^{+0.28}_{-0.08}$ & $5.03^{+0.22}_{-0.03}$ & $1.29^{+0.03}_{-0.03}$ & $2.30^{+0.18}_{-0.03}$  & $1.54^{+0.03}_{-0.03}$ & 935.16/926
		\\
		PKS 1441+25&FSRQ & 0.94 & $21.77^{+0.33}_{-1.17}$ &$-1.80^{+0.96}_{-0.43}$ & $5.18^{+2.62}_{-0.11}$ & $2.15^{+0.26}_{-0.16}$ & $-0.03^{+0.29}_{-0.33}$  &  & 360.76/412
		\\
		4C +04.42&FSRQ & 0.97 & $21.23^{+0.68}_{-0.75}$ &$-0.90^{+0.15}_{-0.95}$ & $5.08^{+2.25}_{-0.07}$ & $1.53^{+0.15}_{-0.24}$ & $-0.30^{+0.30}_{-0.15}$  &  & 505.29/537
		\\
		PKS 0208+512&FSRQ & 1.00 & $21.84^{+0.23}_{-0.11}$ &$-1.13^{+0.07}_{-0.37}$ & $6.65^{+0.15}_{-0.17}$ & $1.60^{+0.02}_{-0.06}$ & $3.28^{+0.35}_{-0.12}$  & $1.84^{+0.02}_{-0.06}$ & 575.64/668
		\\
		PKS 1240-294&FSRQ & 1.13 & $21.93^{+0.22}_{-0.98}$ &$-0.99^{+0.27}_{-0.90}$ & $5.53^{+1.93}_{-0.34}$ & $1.90^{+0.23}_{-0.75}$ & $-0.31^{+0.84}_{-0.26}$  &  & 307.68/374
		\\
		PKS 1127-14&FSRQ & 1.18 & $21.89^{+0.21}_{-0.98}$ &$-1.51^{+0.75}_{-0.79}$ & $5.00^{+0.16}_{-0.18}$ & $2.37^{+1.59}_{-0.75}$ & $0.59^{+0.27}_{-0.25}$  & $1.41^{+0.05}_{-0.06}$ & 649.40/702
		\\
		NRAO 140&FSRQ & 1.26 & $22.19^{+0.15}_{-2.19}$ &$-1.22^{+0.50}_{-0.78}$ & $7.23^{+0.70}_{-0.80}$ & $1.11^{+0.12}_{-0.08}$ & $0.36^{+0.09}_{-0.17}$  &  & 724.79/696
		\\
		OS 319&FSRQ & 1.40 & $21.59^{+0.51}_{-0.65}$ &$-0.81^{+0.02}_{-0.40}$ & $5.87^{+0.42}_{-0.08}$ & $1.86^{+0.11}_{-0.41}$ & $0.20^{+0.31}_{-0.14}$  &  & 366.43/384
		\\
		PKS 2223-05&FSRQ & 1.40 & $22.31^{+0.18}_{-1.13}$ &$-0.85^{+0.01}_{-1.30}$ & $7.27^{+0.61}_{-1.39}$ & $1.42^{+0.19}_{-0.09}$ & $0.28^{+0.13}_{-0.32}$  &  & 433.27/484
		\\
		PKS 2052-47&FSRQ & 1.49 & $22.26^{+0.17}_{-1.30}$ &$-1.47^{+0.73}_{-0.93}$ & $7.20^{+0.74}_{-0.98}$ & $1.33^{+0.13}_{-0.11}$ & $0.20^{+0.18}_{-0.21}$  &  & 374.12/451
		\\
		4C 38.41&FSRQ & 1.81 & $21.94^{+0.27}_{-0.96}$ &$-0.76^{+0.04}_{-1.20}$ & $7.22^{+0.16}_{-0.45}$ & $1.37^{+0.03}_{-0.03}$ & $0.17^{+0.04}_{-0.06}$  &  & 806.01/804
		\\
		PKS 2134+004&FSRQ & 1.93 & $22.37^{+0.05}_{-1.32}$ &$-1.59^{+0.84}_{-0.64}$ & $7.32^{+0.07}_{-1.64}$ & $1.45^{+0.16}_{-0.12}$ & $0.25^{+0.16}_{-0.27}$  &  & 886.53/418
		\\
		PKS 0528+134&FSRQ & 2.06 & $22.32^{+0.08}_{-0.64}$ &$-1.60^{+0.47}_{-0.80}$ & $6.92^{+0.25}_{-1.43}$ & $0.88^{+0.33}_{-0.27}$ & $1.59^{+0.28}_{-0.14}$  & $1.52^{+0.08}_{-0.07}$ & 728.89/760
		\\
		1ES 0836+710&FSRQ & 2.17 & $22.03^{+0.30}_{-0.11}$ &$-2.38^{+1.26}_{-0.12}$ & $5.18^{+1.80}_{-0.30}$ & $1.15^{+0.03}_{-0.06}$ & $2.15^{+0.15}_{-0.46}$  & $1.35^{+0.03}_{-0.04}$ & 972.76/912
		\\
		PKS 2149+306&FSRQ & 2.35 & $22.11^{+0.12}_{-0.21}$ &$-1.08^{+0.34}_{-0.39}$ & $6.40^{+0.37}_{-0.49}$ & $2.13^{+0.42}_{-0.66}$ & $1.02^{+0.15}_{-0.12}$  & $1.27^{+0.03}_{-0.06}$ & 858.87/825
		\\
		J1656-3302 & FSRQ & 2.40 & $22.35^{+0.12}_{-0.03}$ &$-0.72^{+0.02}_{-0.98}$ & $7.19^{+0.65}_{-0.42}$ & $0.12^{+0.46}_{-0.02}$ & $1.50^{+0.62}_{-0.26}$  & $1.36^{+0.13}_{-0.33}$ & 438.94/528
		\\
		PKS 1830-211* & FSRQ & 2.50 & $22.20^{+0.10}_{-0.81}$ &$-0.99^{+0.32}_{-0.55}$ & $6.88^{+0.93}_{-1.84}$ & $0.64^{+0.29}_{-0.23}$ & $0.46^{+0.18}_{-0.25}$  & & 784.65/858
		\\
		TXS0222+185 & FSRQ & 2.69 & $22.28^{+0.14}_{-0.96}$ &$-1.59^{+0.64}_{-0.81}$ & $7.00^{+0.88}_{-1.61}$ & $0.99^{+0.18}_{-0.15}$ & $0.36^{+0.15}_{-0.15}$  & & 699.61/676
		\\
		PKS 0834-20 & FSRQ & 2.75 & $22.30^{+0.22}_{-0.74}$ &$-0.95^{+0.22}_{-0.91}$ & $7.03^{+0.73}_{-0.55}$ & $0.88^{+0.40}_{-0.45}$ & $0.54^{+0.47}_{-0.44}$  & & 320.97/395
		\\
		TXS0800+618 & FSRQ & 3.03 & $22.37^{+0.06}_{-1.22}$ &$-1.85^{+0.83}_{-0.54}$ & $6.40^{+1.30}_{-1.98}$ & $1.77^{+0.34}_{-0.80}$ & $-0.61^{+0.87}_{-0.36}$  & & 297.99/354
		\\
		PKS 0537-286 & FSRQ & 3.10 & $22.21^{+0.07}_{-1.02}$ &$-2.99^{+2.21}_{-0.15}$ & $5.13^{+2.65}_{-0.21}$ & $1.16^{+0.13}_{-0.12}$ & $0.04^{+0.16}_{-0.14}$  & & 620.60/681
		\\
		PKS 2126-158 & FSRQ & 3.27 & $22.30^{+0.08}_{-1.38}$ &$-2.30^{+1.54}_{-0.40}$ & $5.51^{+1.60}_{-1.07}$ & $1.10^{+0.15}_{-0.15}$ & $0.34^{+0.18}_{-0.15}$  & & 732.26/721
		\\
		S50014+81 & FSRQ & 3.37 & $21.70^{+0.76}_{-0.13}$ &$-2.98^{+0.94}_{-0.00}$ & $6.82^{+1.02}_{-1.62}$ & $1.09^{+0.15}_{-0.11}$ & $0.44^{+0.15}_{-0.18}$  & & 492.23/582
		\\
		J064632+445116 & FSRQ & 3.39 & $22.45^{+0.09}_{-0.32}$ &$-1.64^{+0.73}_{-0.66}$ & $7.41^{+0.56}_{-1.35}$ & $1.44^{+0.18}_{-0.17}$ & $5.10^{+2.82}_{-1.20}$  & $3.07^{+0.61}_{-1.65}$& 289.92/309
		\\
		J013126-100931 & FSRQ & 3.51 & $22.22^{+0.16}_{-0.69}$ &$-1.37^{+0.34}_{-0.63}$ & $5.06^{+2.53}_{-0.89}$ & $1.24^{+0.28}_{-0.59}$ & $0.01^{+0.65}_{-0.32}$  & & 256.98/326
		\\
		B3 1428+422 & FSRQ & 4.70 & $22.51^{+0.01}_{-0.87}$ &$-2.80^{+1.76}_{-0.09}$ & $5.33^{+2.44}_{-1.09}$ & $1.76^{+0.34}_{-0.80}$ & $-0.63^{+1.04}_{-0.40}$  & & 249.06/286
		\\
	\hline
	\end{tabular}
	$^* \textrm{Intervening galaxy at}\/\ z = 0.89\/\ \textrm{with}\/\ \mathit{N}\textsc{hx} = 1.94 \times 10^{22}\/\ \textrm{included in fitting using}\/\ \textsc{ztbabs}$
\end{table*}

\begin{table*}
    \renewcommand{\arraystretch}{1.3}
	\centering
	\caption{\textsc{Swift} 2SXPS Catalogue sub-sample for individual observation comparison with co-added spectra results. For each blazar, the columns give Observation ID, redshift, count rate/mean count rate, fitted IGM $\mathit{N}\textsc{hxigm}$ and log-parabolic power law/ mean power law.}
	\label{tab:individual_blazar sub-sample_results}
	\begin{tabular}{cc@{\hspace*{0.7cm}}c@{\hspace*{0.7
	cm}}c@{\hspace*{0.7cm}}c@{\hspace*{0.5cm}}c@{\hspace*{0.5
	cm}}c@{\hspace*{0.2cm}}c}
		\hline

		Blazar & Observation ID & $z$ & $\frac{\mathrm{count\/\ rate}}{\mathrm{mean\/\ rate}}$ & log$(\frac{\mathit{N}\textsc{hxigm}}{\textrm{cm}^{-2}})$  & $\frac{\textrm{logpar\/\ power-law}}{\textrm{mean\/\ power-law}}$  \\
		
		\hline
		3C 454.3 & mean & 0.86 & 1.00 & $21.67^{+0.41}_{-1.19}$ & 1.00 
		\\
		& 00035030001 & 0.86 & 2.65 & $22.06^{+0.22}_{-0.25}$ & 0.96
		\\
		 & 00030024001 & 0.86 & 4.69 &  $21.76^{+0.47}_{-0.98}$ & 0.91
		\\
		 
	     & 00030024002 & 0.86 & 2.96 &  $21.86^{+0.23}_{-1.16}$ & 1.05
		\\
	    & 00035030005 & 0.86 & 3.53 &  $22.30^{+0.17}_{-1.07}$ & 0.98
		\\
		PKS 2149-306& mean & 2.35 & 1.00 &  $21.90^{+0.28}_{-1.12}$ & 1.00
		\\
		& 00031404001 & 2.35 & 0.73 &  $21.90^{+0.53}_{-0.76}$ & 1.11
		\\
		 & 00031404015 & 2.35 & 1.15 &  $22.03^{+0.13}_{-0.06}$ & 0.93
		\\
		 
	     & 00035242001 & 2.35 & 0.93 &  $22.30^{+0.20}_{-0.92}$ & 1.68
		\\
	    & 00031404013 & 2.35 & 1.30 &  $22.36^{+0.11}_{-1.16}$ & 0.89
		\\
		PKS 2126-158 & mean & 3.26 & 1.00 &  $22.30^{+0.08}_{-1.38}$ & 1.00
		\\
		& 00036356001 & 3.26 & 0.92 &  $22.37^{+0.07}_{-0.27}$ & 1.45
		\\
		 & 00036356003 & 3.26 & 0.92 &  $22.37^{+0.01}_{-1.30}$ & 1.19
		\\
		 
	     & 00036356004 & 3.26 & 0.88 &  $22.44^{+0.00}_{-1.32}$ & 1.11
		\\
	    & 00036356002 & 3.26 & 0.96 &  $21.70^{+0.61}_{-0.70}$ & 1.17
		\\
		 PKS 0537-286 & mean & 3.10 & 1.00 &  $22.21^{+0.07}_{-1.02}$ & 1.00
		\\
		& 00035240001 & 3.10 & 0.93 &  $22.15^{+0.33}_{-0.33}$ & 1.04
		\\
		 & 00035240002 & 3.10 & 0.95 &  $22.34^{+0.17}_{-0.60}$ & 0.91
		\\
		 
	     & 00036783001 & 3.10  & 1.25 &  $22.31^{+0.18}_{-0.67}$ & 1.10
		\\
	    & 00030816005 & 3.10 & 1.15 &  $22.00^{+0.46}_{-0.40}$ & 1.22
		\\ 
		1ES 0836+710 & mean & 2.17 & 1.00 &  $21.85^{+0.36}_{-0.98}$ & 1.00
		\\
		& 00035385001 & 2.17 & 1.16 &  $21.90^{+0.39}_{-0.76}$ & 1.16
		\\
		 & 00036376012& 2.17 & 0.93 &  $22.32^{+0.12}_{-0.10}$ & 1.66
		\\
		 
	     & 00080399002 & 2.17 & 1.30 &  $22.27^{+0.09}_{-1.31}$ & 0.95
		\\
	    & 00036376005 & 2.17 & 0.83 &  $22.04^{+0.39}_{-1.26}$ & 1.21
		\\ 
		TXS 0222+185 & mean & 2.69 & 1.00 &  $22.28^{+0.13}_{-0.96}$ & 1.00
		
		\\
		& 00080243001 & 2.69 & 1.15 &  $22.26^{+0.09}_{-0.04}$ & 1.38
		\\
		 &00080243002 & 2.69 & 0.91 &  $22.40^{+0.08}_{-0.14}$ & 1.18
		\\
		 
	     & 00030794003 & 2.69 & 0.95 &  $22.34^{+0.14}_{-0.85}$ & 1.33
		\\
	    & 00030794002 & 2.69 & 0.99 &  $22.27^{+0.16}_{-0.97}$ & 1.01
		\\ 
		4C 38.41 & mean & 1.81 & 1.00 &  $21.94^{+0.22}_{-0.76}$ & 1.00
		\\
		& 00036389050 & 1.81 & 1.23 &  $22.29^{+0.21}_{-1.29}$ & 1.20
		\\
		 &00036389059 & 1.81 & 2.14 &  $21.89^{+0.22}_{-0.64}$ & 0.95
		\\
		 
	     & 00032894004 & 1.81 & 1.84 &  $21.99^{+0.31}_{-0.33}$ & 1.37
		\\
	    & 00036389052 & 1.81 & 1.16 &  $22.35^{+0.12}_{-0.92}$ & 0.97
		\\ 
		PKS 0528+134 & mean & 2.06 & 1.00 &  $22.23^{+0.14}_{-0.86}$ & 1.00
		\\
		& 00035384002 & 2.06 & 1.63 &  $22.35^{+0.05}_{-1.05}$ & 0.44
		\\
		 &00035384003 & 2.06 & 2.12 &  $22.20^{+0.28}_{-1.60}$ & 1.80
		\\
		 
	     & 00035384005 & 2.06 & 2.16 &  $22.30^{+0.14}_{-1.12}$ & 1.30
		\\
	    & 00035384006 & 2.06 & 2.58 &  $22.26^{+0.15}_{-0.17}$ & 2.81
		\\ 
		
		\hline
	\end{tabular}
\end{table*}

\begin{table*}
    \renewcommand{\arraystretch}{1.3}
	\centering
	\caption{\textsc{XMM-Newton}, 0.3 - 10 keV and 0.16 - 13 keV, and \textsc{Swift} sub-sample IGM column density results. For each blazar, the columns give Blazar name, redshift, fitted IGM $\mathit{N}\textsc{hxigm}$ for $\textit{Swift}, \textit{XMM-Newton}$ 0.3-10 keV and 0.16-13 keV respectively.}
	\label{tab:XMM_blazar sub-sample_results}
	\begin{tabular}{cc@{\hspace*{0.7cm}}c@{\hspace*{0.7
	cm}}c@{\hspace*{0.7cm}}c@{\hspace*{0.5cm}}c@{\hspace*{0.5
	cm}}c@{\hspace*{0.2cm}}c}
		\hline
		
		 & & $\textit{Swift}$ 0.3-10 keV & $\textit{XMM-Newton}$ 0.3-10 keV & $\textit{XMM-Newton}$ 0.16-13 keV
		 \\
	
		Blazar &  $z$ &  log$(\frac{\mathit{N}\textsc{hxigm}}{\textrm{cm}^{-2}})$ & log$(\frac{\mathit{N}\textsc{hxigm}}{\textrm{cm}^{-2}})$  & log$(\frac{\mathit{N}\textsc{hxigm}}{\textrm{cm}^{-2}})$  \\
		
		\hline
		3C 454.3  & 0.86 & $21.67^{+0.41}_{-1.19}$ & $21.18^{+0.95}_{-0.57}$ & $21.52^{+0.04}_{-0.12}$
		
		\\
		 PKS2149-306 & 2.35 & $21.90^{+0.28}_{-1.12}$ &  $22.18^{+0.05}_{-0.68}$ & $22.09^{+0.19}_{-0.91}$
		\\
		 
	      PKS2126-158 & 3.26 & $22.30^{+0.08}_{-1.38}$ &  $22.10^{+0.13}_{-1.40}$ & $22.11^{+0.11}_{-1.62}$
		\\
	     PKS0537-286 & 3.10 & $22.21^{+0.07}_{-1.02}$ &  $22.35^{+0.02}_{-0.37}$ & $22.26^{+0.09}_{-0.39}$
		\\
        1ES0836+710 & 2.17 & $21.85^{+0.36}_{-0.98}$ &  $22.23^{+0.17}_{-1.19}$ & $21.64^{+0.37}_{-0.87}$
		\\
		 
	      TXS0222+185 & 2.69 & $22.28^{+0.13}_{-0.96}$ &  $22.44^{+0.04}_{-0.08}$ & $22.18^{+0.04}_{-0.09}$
		\\
	     PKS0528+134& 2.06 & $22.23^{+0.14}_{-0.86}$ &  $22.34^{+0.07}_{-0.12}$ & $22.31^{+0.08}_{-0.14}$
	     \\

\hline
	\end{tabular}
\end{table*}


\bsp	
\label{lastpage}
\end{document}